\documentclass[prd,preprint,preprintnumbers,nofootinbib,eqsecnum,superscriptaddress]{revtex4}

 \usepackage[dvips,final]{graphicx}
  \usepackage{amssymb}
   \usepackage{amsmath}
    \usepackage{amsfonts}
     \usepackage{epsfig}
      \usepackage{bm}

\usepackage{mathpazo}

\usepackage[section]{placeins}

\input epsf.tex
\def\desepsf(#1 width #2){\epsfxsize=#2 \epsfbox{#1}}

\newcommand{\alphas}{\ensuremath{\alpha_\mathrm{s}}}

\newcommand{\PBM}{PB}

\usepackage{multirow}
\usepackage{ctable}
\usepackage{booktabs}
\usepackage{array}
\usepackage{tabularx}
\usepackage{xcolor}
\usepackage{pstricks}
\definecolor{fred}{rgb}{0.90053, 0.00369, 0.00159}  

\newcommand{\dif}{\mathrm{d}}
\newcommand{\diff}[1]{\frac{\mathrm{d}#1}{#1}}

\begin{document}

\title{Consistent treatment of charm production 
in higher-orders at tree-level within \bm{$k_T$}-factorization approach
}


\author{Rafa{\l} Maciu{\l}a}
\email{rafal.maciula@ifj.edu.pl} \affiliation{Institute of Nuclear
Physics, Polish Academy of Sciences, Radzikowskiego 152, PL-31-342 Krak{\'o}w, Poland}

\author{Antoni Szczurek\footnote{also at University of Rzesz\'ow, PL-35-959 Rzesz\'ow, Poland}}
\email{antoni.szczurek@ifj.edu.pl} \affiliation{Institute of Nuclear
Physics, Polish Academy of Sciences, Radzikowskiego 152, PL-31-342 Krak{\'o}w, Poland}

\begin{abstract}
We discuss production of $c \bar c$-pairs within $k_T$-factorization approach
(off-shell initial partons) with unintegrated parton distribution functions (uPDFs).
We present a consistent prescription which merges the standard leading-order (LO)
$k_T$-factorization calculations for this process with tree-level next-to-leading order (NLO)
and next-to-next-to-leading order (NNLO) matrix elements.   
For the first time we include in this framework 2 $\to$ 3 and 2 $\to$ 4 
processes with extra partonic emissions for single particle
distributions as well as for correlation observables.
The use of the KMR uPDF leads to a good description of the existing charm 
($D$-meson) data already at the leading-order. On the other hand, a new Parton-Branching (PB) uPDF strongly
underestimates the same experimental data. A direct inclusion of 
the higher-orders at tree-level leads to an overestimation of the data, especially
for the KMR uPDF. This suggests a significant double-counting.
We propose a simple method how to avoid the double-counting.
Our procedure leads to a much better description of the experimental data
when including the higher-order contributions.
Then with the KMR uPDF we get similar results (both for single particle and correlation observables) as for the standard calculations of the 2 $\to$ 2 processes.
For the PB uPDF inclusion of the higher-orders considerably improves
description of the experimental data. We conclude that the LO calculation
with the KMR uPDF effectively includes the higher-orders which is not
the case for the PB uPDF.
\end{abstract}

\maketitle

\section{Introduction}

The production of heavy flavours is known to be a good example
of perturbative QCD calculations -- the quark mass sets already
a sizeable scale. Charm quark is the lightest heavy quark where
one believes in the pQCD treatment.
Leading-order collinear approach gives much too small cross section,
compared to the experimental cross section for charmed meson production.
Clearly higher-order corrections are needed. Next-to-leading order
approach was developed for inclusive variables only 
(single charm distribution).
In general, one is interested not only in single charm distributions
but also in correlation observables.
Some studies of correlation observables were done \textit{e.g.} in Refs.~\cite{Maciula:2013wg,Karpishkov:2016hnx}
within $k_T$-factorization approach.

In the present paper we will discuss both single particle distributions
(distributions in rapidity or charm transverse momentum) and
correlation observables (distribution in azimuthal angle
between $c \bar c$, invariant mass distribution, transverse momentum
of the $c\bar c$ pair, difference in rapidity between $c$ and $\bar c$).


We propose and discuss a consistent prescription which merges the standard leading-order (LO)
$k_T$-factorization calculations for this process with tree-level next-to-leading order (NLO)
and next-to-next-to-leading order (NNLO) matrix elements. The applied procedure was originally proposed in the context of $B\bar B$ pair production within the Parton-Reggeization-Approach (PRA) in Ref.~\cite{Karpishkov:2017kph}. There, the LO calculations for off-shell initial state partons were supplemented by the NLO corrections from the emission of one additional hard gluon. The consistent merging procedure constructed there can be implemented for the calculation of charm production within the $k_T$-factorization approach. In this paper,    
we basically follow the ideas presented in Ref.~\cite{Karpishkov:2017kph} but we extend those studies to the case of NNLO corrections from the emission of two additional hard gluons.

Our new scheme for the calculations provides a possibility to study 
the charm cross section differentially beyond the NLO collinear approaches, FONLL \cite{Cacciari:2012ny} and GM-VFNS \cite{Kniehl:2012ti}, commonly used as the state of the art in this context. We expect the $2 \to 4$ contributions, that are missing there, to be of the special importance for the large transverse momenta of charm particles studied at the LHC.

\section{Basic formalism}

\subsection{Cross section for charm quark and meson production}

\subsubsection{The standard calculations within the leading-order $2\to 2$ mechanism}

\begin{figure}[!h]
\centering
\begin{minipage}{0.35\textwidth}
  \centerline{\includegraphics[width=1.0\textwidth]{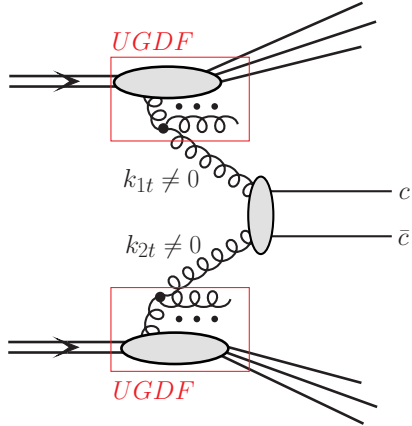}}
\end{minipage}
  \caption{
\small A diagramatic representation of the leading-order mechanism of charm production.
}
\label{fig:diagramLO}
\end{figure}

We recall the standard theoretical formalism for the calculations of the $c\bar{c}$-pair production in the framework of the $k_{T}$-factorization \cite{kTfactorization}. This approach is commonly known to be very efficient not only for inclusive particle distributions but also for studies of kinematical correlations. According to this approach, the transverse momenta $k_{t}$'s (virtualities) of both partons entering the hard process are taken into account. In the case of charm (or in general heavy) flavour production the parton-level cross section is usually calculated via the $2\to 2$ leading-order $g^*g^* \to c\bar c$ fusion mechanism of off-shell initial state gluons that is dominant processs at high energies. Emission of the initial state partons is encoded in the so-called unintegrated parton distribution functions (uPDFs). Then the hadron-level differential cross section at the tree-level for the $c \bar c$-pair production reads:
\begin{eqnarray}\label{LO_kt-factorization} 
\frac{d \sigma(p p \to c \bar c \, X)}{d y_1 d y_2 d^2p_{1,t} d^2p_{2,t}} &=&
\int \frac{d^2 k_{1,t}}{\pi} \frac{d^2 k_{2,t}}{\pi}
\frac{1}{16 \pi^2 (x_1 x_2 s)^2} \; \overline{ | {\cal M}^{\mathrm{off-shell}}_{g^* g^* \to c \bar c} |^2}
 \\  
&& \times  \; \delta^{2} \left( \vec{k}_{1,t} + \vec{k}_{2,t} 
                 - \vec{p}_{1,t} - \vec{p}_{2,t} \right) \;
{\cal F}_g(x_1,k_{1,t}^2,\mu_{F}^2) \; {\cal F}_g(x_2,k_{2,t}^2,\mu_{F}^2) \; \nonumber ,   
\end{eqnarray}
where ${\cal F}_g(x_1,k_{1,t}^2,\mu_{F}^2)$ and ${\cal F}_g(x_2,k_{2,t}^2,\mu_{F}^2)$
are the gluon uPDFs for both colliding hadrons and ${\cal M}^{\mathrm{off-shell}}_{g^* g^* \to c \bar c}$ is the off-shell matrix element for the hard subprocess.
The gluon uPDF depends on gluon longitudinal momentum fraction $x$, transverse momentum
squared $k_t^2$ of the gluons entering the hard process, and in general also on a (factorization) scale of the hard process $\mu_{F}^2$.
The extra integration is over transverse momenta of the initial
partons. Here, one keeps exact kinematics from the very beginning and additional hard dynamics coming from transverse momenta of incident partons. Explicit treatment of the transverse momenta makes the approach very efficient in studies of correlation observables. The two-dimensional Dirac delta function assures momentum conservation.
The unintegrated (transverse momentum dependent) gluon distributions must be evaluated at:
\begin{equation}
x_1 = \frac{m_{1,t}}{\sqrt{s}}\exp( y_1) 
     + \frac{m_{2,t}}{\sqrt{s}}\exp( y_2), \;\;\;\;\;\;
x_2 = \frac{m_{1,t}}{\sqrt{s}}\exp(-y_1) 
     + \frac{m_{2,t}}{\sqrt{s}}\exp(-y_2), \nonumber
\end{equation}
where $m_{i,t} = \sqrt{p_{i,t}^2 + m_c^2}$ is the quark/antiquark transverse mass. In the case of charm quark production at the LHC energies, especially in the forward rapidity region, one tests very small gluon longitudinal momentum fractions $x < 10^{-5}$ \cite{Maciula:2013wg}.  

The off-shell matrix elements are known explicitly only in the LO and only for limited types of QCD $2 \to 2$ processes (see e.g. heavy quarks \cite{Catani:1990eg}, dijet \cite{Nefedov:2013ywa}, Drell-Yan \cite{Nefedov:2012cq}). 
The calculation of higher-order corrections in the $k_T$-factorization is much more complicated than in the collinear approximation approach. Some first steps to calculate NLO corrections in the $k_{T}$-factorization framework have been tried only very recently for diphoton production \cite{Nefedov:2015ara,Nefedov:2016clr}.
There are ongoing intensive works on construction of the full NLO Monte Carlo generator for off-shell initial state partons that are expected to be finished in near future \cite{private-Hameren}. Another method for calculation of higher multiplicity final states is to supplement the QCD $2 \to 2$ processes with parton shower. For the off-shell initial state partons it was done only with the help of full hadron level Monte Carlo generator CASCADE \cite{Jung:2010si}. However, this method can be consistently used only with dedicated models of uPDFs.   

On the other hand, the popular statement is that actually in the $k_{T}$-factorization approach with $2\to 2$ tree-level off-shell matrix elements some part of real higher-order corrections can be effectively included.
This is due to possible extra emissions of soft and even hard partons encoded in the uPDFs. In this sense, when calculating the charm production cross section via the $g^* g^* \to c \bar c$ mechanism one could expect
to include some contributions related to an additional one or even two extra partonic emissions, effectively taking into account, as an example, the $gc \bar c$ and $ggc \bar c$ final states. However, the presence and a size of the extra emissions strongly depends on the internal construction of the unintegrated parton distributions. The extra emissions (from the uPDFs) are expected to be very important, especially for studies of kinematical correlations. The correlation observables, such as the azimuthal angle difference of $c$ and $\bar{c}$ or transverse momentum of the produced system are very useful for testing transverse momenta of initial partons and may be helpful in limiting uPDFs uncertainties and in understanding their evolution.

Some time ago we showed that in the case of charm production at the LHC, within the above formalism only the Kimber-Martin-Ryskin (KMR) uPDF leads to a reasonable decription of the experimental data for $D$-meson and $D\bar D$-pair production \cite{Maciula:2013wg}. This was further confirmed by other authors \cite{Karpishkov:2016hnx}. As also discussed in Ref.~\cite{Maciula:2016kkx} the $k_T$-factorization approach at leading-order with the KMR uPDF leads to results well consistent with collinear NLO approach.
The KMR uPDF is known to allow by its construction for a large contribution from the $k_{T} > \mu_{F}$ kinematical regime. Effectively, this extra emission of hard partons (gluons) from the uPDF corresponds to higher-order contributions. As reported in Ref.~\cite{Maciula:2013wg} the rest of the commonly-used models of the uPDFs from the literature is rather missing those contributions and significantly underestimates the experimental data on charm production at the LHC.
  
In the numerical calculation below, based on the standard $k_{T}$-factorization framework, we apply the KMR uPDF in its original form. The KMR distributions are calculated from the MMHT2014 \cite{Harland-Lang:2014zoa} and CT14 \cite{Dulat:2015mca} gluon PDFs. As a default set of the calculations we use 
the renormalization and factorization scales $\mu^2 = \mu_{R}^{2} =
\mu_{F}^{2} = \sum_{i=1}^{n=2} \frac{m^{2}_{iT}}{n}$ and charm quark mass 
$m_{c} = 1.5$ GeV. The uncertainties related to the choice of the collinear PDF and of the renormalization/factorization scales will be discussed when presenting numerical results.

The transition of charm quarks to open charm mesons is done in the framework of the independent parton fragmentation picture (see \textit{e.g.} Ref.~\cite{Maciula:2015kea}).
Here we follow the standard prescription, where the inclusive distributions of open charm meson are obtained through a convolution of inclusive distributions of charm quarks/antiquarks and $c \to D$ fragmentation functions:
\begin{equation}
\frac{d \sigma(pp \rightarrow D \bar{D} X)}{d y_D d^2 p_{t,D}} \approx
\int_0^1 \frac{dz}{z^2} D_{c \to D}(z)
\frac{d \sigma(pp \rightarrow c \overline{c} X)}{d y_c d^2 p_{t,c}}
\Bigg\vert_{y_c = y_D \atop p_{t,c} = p_{t,D}/z} \;,
\label{Q_to_h}
\end{equation}
where $p_{t,c} = \frac{p_{t,D}}{z}$ and $z$ is the fraction of
longitudinal momentum of charm quark $c$ carried by a meson $D$.
In the numerical calculations we take the Peterson fragmentation function \cite{Peterson:1982ak}, often used in the context of hadronization of heavy flavours. 
Then, the hadronic cross section is normalized by the relevant charm fragmentation fractions for a given type of $D$ meson \cite{Lisovyi:2015uqa}.

\subsubsection{A new scheme of the calculations with the higher-order $2\to 3$ and $2\to 4$ mechanisms}

Here we describe our proposal for an alternative scheme of the calculation of the heavy flavour cross sections within the $k_{T}$-factorization approach. The main idea is to include higher-order corrections at the level of hard matrix elements with simultaneous limiting of the corresponding contributions incorporated in the uPDFs.
The limitations of the emissions from uPDFs are consequences of merging LO, NLO and NNLO contributions. This is a direct analogy to the issue of merging
hard emissions from higher-order matrix elements with soft emissions from the parton showers \cite{Catani:2001cc}.    
Due to the lack of the full NLO and/or NNLO framework of the $k_{T}$-factorization, within the present methods this can be done only at tree-level. In the proposed scheme, we include and sum up the dominant $2\to 2$ , $2\to 3$ and even $2\to 4$ contributions to heavy quark-antiquark pair production under a special conditions introduced to avoid a possible double-counting. In this model, the higher-orders with hard extra emissions come from the higher-order matrix elements, while
only the softer extra emissions are included via the uPDF. Therefore, within this method for studies of heavy flavour production one could apply different models of uPDFs that do not include in their evolution a sufficiently hard extra emissions.

\begin{figure}[!h]
\centering
\begin{minipage}{0.35\textwidth}
  \centerline{\includegraphics[width=1.0\textwidth]{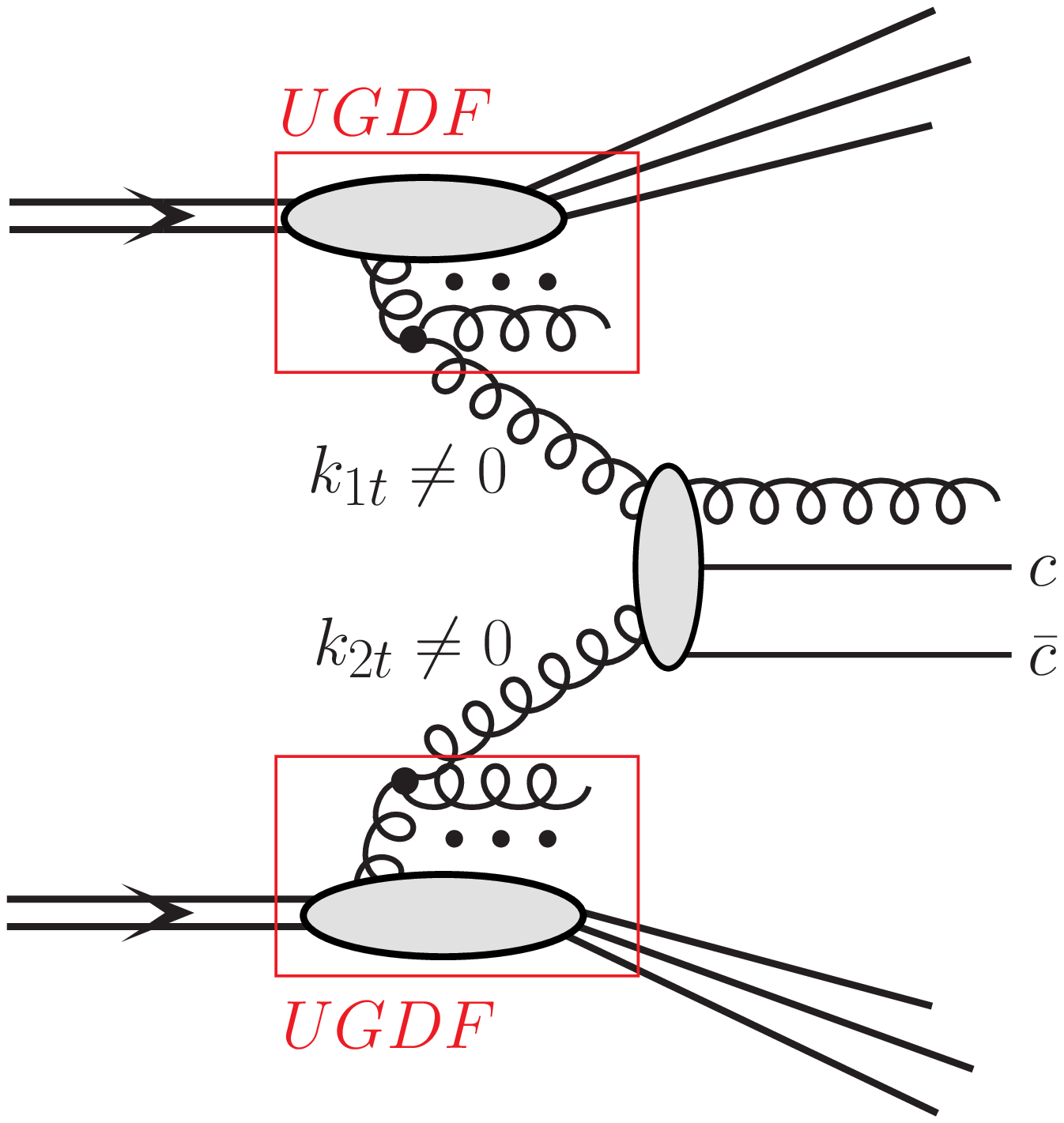}}
\end{minipage}
\begin{minipage}{0.35\textwidth}
  \centerline{\includegraphics[width=1.0\textwidth]{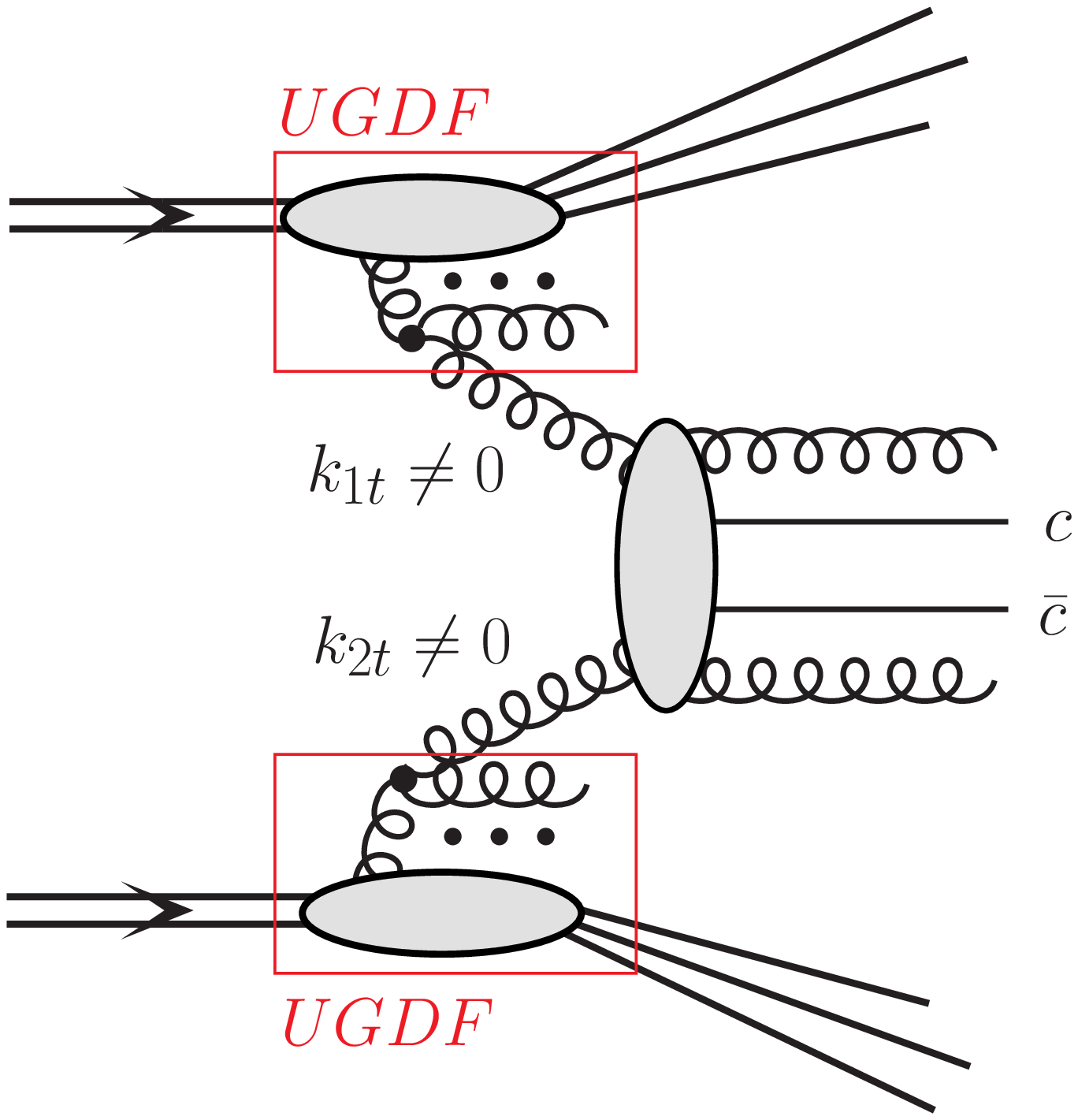}}
\end{minipage}
  \caption{
\small A diagramatic representation of the higher-order mechanisms of charm production.
}
\label{fig:diagramHO}
\end{figure}

In this model we calculate the $g^*g^* \to c \bar c$, $g^*g^* \to g c \bar c$ and $g^*g^* \to ggc \bar c$ mechanisms for off-shell initial state partons. We have check numerically, that for the LHC energy the 
channels driven by gluon-gluon fusion are the dominant ones for each of the considered reactions. For the present studies (high-energy collisions) the contributions from the quark-induced processes can be safely neglected.

The numerical calculations for the considered higher-order contributions (see Fig.~\ref{fig:diagramHO}) are performed in the framework of the $k_{T}$-factorization approach within the methods adopted in the KaTie Monte Carlo generator \cite{vanHameren:2016kkz}. The off-shell matrix elements for higher final state parton multiplicities, at the tree-level can be calculated analytically applying
well defined Feynman rules \cite{vanHameren:2012if} or recursive methods, like  generalised  BCFW recursion \cite{vanHameren:2014iua}, or numerically with the help of methods of numerical BCFW recursion 
\cite{Bury:2015dla}. The latter method was already applied for $2 \to 3$ production mechanisms in the case of $c\bar c + \mathrm{jet}$ \cite{Maciula:2016kkx} and even for $2 \to 4$ processes in the case of $c\bar c c\bar c$ \cite{vanHameren:2015wva}, four-jet \cite{Kutak:2016mik} and $c\bar c + \mathrm{2jets}$ \cite{Maciula:2017egq} final states within the KaTie enviroment.

In general, the cross secton for $pp \to g(g)c\bar c\, X$ reaction in the $k_T$-factorization approach can be written as
\begin{equation}
d \sigma_{p p \to g(g)c\bar c \; X} =
\int d x_1 \frac{d^2 k_{1t}}{\pi} d x_2 \frac{d^2 k_{2t}}{\pi}
{\cal F}_{g}(x_1,k_{1t}^2,\mu_F^2) {\cal F}_{g}(x_2,k_{2t}^2,\mu_F^2)
d {\hat \sigma}_{g^*g^* \to g(g)c\bar c}
\; .
\label{cs_formula}
\end{equation}
Then, the elementary cross section from the above can be written
somewhat formally as:
\begin{equation}
d {\hat \sigma}_{gg \to g^*g^* \to g(g)c\bar c } =
\prod_{l=1}^{n}
\frac{d^3 p_l}{(2 \pi)^3 2 E_l} 
(2 \pi)^n \delta^{n}(\sum_{l=1}^{n} p_l - k_1 - k_2) \times\frac{1}{\mathrm{flux}} \overline{|{\cal M}_{g^*g^* \to g(g)c\bar c}(k_{1},k_{2})|^2}
\; ,
\label{elementary_cs}
\end{equation}
with $n=3$ and $n=4$ for $g^*g^* \to gc\bar c$ and $g^*g^* \to ggc\bar c$, respectively, where $E_{l}$ and $p_{l}$ are energies and momenta of final state gluon(s) and charm quarks. Above only dependence of the matrix element on four-vectors of incident partons $k_1$ and $k_2$ is made explicit. In general, all four-momenta associated with partonic legs enter.
Also in this case, the matrix element takes into account that both gluons entering the hard
process are off-shell with virtualities $k_1^2 = -k_{1t}^2$ and $k_2^2 = -k_{2t}^2$.
Here, as a default choice we set the renormalization/factorization scale to be equal to the averaged sum of the transverse mass squared of the final state particles $\mu^2 = \sum_{i=1}^n \frac{m_{iT}^2}{n}$, where $m_{iT} = \sqrt{m_{i}^2+p_{iT}^2}$ with $n=3$ and $n=4$ for $2 \to 3$ and $2 \to 4$ cases, respectively.  

Calculating the minijets at tree-level requires some technical methods for regularization of the cross section. For this purpose, we follow the method adopted \textit{e.g.} in PYTHIA \cite{Sjostrand:2014zea} for the calculations of the $2\to 2$ pQCD processes with light quarks and gluons in the final states. This procedue was also applied recently in the context of $D$-meson production via unfavoured fragmentation \cite{Maciula:2017wov} or $J\!/\!\psi$-meson production in the color-evaporation model \cite{Maciula:2018bex}. Here, we introduce a special suppression factor $F_{\mathrm{sup}}(p_{T}) = p_{T}^4/(p^{2}_{T0}+p_{T}^2)^2$ for the $g^*g^* \to g c \bar c$ and $g^*g^* \to ggc \bar c$ cross sections with $p_{T}$ being the outgoing minijet transverse momentum and with $p_{T0}$ being a free parameter. This parameter could, in principle, be fitted to total charm cross section measured experimentally or calculated in the NLO/NNLO collinear calculations. Usually, the values $p_{T0} = 1 - 3$ GeV are taken in phenomenological applications. As a default set in our calculations here we use $p_{T0} = 1$ GeV. Within the referred range, we expect only a small sensitivity of the calculated charm quark distributions on the value of this parameter. The uncertainties could be visible only at very small transverse momenta of charm quark, i.e. $p_{T}^{c} < 3 - 4$ GeV. This is the region where still the leading-order mechanism should dominate. At larger charm quark transverse momenta, where higher-order contributions should play the most important role, our calculations are not sensitive to the choice of the regularization parameter. The calculated transverse momentum distributions of $D$-meson can be treated as the approximate attempt to study the differential charm cross section beyond the NLO.

Within the proposed scheme we sum together the three contributions $g^*g^* \to c \bar c$, $gc \bar c$, and $ggc \bar c$. It is known, when mixing different final state multiplicities that a problem of possible double counting appears. The double-counting effects shall also appear in the case under consideration. Their consistent treatment is not an easy task since there is a lack of well established theoretical techniques. However, very recently the consistent prescription for merging leading- and next-to-leading-order calculations for off-shell initial state partons were established \cite{Karpishkov:2017kph}. Here, for the first time a similar procedure is adopted also for the case of NNLO corrections.
According to this approach, we introduce for each of the three reactions, a set of the double-counting-exclusion (DCE) cuts.
For the three reactions, transverse momenta of (mini)jets from the uPDF are constrained to be subleading. A similar constrain was also used in the PRA studies of dijet azimuthal decorrelations \cite{Nefedov:2013ywa}.  
Therefore, the proposed DCE conditions set the following extra limitations on transverse momenta of incident partons:
\begin{enumerate}
\item $k_{T} < \mu_{F}$ for $g^*g^* \to c \bar c$, where $\mu_F$ is the factorization scale,
\item $k_{T} < p_{T}$ of the minijet for $g^*g^* \to g c \bar c$,
\item $k_{T} < p_{T}^{\mathrm{min}}$ of the two minijets for $g^*g^* \to g g c \bar c$.
\end{enumerate}
As was shown in Ref.~\cite{Karpishkov:2017kph} the above kinematic cuts shall provide a clear separation
of events that correspond to the $2\to 2$, $2\to 3$ and $2\to 4$ reactions.
The first condition excludes the possible extra hard emissions from the uPDF in the $2\to 2$ case,
that are not under full theoretical/kinematical control. It reduces this contribution rather to the leading-order collinear calculations with $c$ and $\bar c$ being the leading (mini)jets. The second condition assures that the hardest (mini)jet in the $2 \to 3$ event comes always from the hard matrix element. It removes the contributions that correspond rather to the mechanisms explicitly present in the $2\to 4$ calculations. Similarly, the third condition assures that the two hardest (mini)jets in the $2 \to 4$ event also do not originate from the uPDF.
Including one or two hardest minijets from the uPDF in the $2\to 4$ case in association with soft minijets 
from the matrix element may also lead to contributions already present in the case of the $2 \to 3$ processes.   
The additional hard emissions associated with charm determine the kinematics of the $c$ and $\bar c$ and their correlations.
In this context, having them at the level of hard matrix elements seems to be more accurate.
Within the presented framework the leading- and higher-order contributions can be consistently taken into account together
without additional double-counting subtractions 
to describe \textit{e.g.} correlation observables. 
   
\subsection{Unintegrated gluon distributions} 

\subsubsection{Kimber-Martin-Ryskin uPDF}

As a default, in the calculations below we use the leading-order Kimber-Martin-Ryskin (KMR) approach \cite{Kimber:2001sc,Watt:2003vf}
with the angular (or rapidity) ordering constraints imposed, which comes from inclusion of coherence effects in gluon emission.
According to this approach the unintegrated gluon distribution is given
by the following formula
\begin{eqnarray} \label{eq:UPDF}
  f_g(x,k_t^2,\mu^2) &\equiv& \frac{\partial}{\partial \log k_t^2}\left[\,g(x,k_t^2)\,T_g(k_t^2,\mu^2)\,\right]\nonumber \\ &=& T_g(k_t^2,\mu^2)\,\frac{\alpha_S(k_t^2)}{2\pi}\,\sum_{b }\,\int_x^1\! d z\,P_{gb}(z)\,b \left (\frac{x}{z}, k_t^2 \right).
\end{eqnarray}
This formula makes sense for $k_t > \mu_0$, where $\mu_0\sim 1$ GeV
is the minimum scale for which DGLAP evolution of the conventional
collinear gluon PDF, $g(x,\mu^2)$, is valid.

The virtual (loop) contributions may be resummed to all orders by the Sudakov form factor
\begin{equation} \label{eq:Sudakov}
  T_g (k_t^2,\mu^2) \equiv \exp \left (-\int_{k_t^2}^{\mu^2}\!\diff{\kappa_t^2}\,\frac{\alpha_S(\kappa_t^2)}{2\pi}\,\sum_{b}\,\int_0^1\!\dif{z}\;z \,P_{b g}(z) \right ),
\end{equation}
which gives the probability of evolving from a scale $k_t$ to a scale $\mu$ without parton emission.
The exponent of the gluon Sudakov form factor can be simplified using the following identity: $P_{qg}(1-z)=P_{qg}(z)$.  Then the gluon Sudakov form factor reads
\begin{equation}
  T_g(k_t^2,\mu^2) = \exp\left(-\int_{k_t^2}^{\mu^2}\!\diff{\kappa_t^2}\,\frac{\alpha_S(\kappa_t^2)}{2\pi}\,\left( \int_{0}^{1-\Delta}\!\dif{z}\;z \,P_{gg}(z) + n_F\,\int_0^1\!\dif{z}\,P_{qg}(z)\right)\right),
\end{equation}
where $n_F$ is the quark--antiquark active number of flavours into which the gluon may split. Due to the presence of the Sudakov form factor in the KMR prescription only last emission generates transverse momentum of incoming gluons. Here, the variable $\Delta = k_{t}/(k_{t} + \mu)$ introduces a restriction of the phase space for gluon emission due to the angular-ordering condition. This constraints translate into the permission for hard emissions from the uPDF, that correspond to the $k_{t} > \mu$ kinematical regime.

Taking all together, the precise expression for the unintegrated gluon distribution reads
\begin{eqnarray} \label{eq:a12}
f_g(x,k_t^2,\mu^2) 
&& = T_g(k_t^2,\mu^2)\,\frac{\alpha_S(k_t^2)}{2\pi}\, \times \nonumber \\  
&& \int_x^1\! d z \left[\sum_q P_{gq}(z)\frac{x}{z}q\left(\frac{x}{z},k_t^2\right) + P_{gg}(z)\frac{x}{z}g\left(\frac{x}{z},k_t^2\right)\Theta\left(\frac{\mu}{\mu+k_t}-z\right)\right]. \;\;\;\;\;
\end{eqnarray}
This prescription was found to be consistent with the Multi-Regge-Kinematics limit of the QCD amplitudes \cite{Karpishkov:2017kph}.

In numerical calculations below we apply different sets of the KMR gluon uPDF, obtained from different collinear PDFs, including LO, NLO, and even NNLO fits. 
As was discussed in Ref.~\cite{Martin:2009ii}, the LO KMR model together with NLO PDFs leads to gluon distributions compatible with their counterparts calculated within full NLO KMR approach. Thus, in phenomenological studies one can safely neglect effects related to the higher-order perturbative-splitting functions and concentrate only on the collinear PDF input.

\subsubsection{Parton-Branching uPDF}

The Parton Branching (PB) method, introduced in Refs.~\cite{Hautmann:2017fcj,Martinez:2018jxt}, provides an iterative solution for the evolution of both collinear and transverse momentum dependent parton distributions. Within this novel method the splitting kinematics at each branching vertex stays under full control during the QCD evolution.
Here, soft-gluon emission in the region $z\to 1$ and transverse momentum recoils in the parton branchings along the QCD cascade are taken into account simultaneously. Therefore the PB approach allows for a natural determination of the uPDFs, as the transverse momentum at every branching vertex is known. It agrees with the usual methods to solve the DGLAP equations, but provides in addition a possibility to apply angular ordering instead of the standard ordering in virtuality.

Within the PB method, a soft-gluon resolution scale parameter $z_M$ is introduced into the QCD evolution equations that distinguish between non-resolvable and resolvable emissions. These two types of emissions are further treated with the help of the Sudakov form factors 
\begin{equation}
\label{sud-def}
 \Delta_a ( z_M, \mu^2 , \mu^2_0 ) = 
\exp \left(  -  \sum_b  
\int^{\mu^2}_{\mu^2_0} 
{{d \mu^{\prime 2} } 
\over \mu^{\prime 2} } 
 \int_0^{z_M} dz \  z 
\ P_{ba}^{(R)}\left(\alphas , 
 z \right) 
\right) 
  \;\; ,   
\end{equation}
and with the help of resolvable splitting probabilities $P_{ba}^{(R)} (\alphas,z)$, respectively.
Here $a , b$ are flavor indices, 
$\alphas$ is the strong coupling at a scale being a function of ${\mu}^{\prime 2}$,   
$z$ is 
the longitudinal momentum 
splitting variable, and   
$z_M < 1 $ is the soft-gluon resolution parameter.
Then, by connecting the evolution variable $\mu$ in the splitting process $b \to a c$
with the angle $\Theta$ of the momentum of particle $c$ with respect to the beam direction, 
the known angular ordering relation $\mu = | q_{t,c} | / (1 - z)$ is obtained, that ensures quantum coherence of softly radiated partons.

The \PBM\ evolution  equations with angular ordering condition for unintegrated parton densities    
$ {\cal F}_a ( x , k_{t} , \mu^2) $ 
are given 
by~\cite{Hautmann:2017fcj}   
\begin{eqnarray}
\label{evoleqforA}
   { {\cal F}}_a(x,k_t, \mu^2) 
 &=&  
 \Delta_a (  \mu^2  ) \ 
 { {\cal F}}_a(x,k_t,\mu^2_0)  
 + \sum_b 
\int
{{d^2 q_{t}^{\prime } } 
\over {\pi q_{t}^{\prime 2} } }
 \ 
{
{\Delta_a (  \mu^2  )} 
 \over 
{\Delta_a (  q_{t}^{\prime 2}  
 ) }
}
\ \Theta(\mu^2-q_{t}^{\prime 2}) \  
\Theta(q_{t}^{\prime 2} - \mu^2_0)
 \nonumber\\ 
&\times&  
\int_x^{z_M} {{dz}\over z} \;
P_{ab}^{(R)} (\alphas 
,z) 
\;{ {\cal F}}_b\left({x \over z}, k_{t}+(1-z) q_{t}^\prime , 
q_{t}^{\prime 2}\right)  
  \;\;  .     
\end{eqnarray}
These equations can be  solved by an iterative Monte Carlo method. In this method every resolvable branching is reconstructed explicitly and the full kinematics at each branching is taken into account. Here, the starting disitribution for the uPDF evolution is taken in the factorized form as a product of collinear PDF fitted to the precise DIS data and an intrinsic transverse momentum distribution in a simple gaussian form.

There are two sets available of the parton-branching uPDFs - PB-NLO-2018-set1 and PB-NLO-2018-set2, that correspond to different choice of the parameters
of the initial distributions \cite{Martinez:2018jxt}. Both of them are based on the HERAPDF2.0 collinear parton densities at NLO.   
In the numerical calculations below we use the PB-NLO-2018-set1 uPDF. 
The resulting unintegrated parton densities, PB-NLO-2018-set1 and PB-NLO-2018-set2, including uncertainties are available in TMDLIB \cite{Hautmann:2014kza}.

\subsubsection{Comparison of the uPDF distributions}

\begin{figure}[!h]
\centering
\begin{minipage}{0.32\textwidth}
  \centerline{\includegraphics[width=1.0\textwidth]{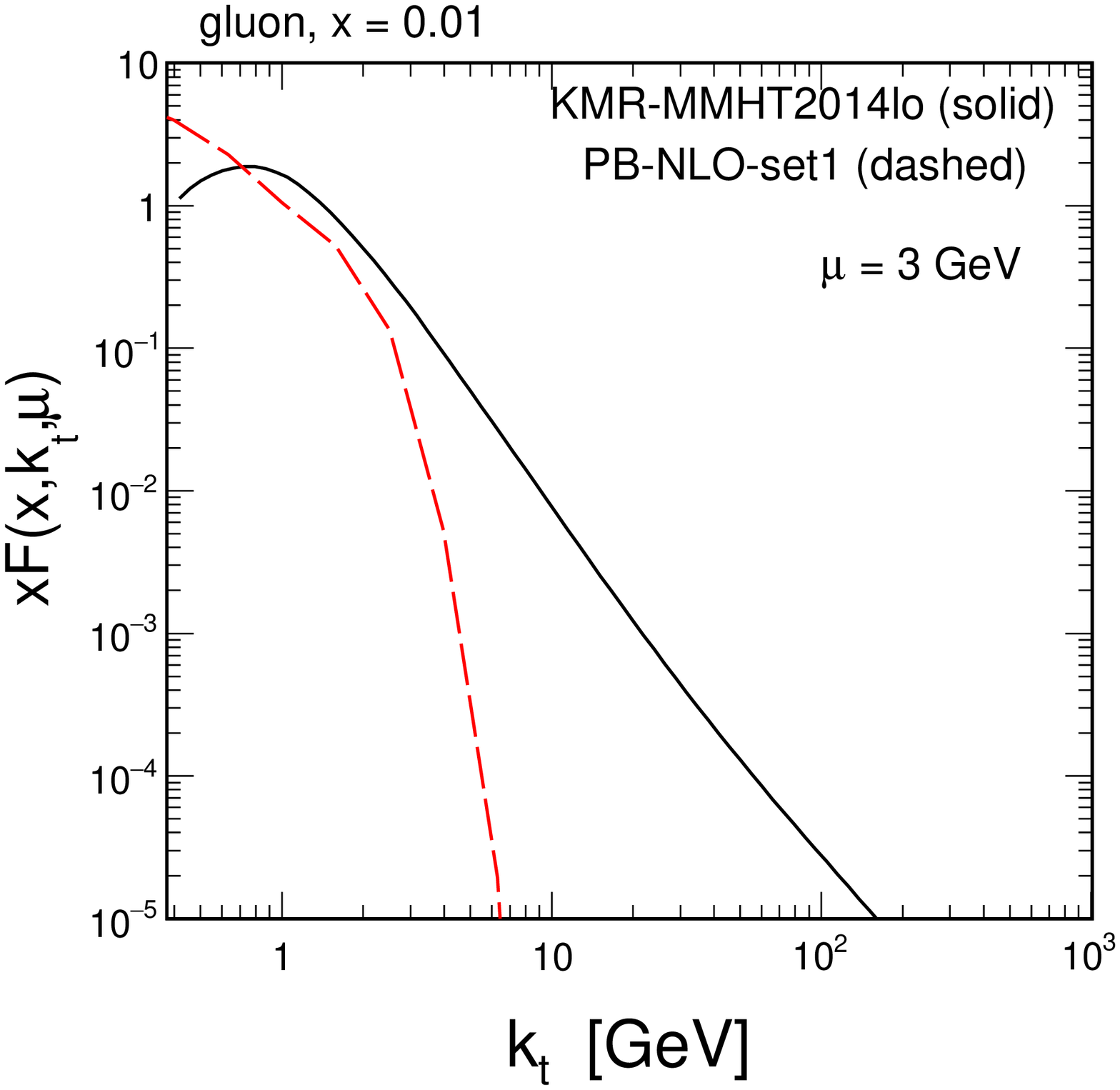}}
\end{minipage}
\begin{minipage}{0.32\textwidth}
  \centerline{\includegraphics[width=1.0\textwidth]{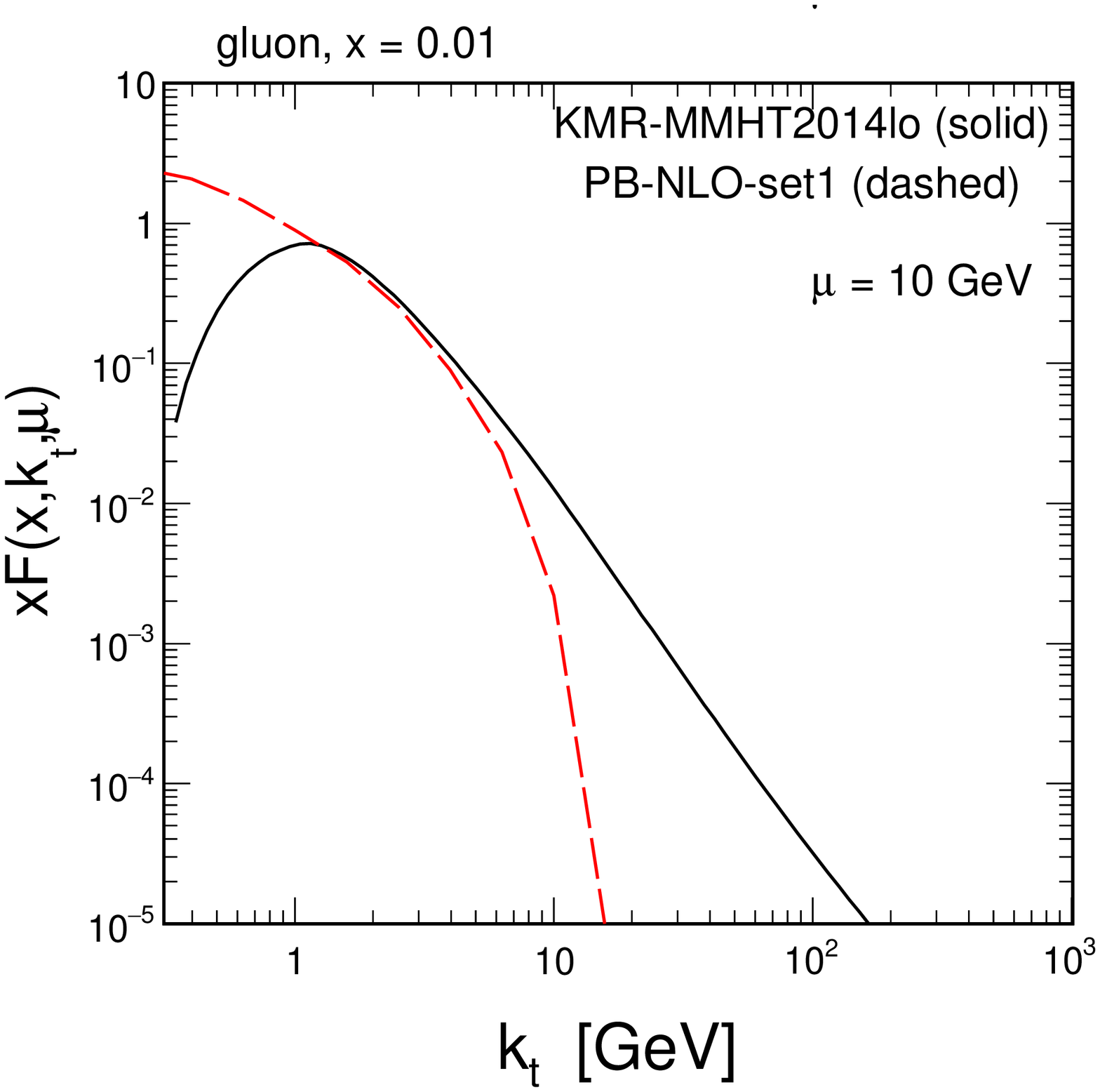}}
\end{minipage}
\begin{minipage}{0.32\textwidth}
  \centerline{\includegraphics[width=1.0\textwidth]{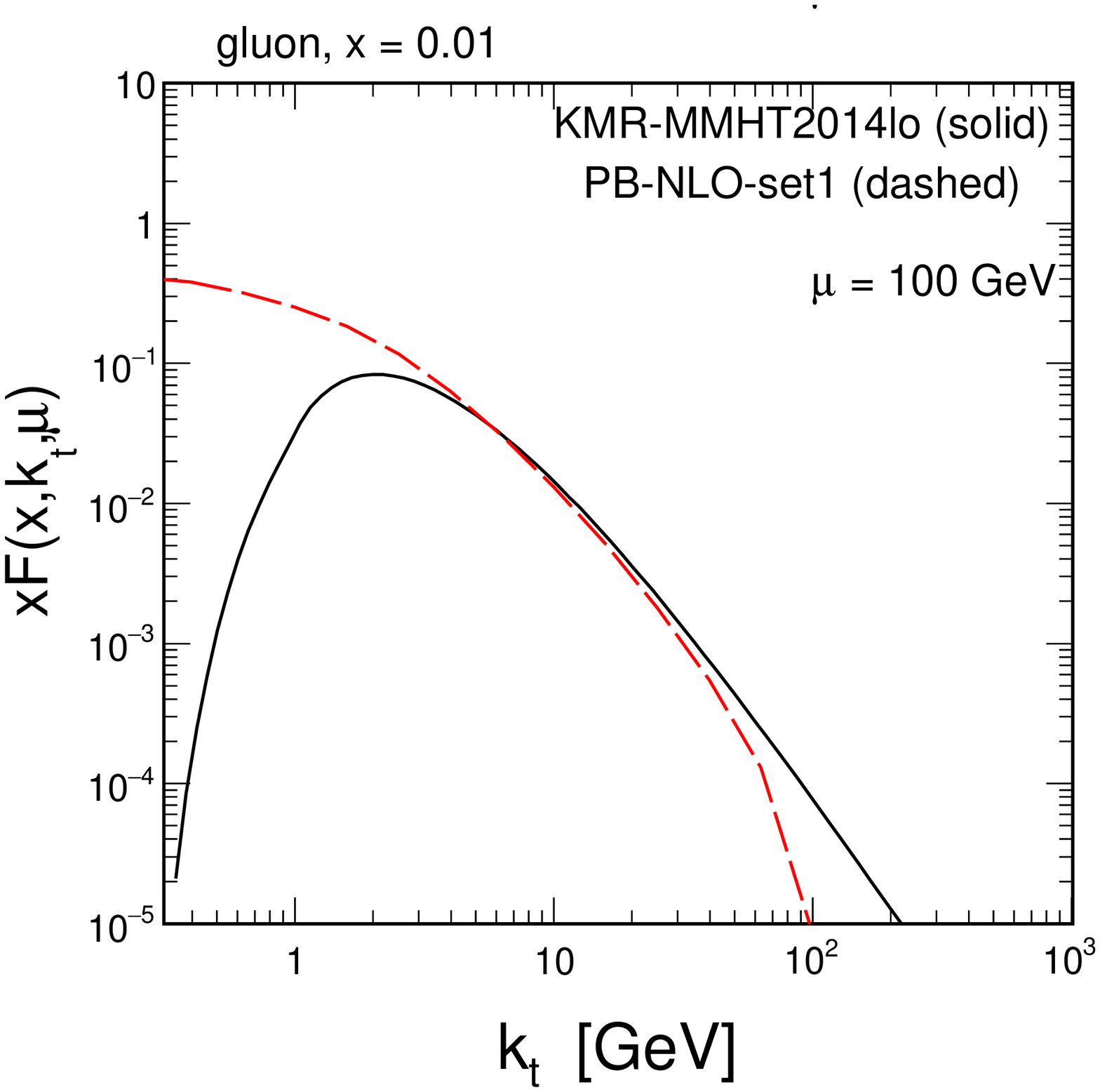}}
\end{minipage}\\
\begin{minipage}{0.32\textwidth}
  \centerline{\includegraphics[width=1.0\textwidth]{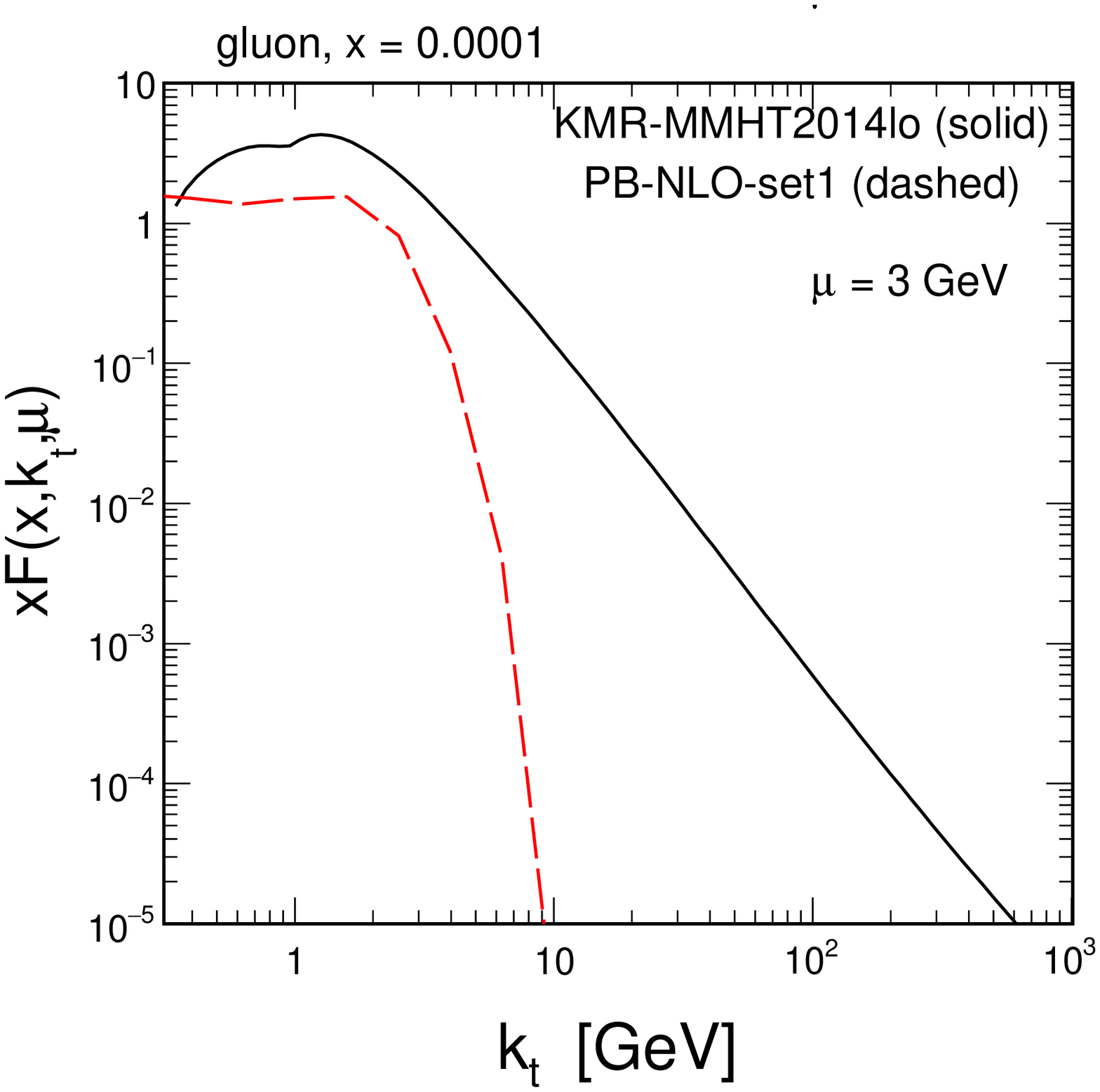}}
\end{minipage}
\begin{minipage}{0.32\textwidth}
  \centerline{\includegraphics[width=1.0\textwidth]{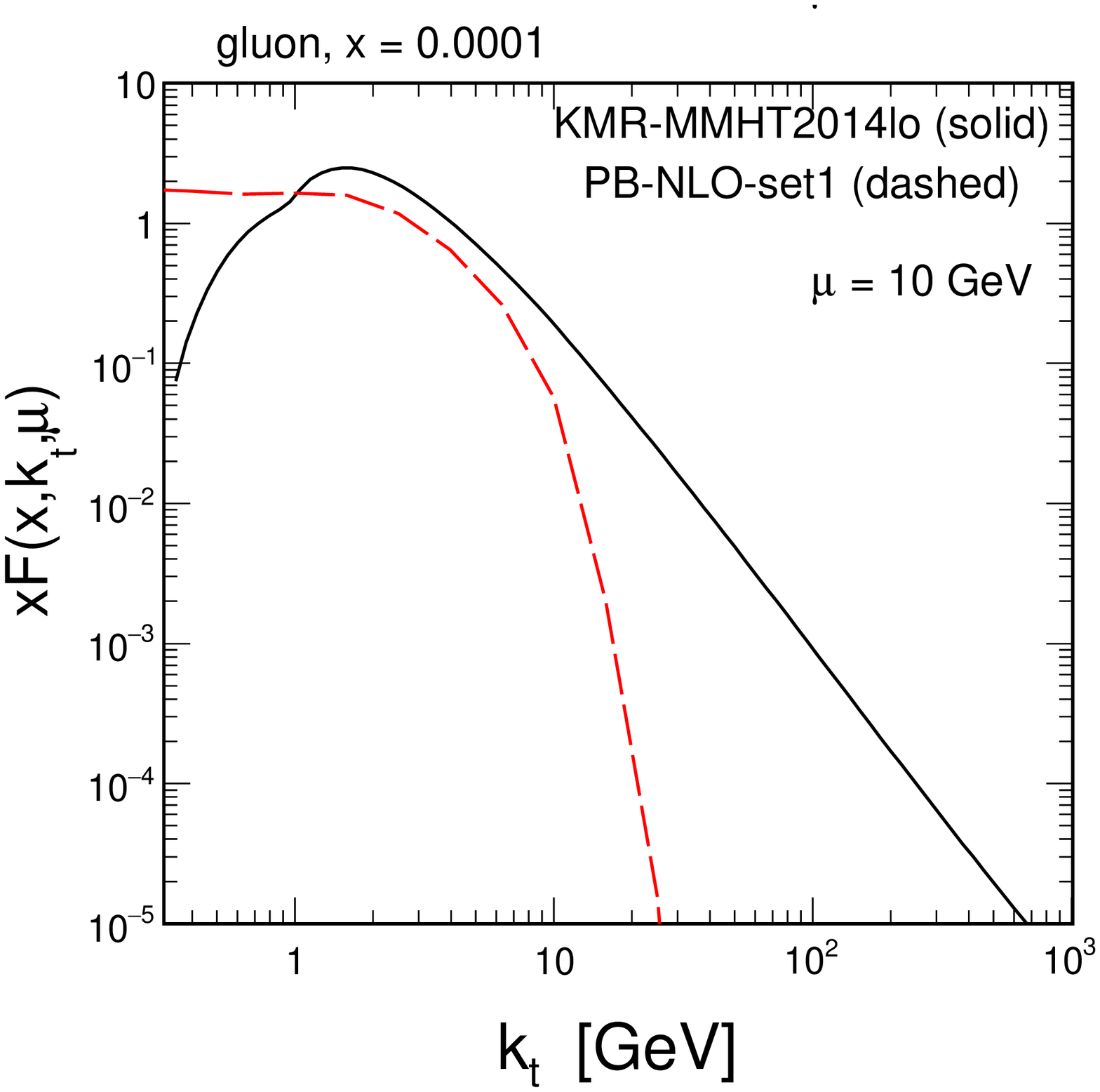}}
\end{minipage}
\begin{minipage}{0.32\textwidth}
  \centerline{\includegraphics[width=1.0\textwidth]{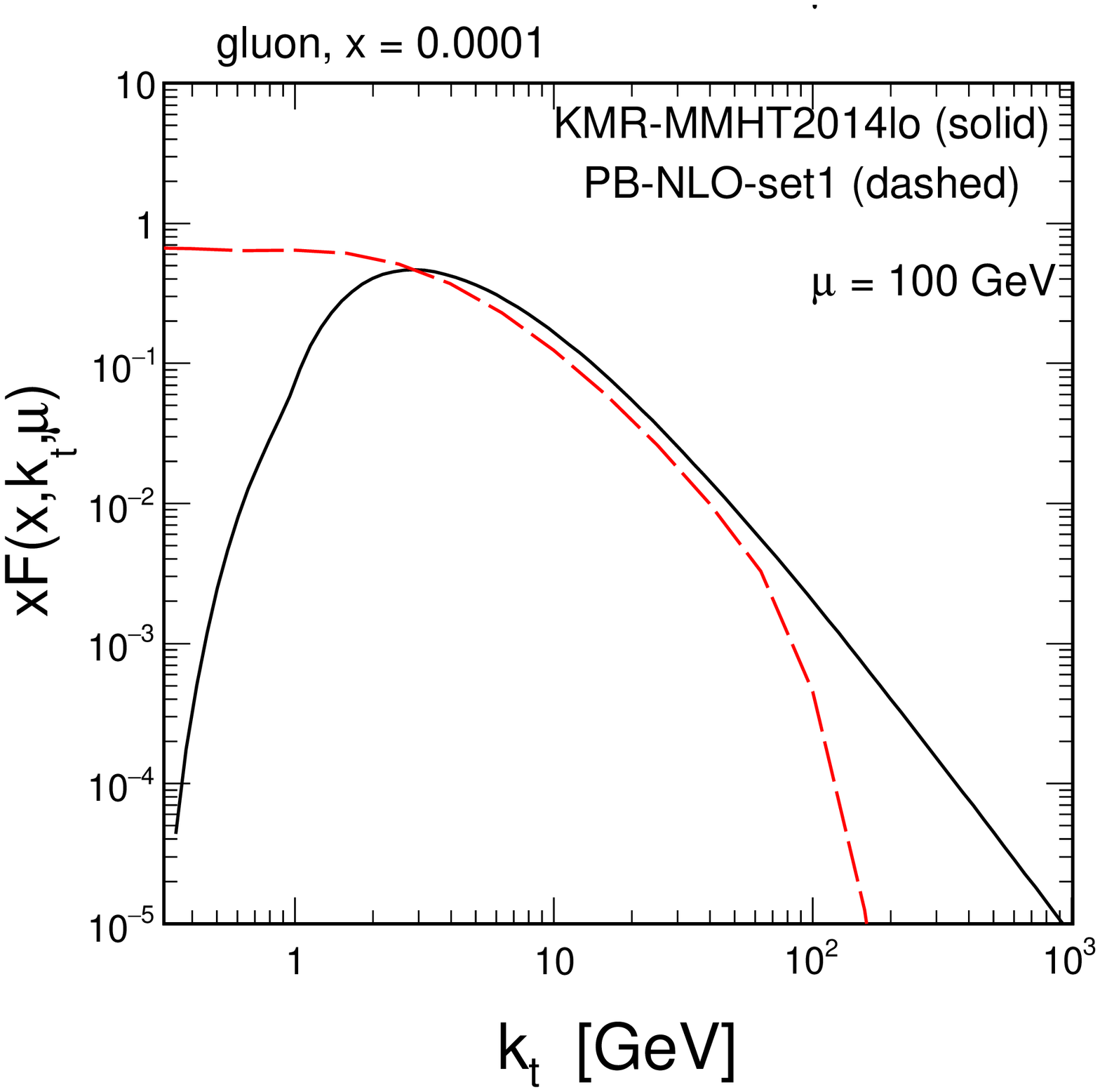}}
\end{minipage}
  \caption{
\small A comparison of transverse momentum dependence of the KMR-MMHT2014lo and the PB-NLO-set1 gluon uPDFs for different scales (left, middle and right panels) and for different longitudinal momentum fractions (upper and bottom panels).
}
\label{fig:uPDFs}
\end{figure}

Let us present now a numerical comparison between the two uPDF models used in the calculations below.
Here, we compare the KMR-MMHT2014lo and PB-NLO-set1 gluon unintegrated distributions. In Fig.~\ref{fig:uPDFs} we show transverse momentum dependence of the uPDFs
for two different values of longitudinal momentum fractions: $x=0.01$ (upper panels) and $x=0.0001$ (lower panels) as well as for three different values of the factorization scale: $\mu= 3, 10, 100$ GeV (left, middle and right panels). We observe significant differences between the two models of uPDF at both, very small and very large transverse momenta of gluons. The differences for $k_{t} \lesssim 1$ GeV come from miscellaneous treatment of the non-perturbative regime in the considered uPDFs, which in both cases is rather uncertain. However, the main visible difference appears in the region of large transverse momenta. The KMR uPDF (solid lines) has long tails which is a consequence of $k_{t} > \mu$ contributions allowed for the gluon emissions. In contrast, the PB-NLO-set1 uPDF (dashed lines) is strongly suppressed in this kinematical regime. As will be discussed in the following, this behaviour of the two uPDFs has a crucial meaning for valueable predictions of charm hadroproduction at the LHC.

\section{Numerical results}

\subsection{The standard \bm{$k_{T}$}-factorization calculations of the \bm{$D$}-meson cross section including only \bm{$g^*g^* \to c \bar c$} mechanism with the KMR uPDFs}

We start presentation of our numerical results with the inclusive distributions of $D$-meson and with correlation observables for $D\bar D$ meson-antimeson pair production. We compare our theoretical predictions with the LHCb open charm data from $pp$-scattering at $\sqrt{s} = 7$ TeV \cite{Aaij:2013mga,Aaij:2012dz}. Here, we follow the standard $k_{T}$-factorization approach and calculate the cross section for $c\bar c$-pair production by taking into account the $g^*g^* \to c\bar c$ mechanism. We use the KMR gluon uPDF in the original form, that allows
for extra hidden hard emissions at the very last step of its evolution, i.e. including contributions from the $k_{T} > \mu_{F}$ kinematical regime. In this way, a part of real higher-order corrections is effectively
taken into account in the calculations. This was already discussed in the case of charm production, \textit{e.g.} in Ref.~\cite{Maciula:2013wg}. Here, we wish to extend those studies by discussing some important details of the calculation, relevant to estimate overall uncertainties of the model.  

\begin{figure}[!h]
\centering
\begin{minipage}{0.47\textwidth}
  \centerline{\includegraphics[width=1.0\textwidth]{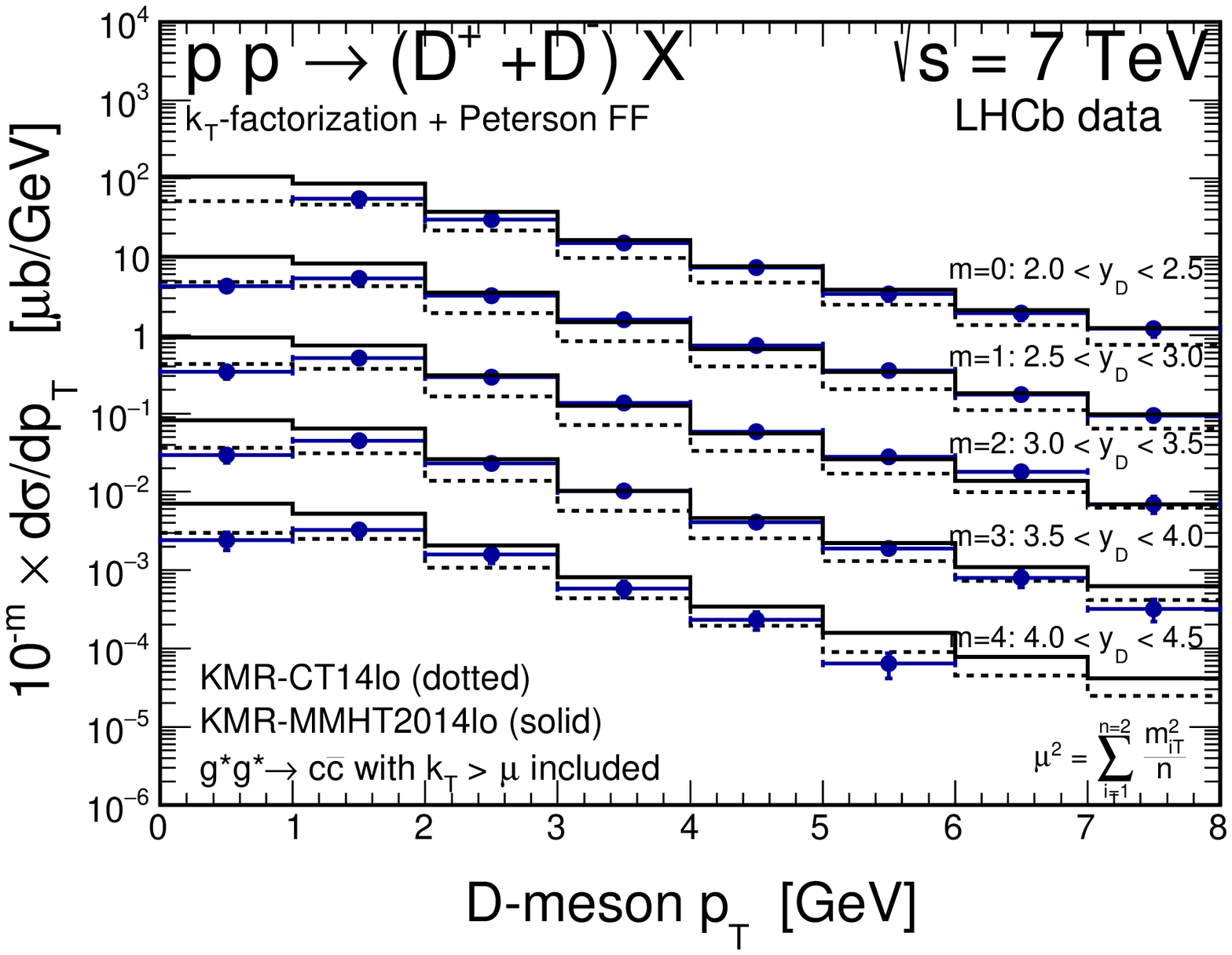}}
\end{minipage}
\begin{minipage}{0.47\textwidth}
 \centerline{\includegraphics[width=1.0\textwidth]{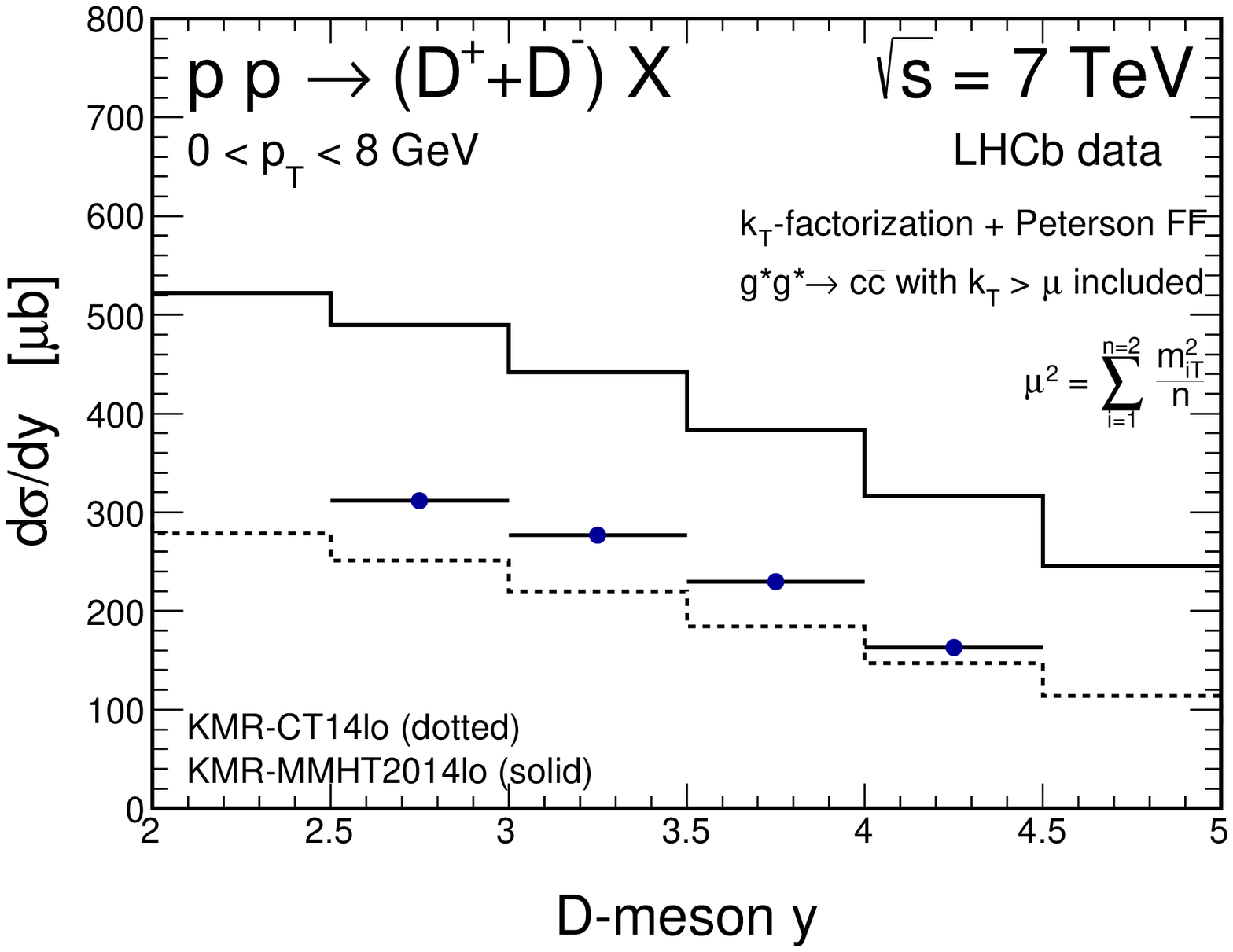}}
\end{minipage}
  \caption{
\small Transverse momentum (left) and rapidity distributions of charged $D$-meson for $\sqrt{s}= 7$ TeV together with the LHCb data \cite{Aaij:2013mga}. Here, we compare the results for different collinear PDFs used for calculating the KMR uPDFs.
}
\label{fig:1}
\end{figure}

In Fig.~\ref{fig:1} we show the transverse momentum distributions for different rapidity bins (left panel) and the rapidity distribution (right panel) of the charged $D$-meson measured by the LHCb experiment in the kinematical range: $0 < p_{T} < 8$ GeV and $2 < y < 4.5$. We compare results obtained with two different sets of the KMR gluon uPDF: KMR-MMHT2014lo (solid) and KMR-CT14lo (dotted). Both of them lead to a reasonable description of the LHCb data for larger transverse momenta, however, we observe some visible differences of results for small transverse momenta. The KMR-MMHT2014lo results overestimate the $D$ meson $p_T$ dsitributions at very small transverse momenta and give a very good description of the larger $p_T$'s. The
KMR-CT14lo gluon uPDF leads to slightly smaller cross sections, which improves the description of the data at very small $p_{T}$'s and in the consequence of rapidity distribution. Except of the first bin, for the transverse momentum distribution the uncertainty 
related to the choice of the collinear gluon uPDFs can be estimated at the level of less than a factor of 2.
The calculated rapidity distributions reflect the behaviour of the cross section in the first bin in transverse momentum. The KMR-MMHT2014lo overestimates the experimental points while the KMR-CT14lo result lies much closer to the experimental data.

\begin{figure}[!h]
\centering
\begin{minipage}{0.47\textwidth}
  \centerline{\includegraphics[width=1.0\textwidth]{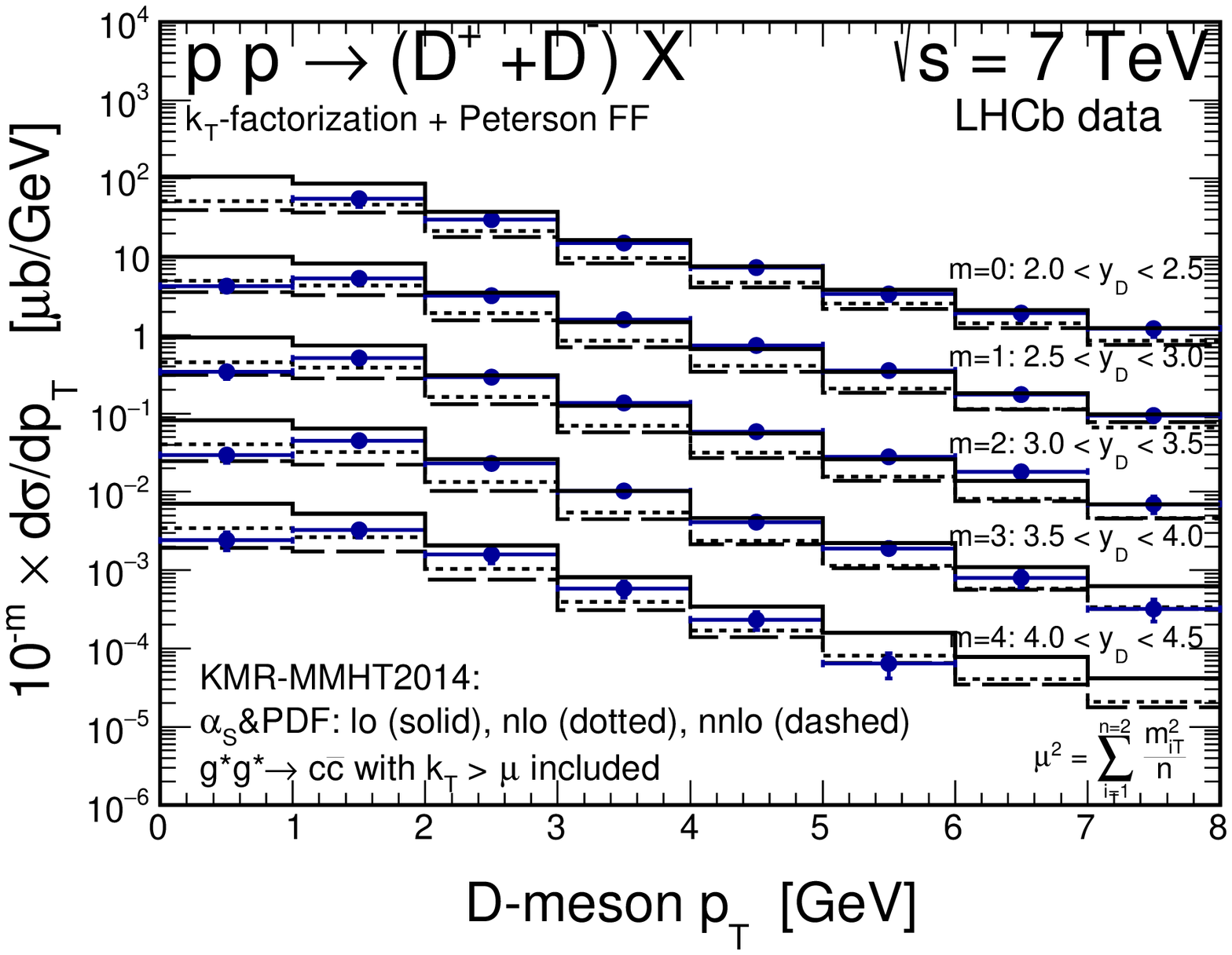}}
\end{minipage}
\begin{minipage}{0.47\textwidth}
 \centerline{\includegraphics[width=1.0\textwidth]{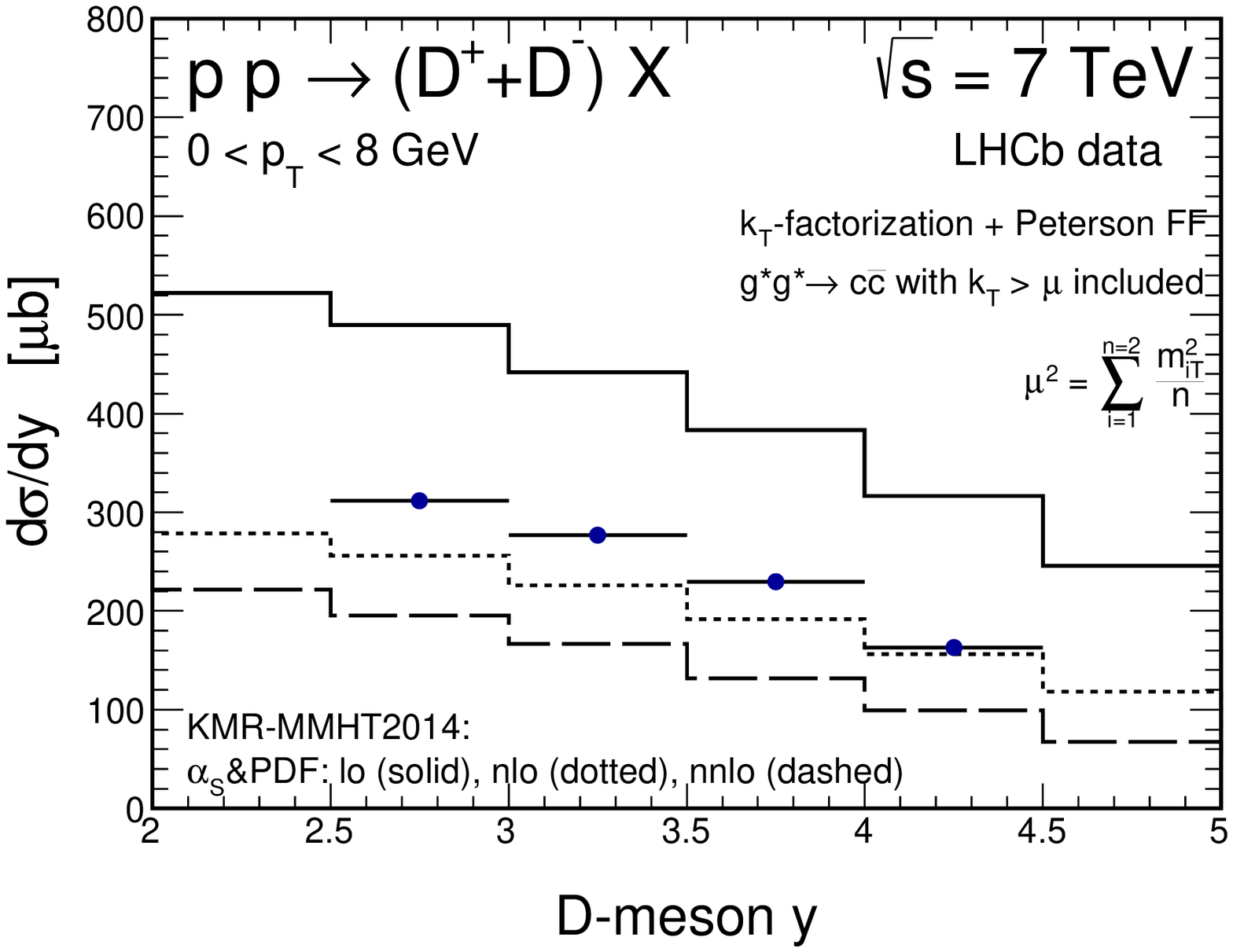}}
\end{minipage}
  \caption{
\small Transverse momentum (left) and rapidity (right) distributions of charged $D$-meson for $\sqrt{s}= 7$ TeV together with the LHCb data \cite{Aaij:2013mga}. Here, we compare results for the LO, NLO and NNLO collinear MMHT2014 PDFs used when calculating the KMR uPDFs.
}
\label{fig:2}
\end{figure}

In Fig.~\ref{fig:2} we discuss theoretical uncertainties related to the perturbative order of the strong coupling $\alpha_{S}$ and simultaneously of the choice of the collinear gluon PDF used in the calculations. Here we use the KMR-MMHT2014 gluon uPDF and consider three different choices of $\alpha_{S}$ and collinear PDFs: LO (solid), NLO (dotted), and NNLO (dashed). Again, we observe a visible sensitivity of our results to the choice of these  basic ingredients. The higher-order sets lead to a better agreement with the data for the rapidity distribution and at small transverse momenta, while the larger $p_{T}$ bins prefer the leading-order $\alpha_{S}$ and PDF.          

\begin{figure}[!h]
\centering
\begin{minipage}{0.47\textwidth}
  \centerline{\includegraphics[width=1.0\textwidth]{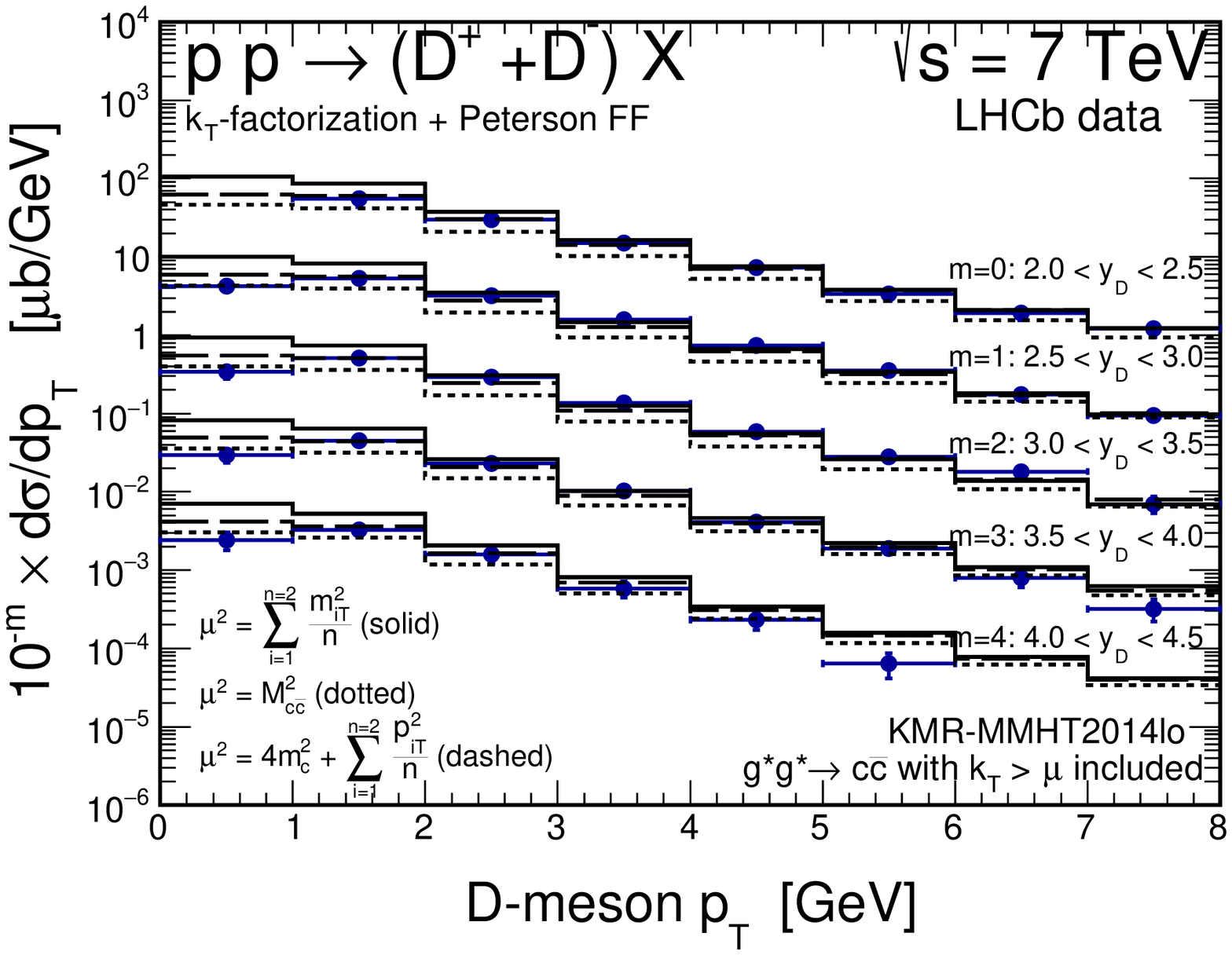}}
\end{minipage}
\begin{minipage}{0.47\textwidth}
 \centerline{\includegraphics[width=1.0\textwidth]{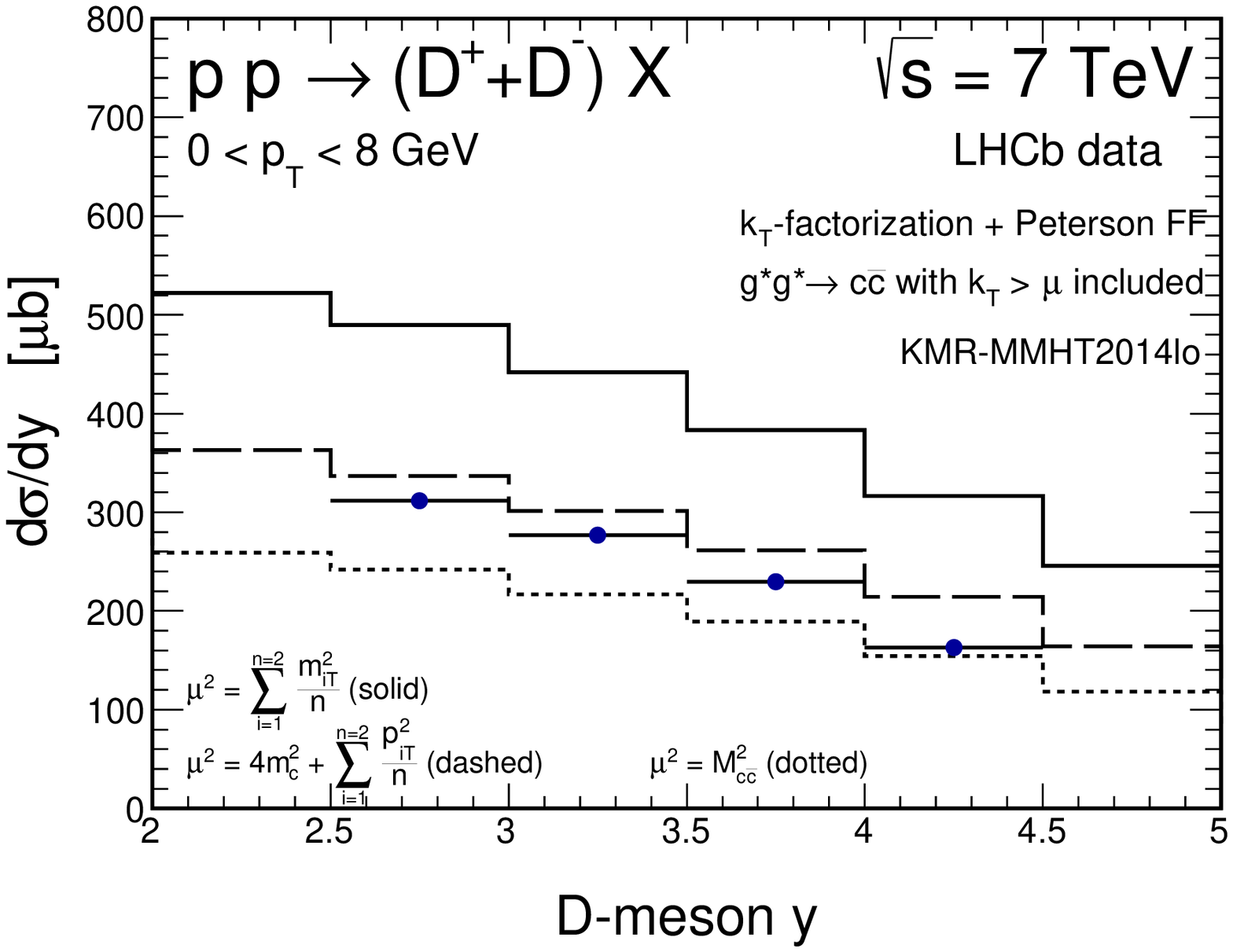}}
\end{minipage}
  \caption{
\small Transverse momentum (left) and rapidity (right) distributions of charged $D$-meson for $\sqrt{s}= 7$ TeV together with the LHCb data \cite{Aaij:2013mga}. Here, we compare the KMR-MMHT2014lo results for different renormalization/factorizaton scales.
}
\label{fig:3}
\end{figure}

Another important source of uncertainties of the pQCD calculation is the choice of the renormalization and factorization scale. Typically, both of them are set to be equal $\mu = \mu_{R} = \mu_F$. Usually, in the case of heavy flavours the scales are connected with the transverse mass (or momentum) of the produced particles. In the following, as a default set of the calculations we take the averaged sum of transverse mass squared of the final state particles $\mu^2 = \sum_{i=1}^n \frac{m_{iT}^2}{n}$, where $m_{iT} = \sqrt{m_{i}^2+p_{iT}^2}$.  
Here, instead of varying the default set to produce the scale uncertainty band, we also consider two different sets: $\mu^2 = M_{c\bar c}^2 = \hat{s}$ and $\mu^2 = 4m_{c}^2 + \sum_{i=1}^n \frac{p_{iT}^2}{n}$. A comparison of corresponding results for different scales is shown in Fig.~\ref{fig:3}. These three sets of the scales lead to visible differences only for very small meson $p_{T}$'s. The overall uncertainty related to the scales is of the same order as those discussed above. 
     
\begin{figure}[!h]
\centering
\begin{minipage}{0.47\textwidth}
  \centerline{\includegraphics[width=1.0\textwidth]{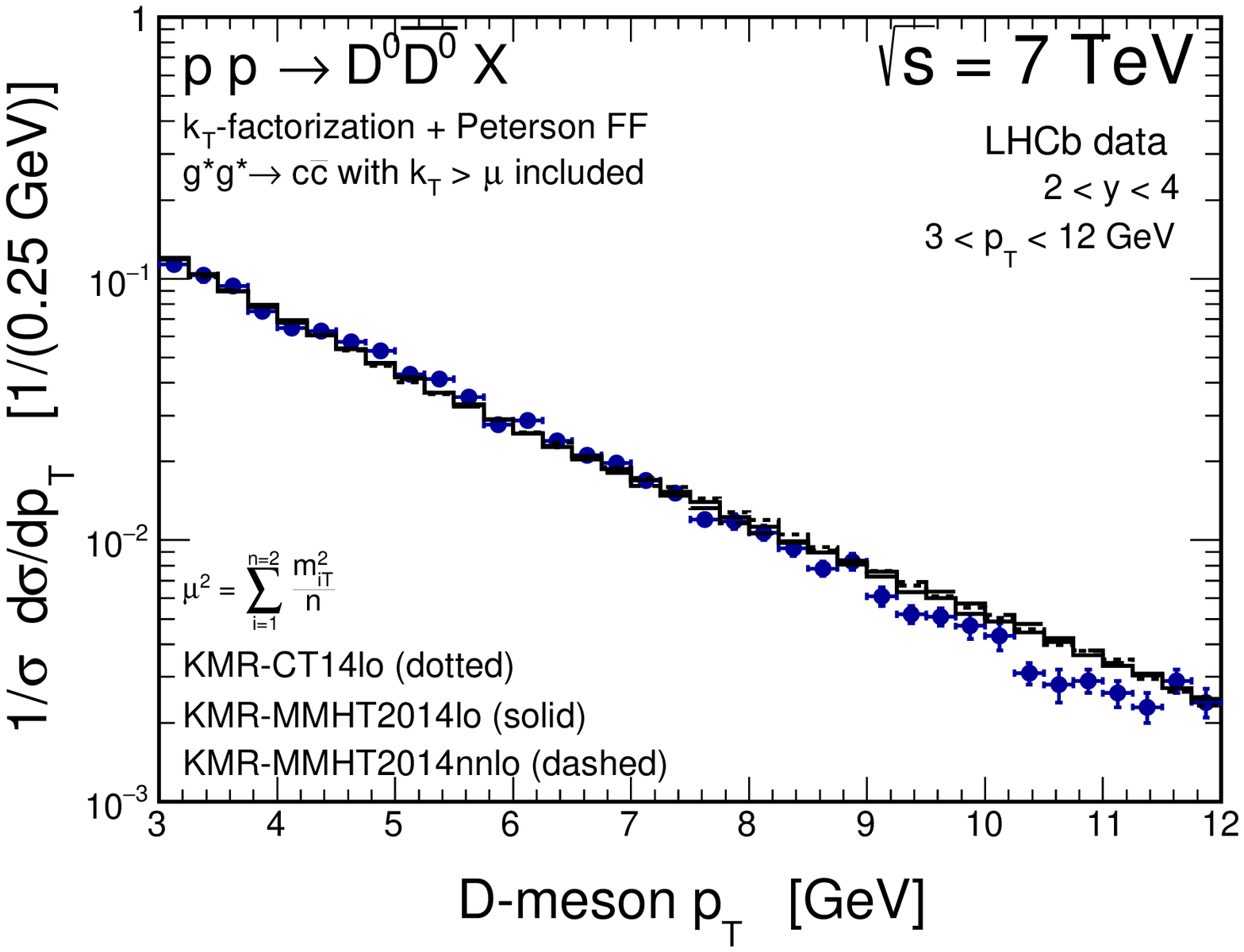}}
\end{minipage}
\begin{minipage}{0.47\textwidth}
 \centerline{\includegraphics[width=1.0\textwidth]{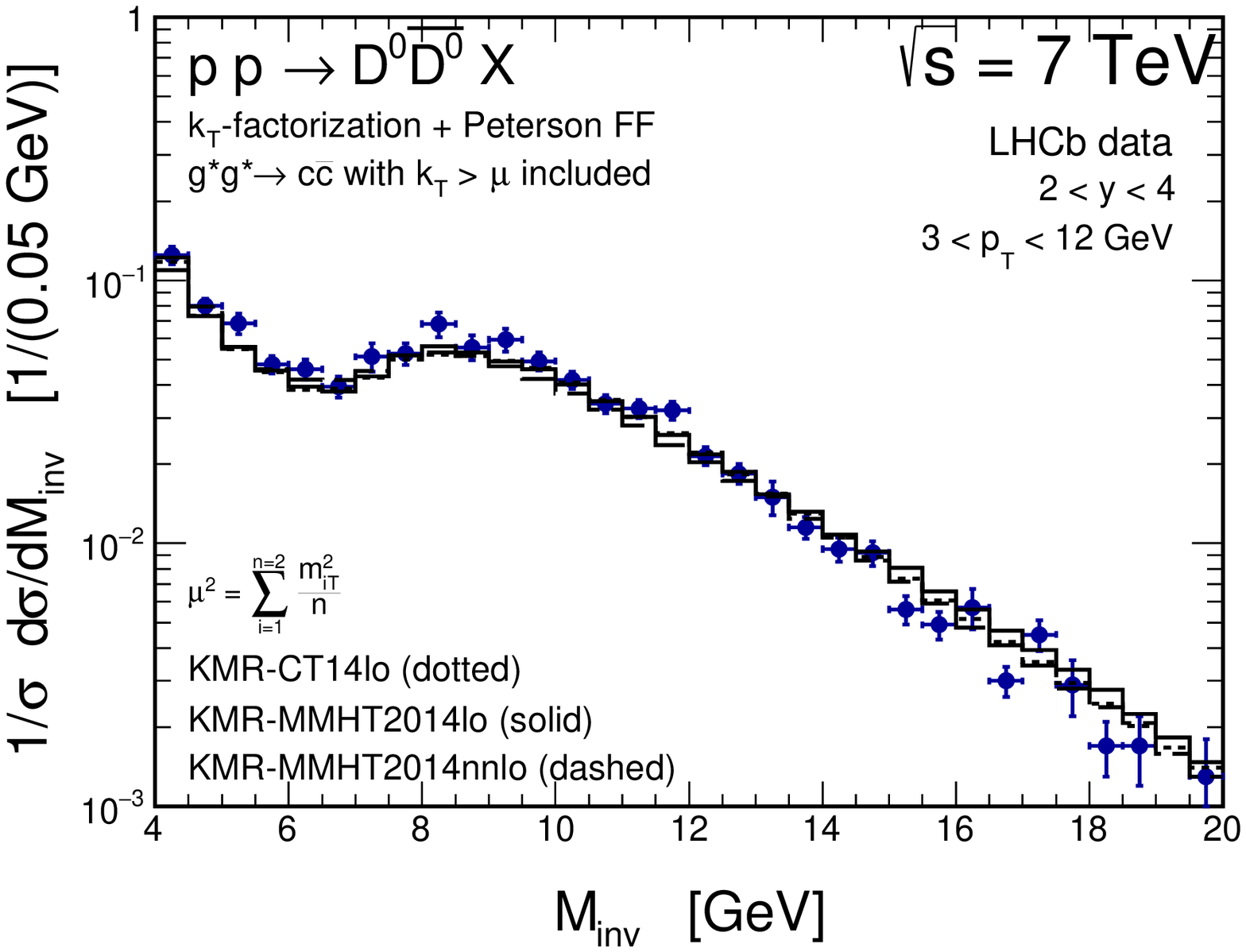}}
\end{minipage}\\
\begin{minipage}{0.47\textwidth}
  \centerline{\includegraphics[width=1.0\textwidth]{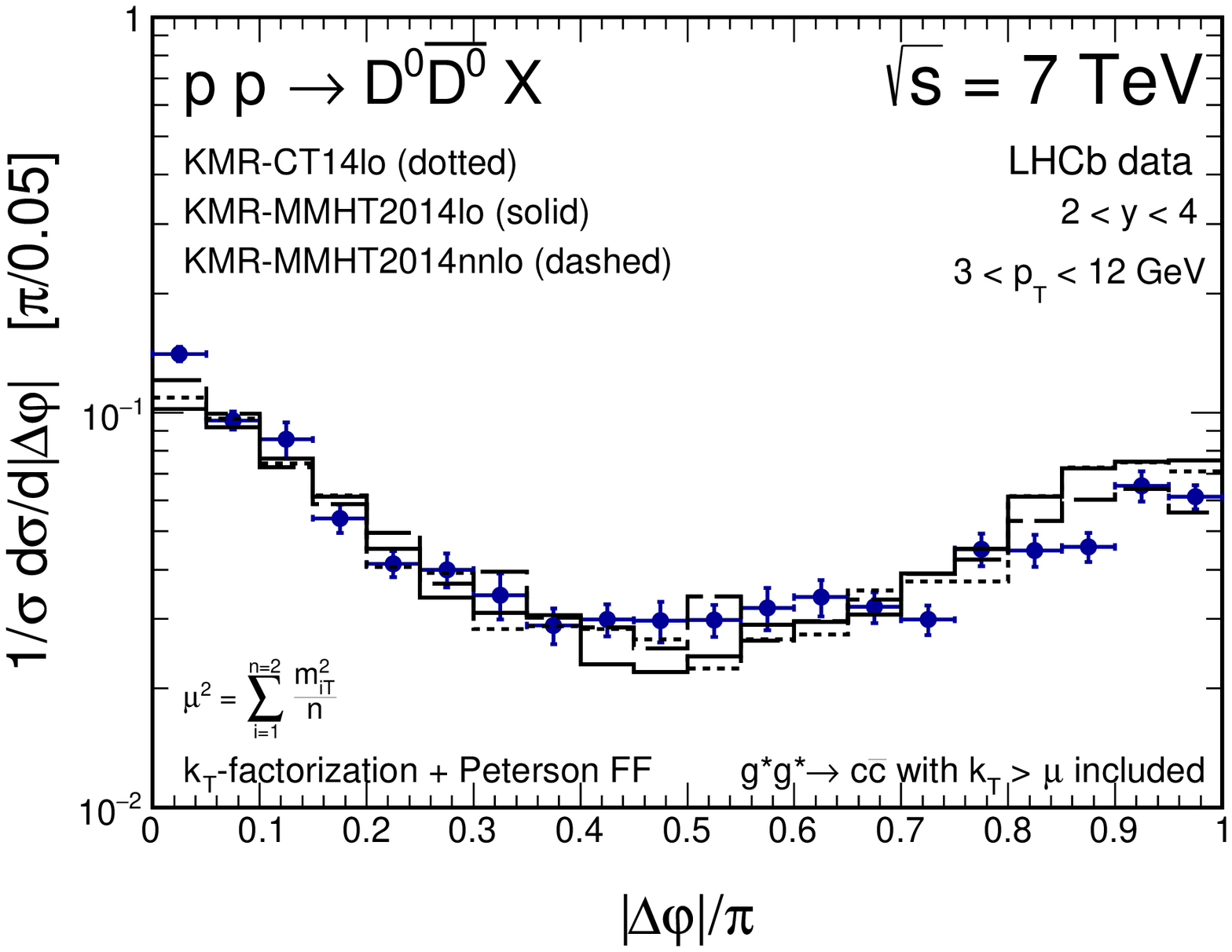}}
\end{minipage}
\begin{minipage}{0.47\textwidth}
 \centerline{\includegraphics[width=1.0\textwidth]{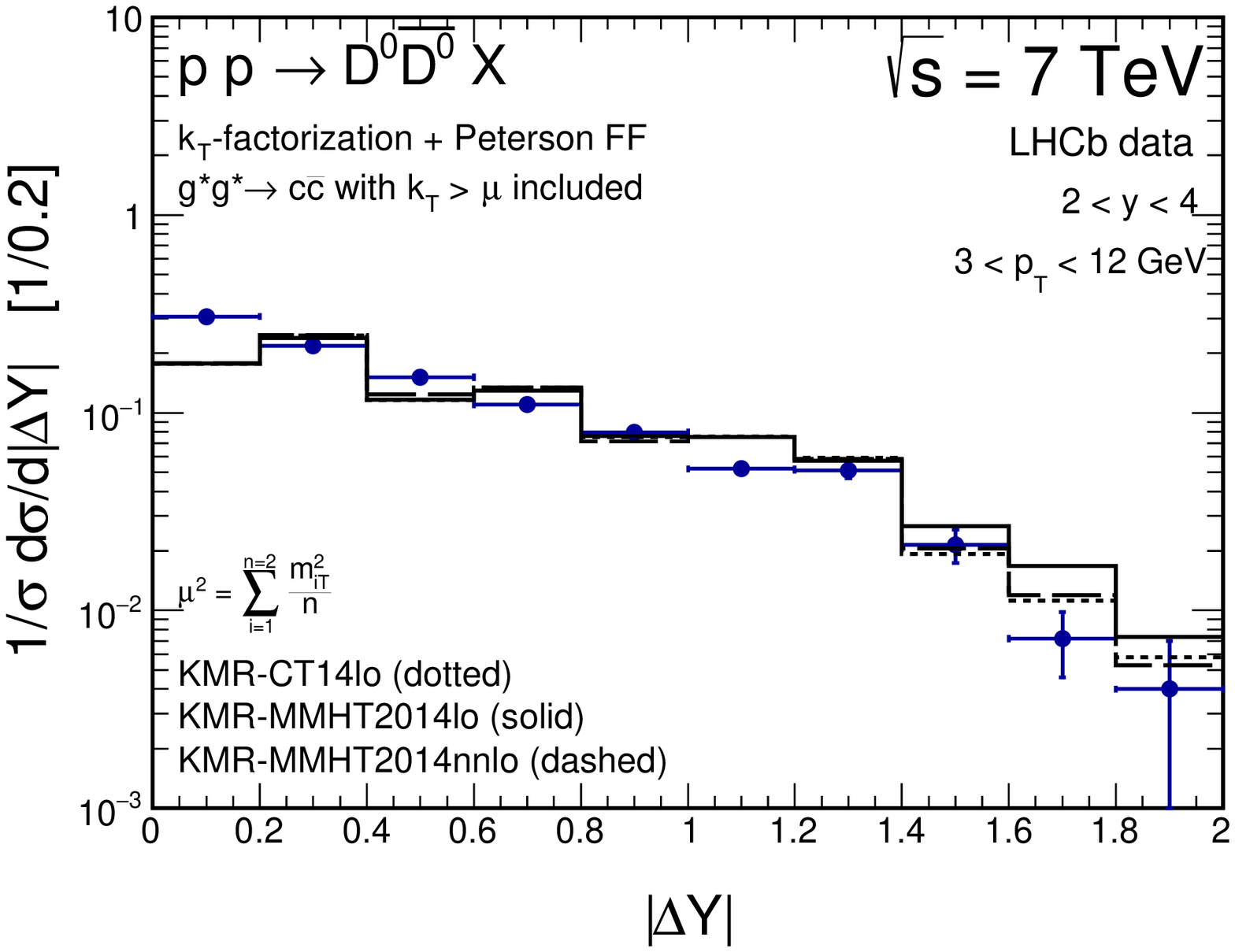}}
\end{minipage}
  \caption{
\small Transverse momentum (top-left), invariant mass (top-right), azimuthal angle (bottom-left) and rapidity distance (bottom-right) distributions for charged $D \bar D$ meson-antimeson pair production at $\sqrt{s}= 7$ TeV together with the LHCb data \cite{Aaij:2012dz}. Here, we compare results for different collinear gluon PDFs used in calculating the KMR uPDF.
}
\label{fig:4}
\end{figure}

\begin{figure}[!h]
\centering
\begin{minipage}{0.47\textwidth}
  \centerline{\includegraphics[width=1.0\textwidth]{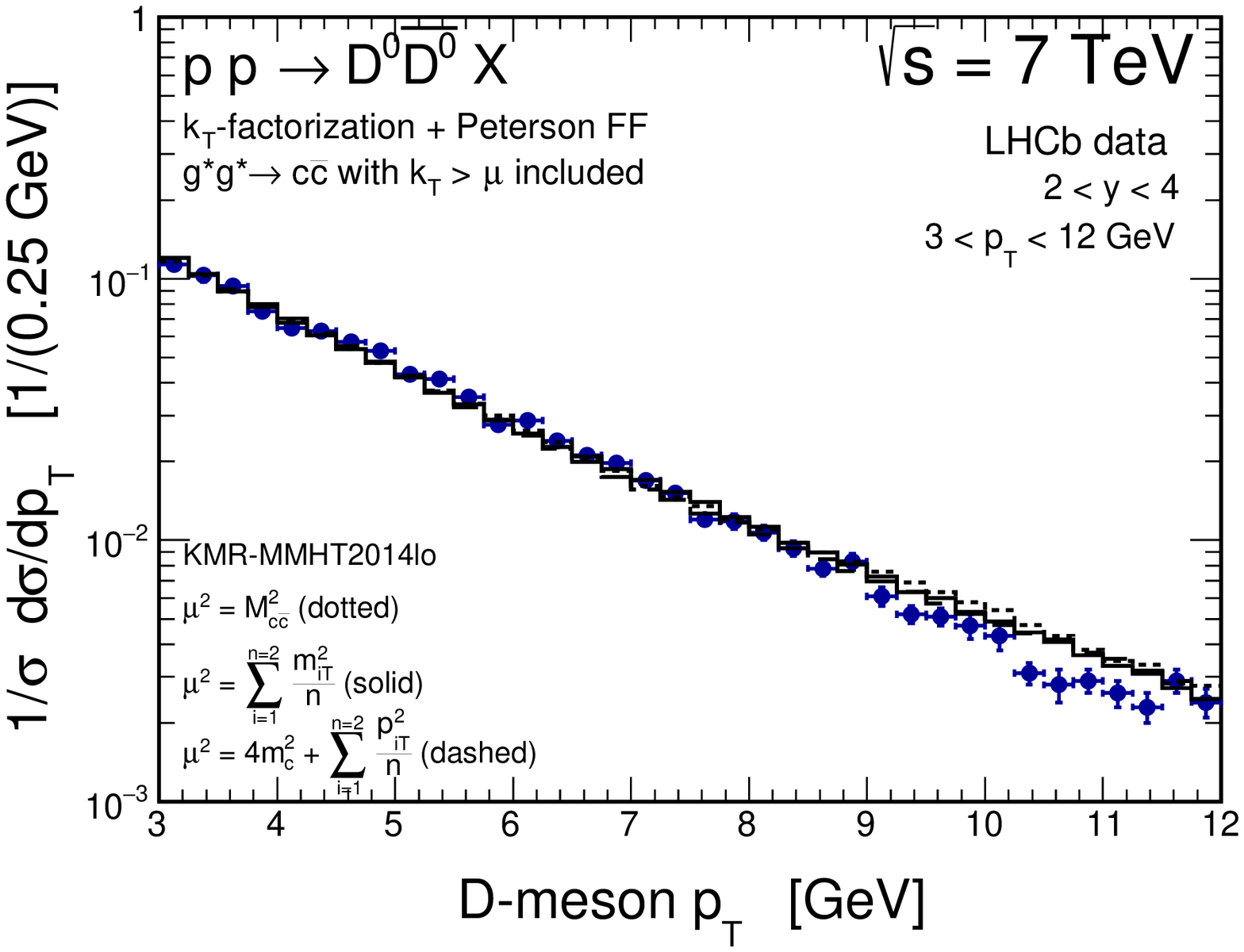}}
\end{minipage}
\begin{minipage}{0.47\textwidth}
 \centerline{\includegraphics[width=1.0\textwidth]{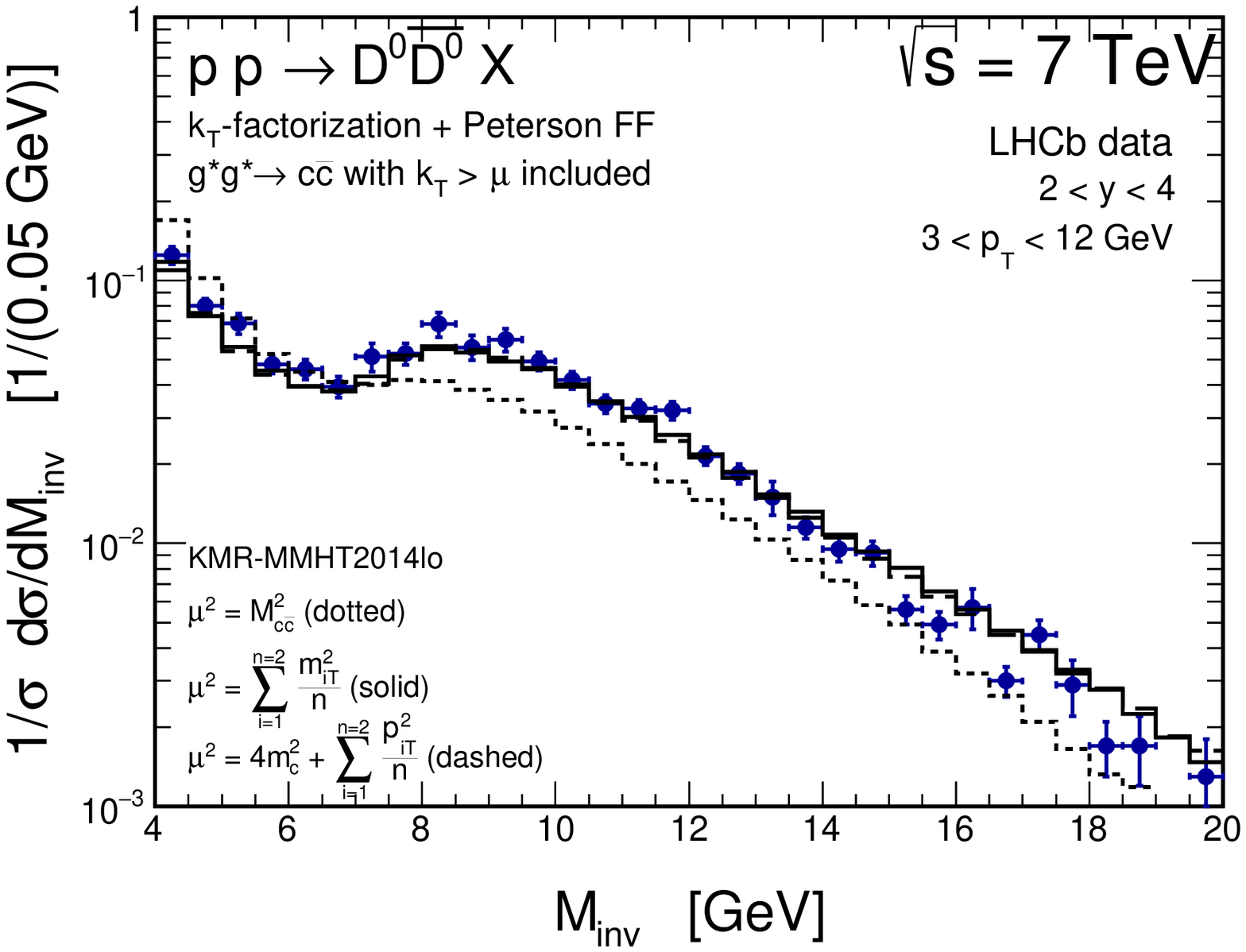}}
\end{minipage}\\
\begin{minipage}{0.47\textwidth}
  \centerline{\includegraphics[width=1.0\textwidth]{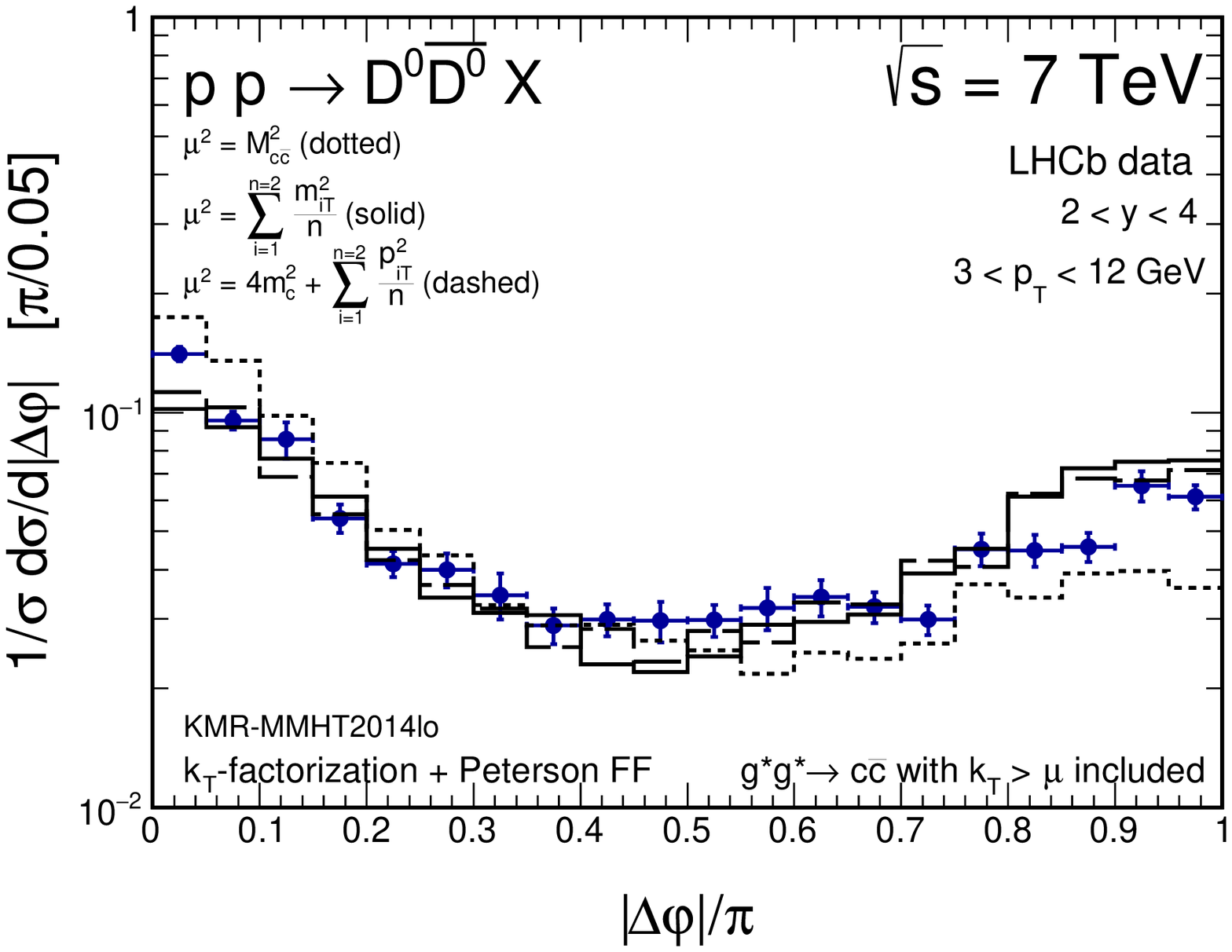}}
\end{minipage}
\begin{minipage}{0.47\textwidth}
 \centerline{\includegraphics[width=1.0\textwidth]{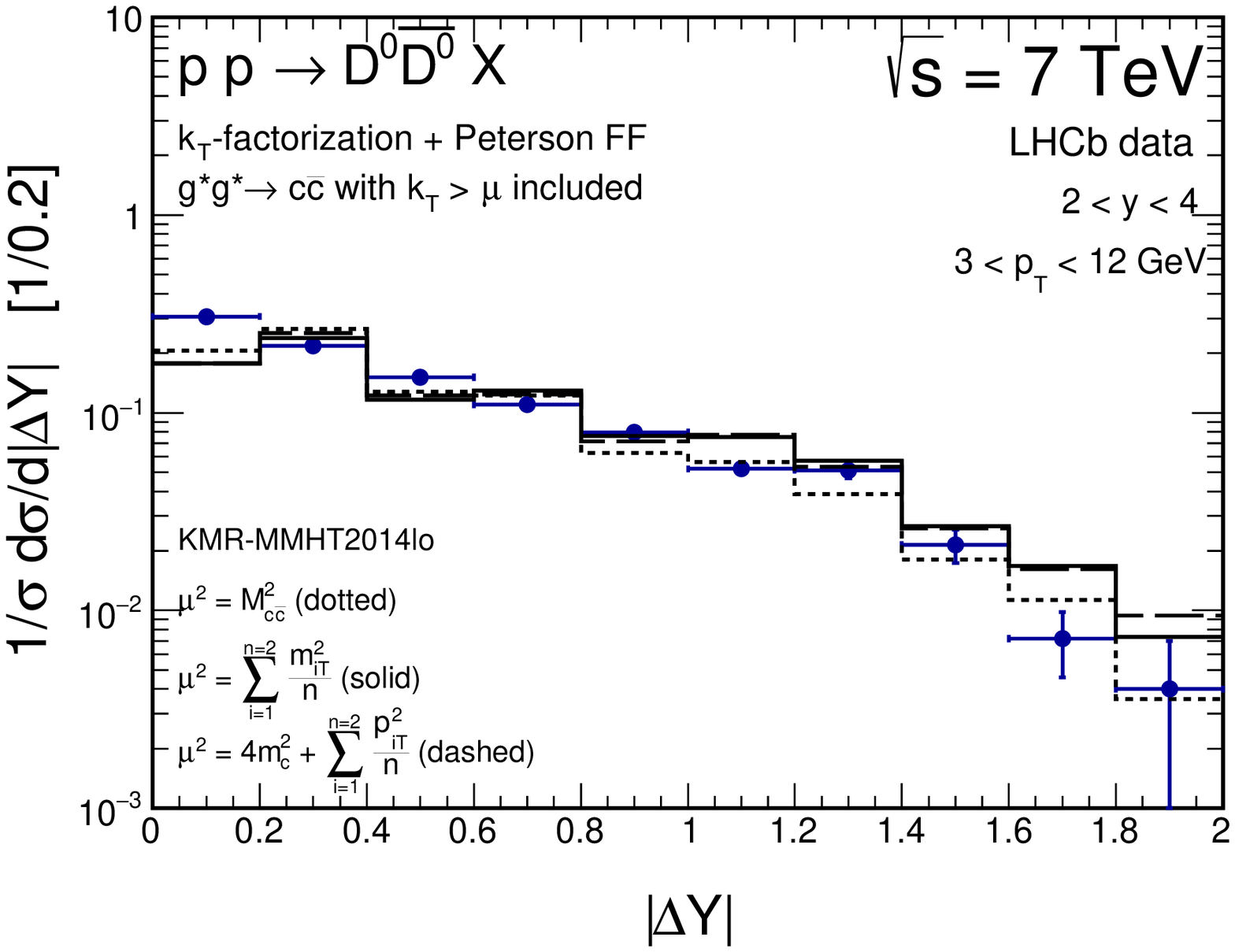}}
\end{minipage}
  \caption{
\small Transverse momentum (top-left), invariant mass (top-right), azimuthal angle (bottom-left) and rapidity distance (bottom-right) distributions for charged $D \bar D$ meson-antimeson pair production at $\sqrt{s}= 7$ TeV together with the LHCb data \cite{Aaij:2012dz}. Here, we compare the KMR-MMHT2014lo results for different renormalization/factorizaton scales.
}
\label{fig:5}
\end{figure}

The framework of the $k_{T}$-factorization is known to be very efficient in studying kinematical correlations between produced particles (charmed mesons).
It allows for a direct calculation of the less inclusive distributions already at the leading-order. The correlation observables are fully determined by the transverse momenta of the initial state partons.
In Figs.~\ref{fig:4} and \ref{fig:5} we extend the above studies of the inclusive $D$-meson spectra to the case of the $D\bar D$-pair production. We present differential distributions as a function of transverse momentum of $D$-meson (or $\bar D$-antimeson) for the pair (top-left panels), di-meson invariant mass $M_{inv}=M_{D\bar{D}}$ (top-right panels), azimuthal angle $\Delta \varphi = \varphi_{D\bar{D}}$ (bottom-left panels) and rapidity difference $\Delta Y = |y_{D} - y_{\bar D}|$ (bottom right panels). Again, we show uncertainties due to the choice of collinear gluon PDFs (Fig.~\ref{fig:4}) and due to the choice of scales (Fig.~\ref{fig:5}).
Here the LHCb correlation data \cite{Aaij:2012dz} are not absolutely normalized and we consider only shapes of the distributions. The more exclusive observables bring more usefull informations about the model calculations. The shapes of the calculated distributions are almost insensitive to the choice of the collinear gluon PDF used in calculating the KMR uPDF. On the other hand, one of the used sets of the scales $\mu^2 = M_{c\bar c}^2 = \hat{s}$ is clearly not supported by the experimental data.        
 
Summarizing this subsection, we conclude that within the typical pQCD uncertainties we are able to get a satisfactory description of the LHCb charm data. The statement is valid for both, the absolutely normalized inclusive $D$-meson distributions as well as for the shapes of the $D\bar D$ correlation observables.
The framework of the $k_{T}$-factorization together with the KMR gluon uPDF allows to describe the LHCb charm data already within the leading-order $g^*g^* \to c\bar c$ mechanism. This is completely opposite to the calculations within the collinear-approximation. There, only the NLO framework is able to obtain the same level of quality of the description of the LHC heavy flavour data \cite{Cacciari:2012ny,Kniehl:2012ti}. This clearly shows that within the $k_{T}$-factorization approach we effectively include higher-order contributions. However, the fact and the size of the effective resummation strictly depends on the construction of the used uPDF. The KMR model is unique and very useful in this context. It allows even for two extra emissions of hard partons from the uPDFs, that correspond 
to the $gg \to g c\bar c$ and the $gg \to gg c\bar c$ contributions. As we have shown for charm production at the LHC this model seems to work very well, however, the overall picture is more complicated.

The role of the extra emissions from the KMR uPDF can dramatically change when going to the lower energies. The emissions are not under full kinematical control and in the case of some processes, \textit{e.g.} for dijets production, they may even lead to a problematic double counting. Moreover, other models of the uPDFs from the literature do not contain such large contributions from the $k_{T} > \mu_{F}$ regime and are useless at leading-order calculations for processes where higher-orders are of the special importance. Due to the lack of the full NLO/NNLO formalism with off-shell initial state partons we propose a simplified scheme for the calculation of heavy flavour cross section in the $k_T$-factorization with higher-order mechanisms taken into account at the tree level.     
       
\subsection{A new scheme of calculations in the framework of the \bm{$k_{T}$}-factorization with higher-order effects at tree level}

\subsubsection{The Kimber-Martin-Ryskin uPDF with limited hard emissions}

The idea of the proposed scheme is to exclude the extra hard emissions from the uPDF and include  
the higher-order contributions $g^*g^* \to g c\bar c$ and $g^*g^* \to gg c\bar c$ explicitly at the level of hard matrix elements. 

\begin{figure}[!h]
\centering
\begin{minipage}{0.47\textwidth}
  \centerline{\includegraphics[width=1.0\textwidth]{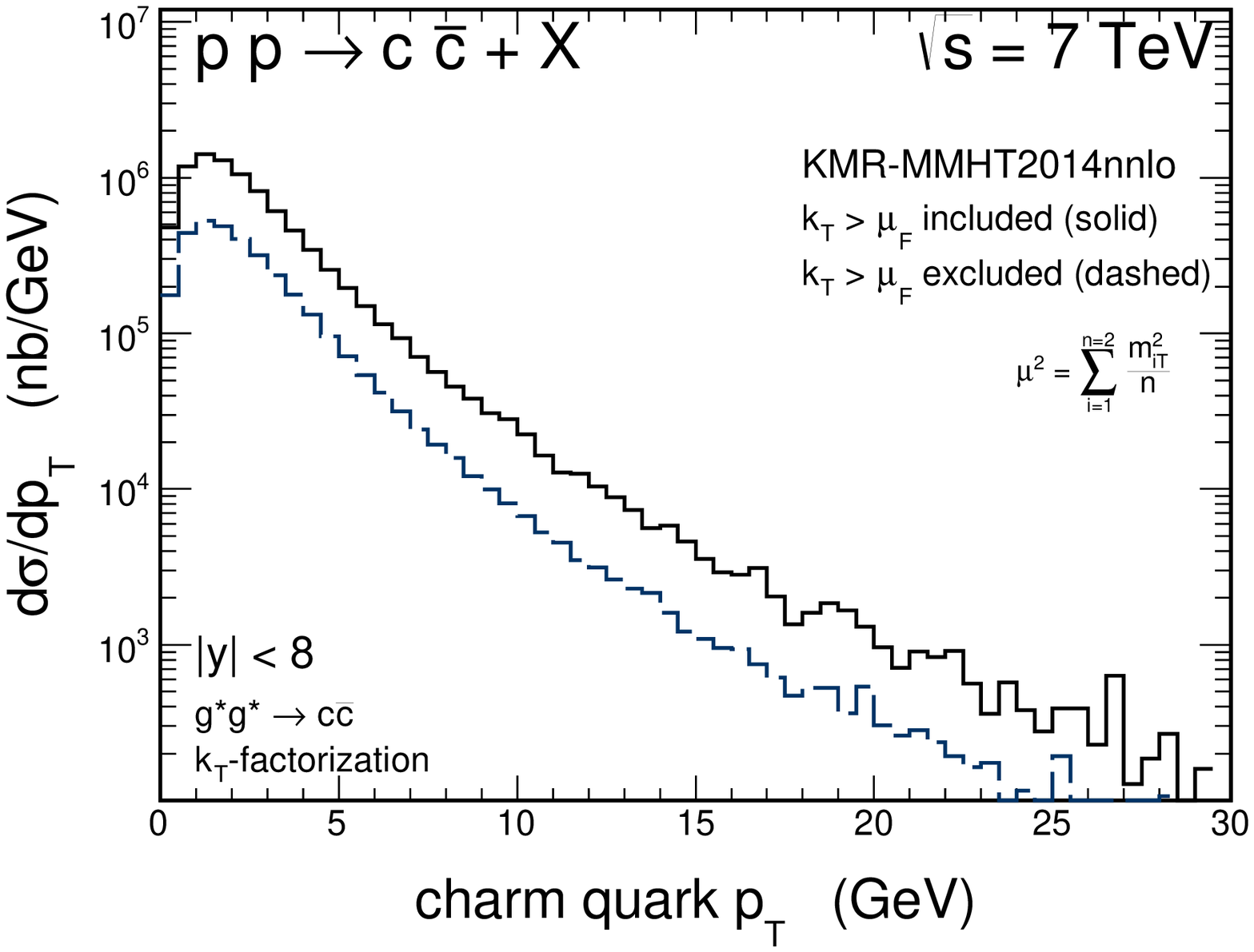}}
\end{minipage}
\begin{minipage}{0.47\textwidth}
 \centerline{\includegraphics[width=1.0\textwidth]{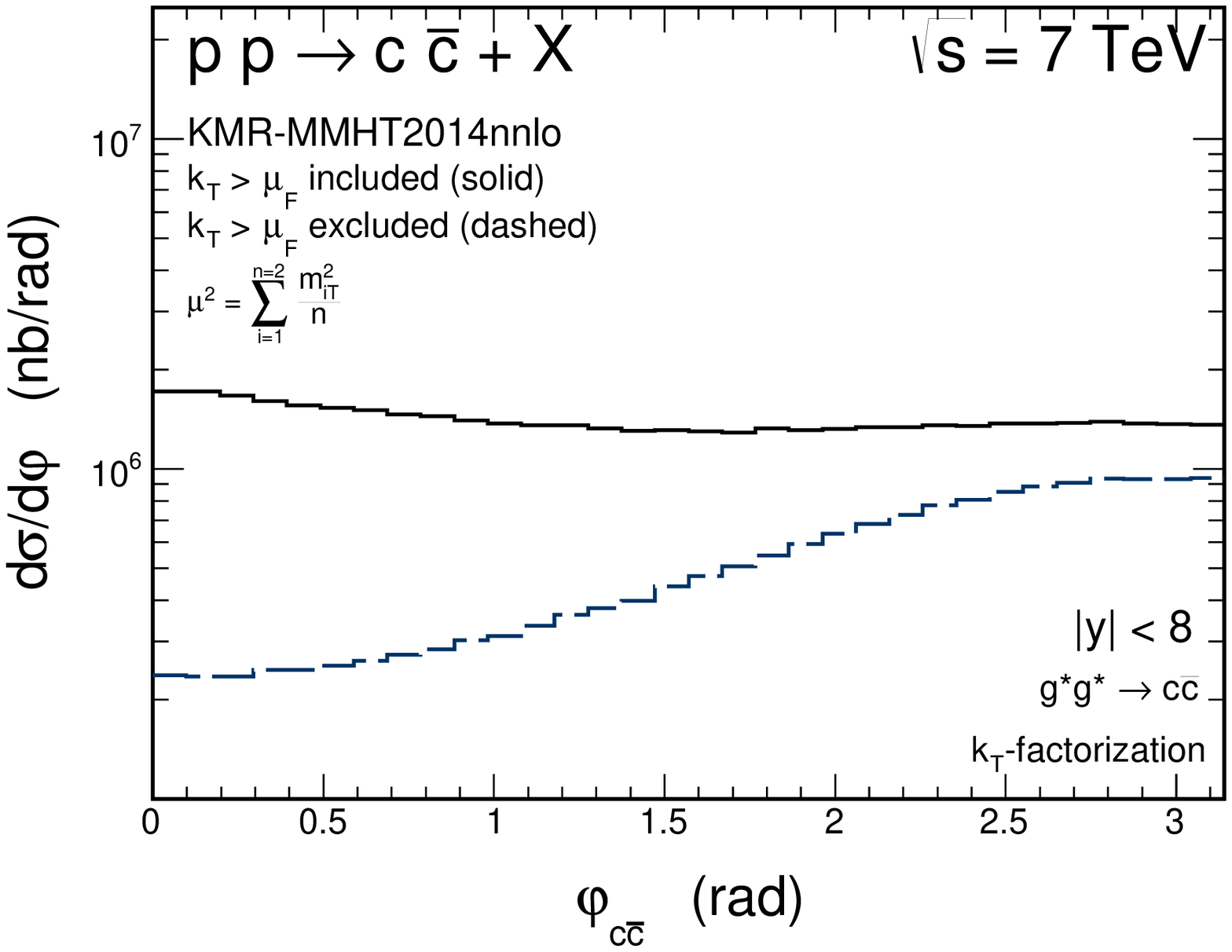}}
\end{minipage}
  \caption{
\small Transverse momentum distribution of charm quark (left) and azimuthal angle distributions $\varphi_{c\bar c}$ between $c$ quark and $\bar c$ antiquark (right) for $\sqrt{s}= 7$ TeV. Here, we compare the KMR-MMHT2014nnlo results with and without the contributions from the $k_{T} > \mu_{F}$ region for the basic $g^*g^* \to c\bar c$ mechanism.
}
\label{fig:6}
\end{figure}

First of all we wish to show the importance of the $k_{T} > \mu_{F}$ contributions in the case of the KMR uPDF for the leading-order $g^*g^* \to c\bar c$ mechanism. In Fig.~\ref{fig:6} we present the $c$-quark transverse momentum (left) and $\varphi_{c\bar c}$ azimuthal angle (right) distributions for $c\bar c$-pair production at $\sqrt{s}=7$ TeV. The solid histograms correspond to the standard KMR calculations with the $k_{T} > \mu_{F}$ limitation included and the dashed histograms are for the calculations with excluded contributions from the $k_{T} > \mu_{F}$ region. We observe a significant differences between the both results. The $k_{T} > \mu_{F}$ contribution is very important for the whole considered distribution of the transverse momenta and concentrated especially at small azimuthal angles. The $k_{T} <  \mu_{F}$ limitation of the KMR uPDF that allows only for soft extra emissions and as a consequence significantly reduces the predicted cross section.  

\begin{figure}[!h]
\centering
\begin{minipage}{0.47\textwidth}
  \centerline{\includegraphics[width=1.0\textwidth]{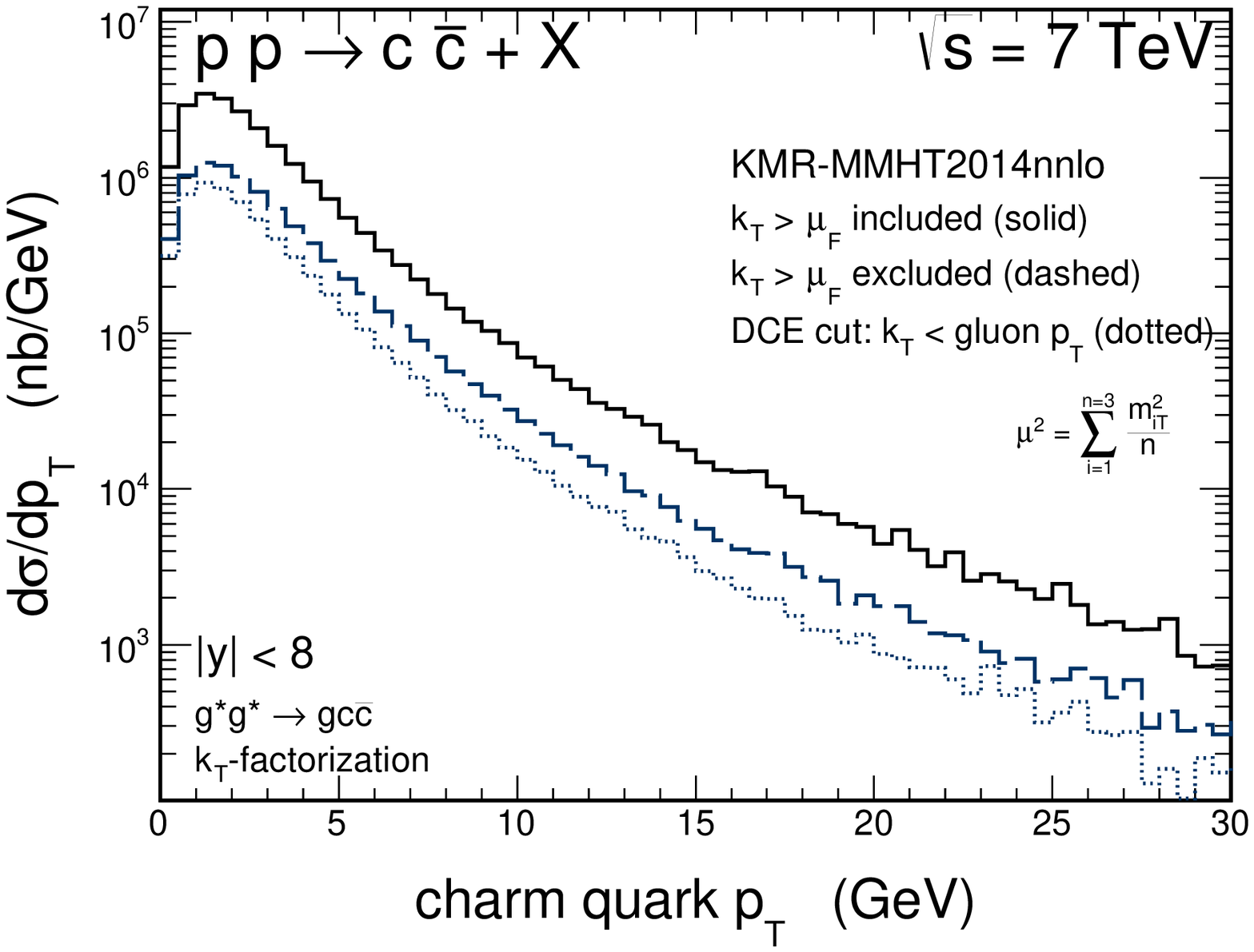}}
\end{minipage}
\begin{minipage}{0.47\textwidth}
 \centerline{\includegraphics[width=1.0\textwidth]{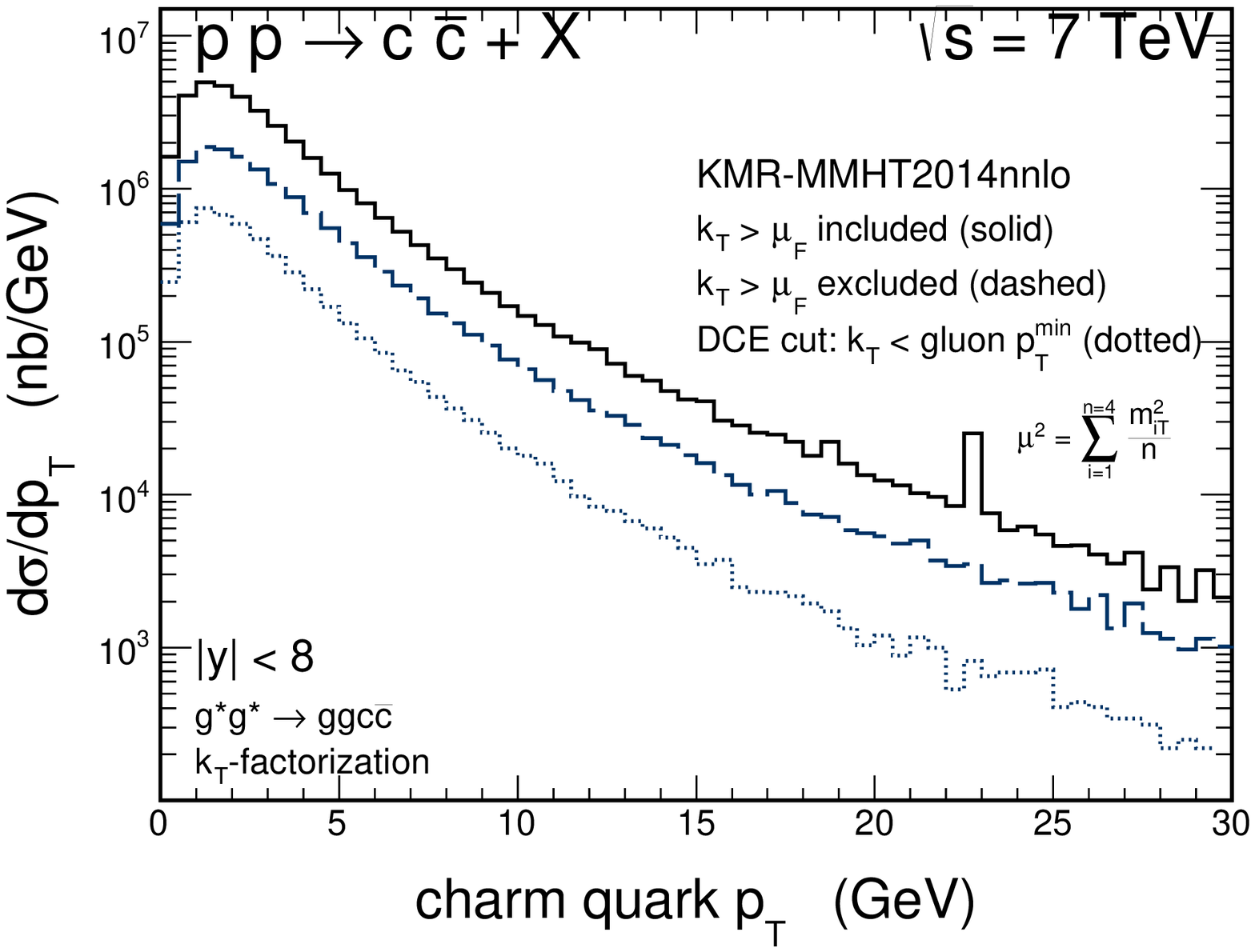}}
\end{minipage}
  \caption{
\small Transverse momentum distribution of charm quark for $\sqrt{s}= 7$ TeV. Here, we compare the KMR-MMHT2014nnlo results with and without the contributions from $k_{T} > \mu_{F}$ region as well as with the double-counting-exclusion cuts included in addition for the $g^*g^* \to gc\bar c$ (left) and $g^*g^* \to ggc\bar c$ (right) mechanisms.
}
\label{fig:7}
\end{figure}

In Fig.~\ref{fig:7} we present a similar analysis as the above one for the higher-order components. Here we consider the role of the $k_{T} > \mu_{F}$ contribution in the KMR uPDF both for the $g^*g^* \to gc\bar c$ (left panel) and $g^*g^* \to ggc\bar c$ (right panel) mechanism. Also in the case of the higher-order processes the $k_{T} > \mu_{F}$ kinematical region in the KMR uPDF significantly contributes to the charm quark production cross sections in the whole range of considered $p_{T}$'s. As we already argued, in the case of higher-order processes the $k_{T} < \mu_{F}$ limitation is not enough to fully avoid double-counting effects when summing up the leading- and higher-order contributions. Therefore, we also plot here the contributions that correspond to the case of the proposed double-counting-exclusion (DCE) cuts (see the dotted histograms). The effects related to these cuts are very important for both, the $2\to 3$ and $2\to 4$ processes. In the latter case the DCE cut significantly reduces the basic cross section by about one order of magnitude.   

\begin{figure}[!h]
\centering
\begin{minipage}{0.47\textwidth}
  \centerline{\includegraphics[width=1.0\textwidth]{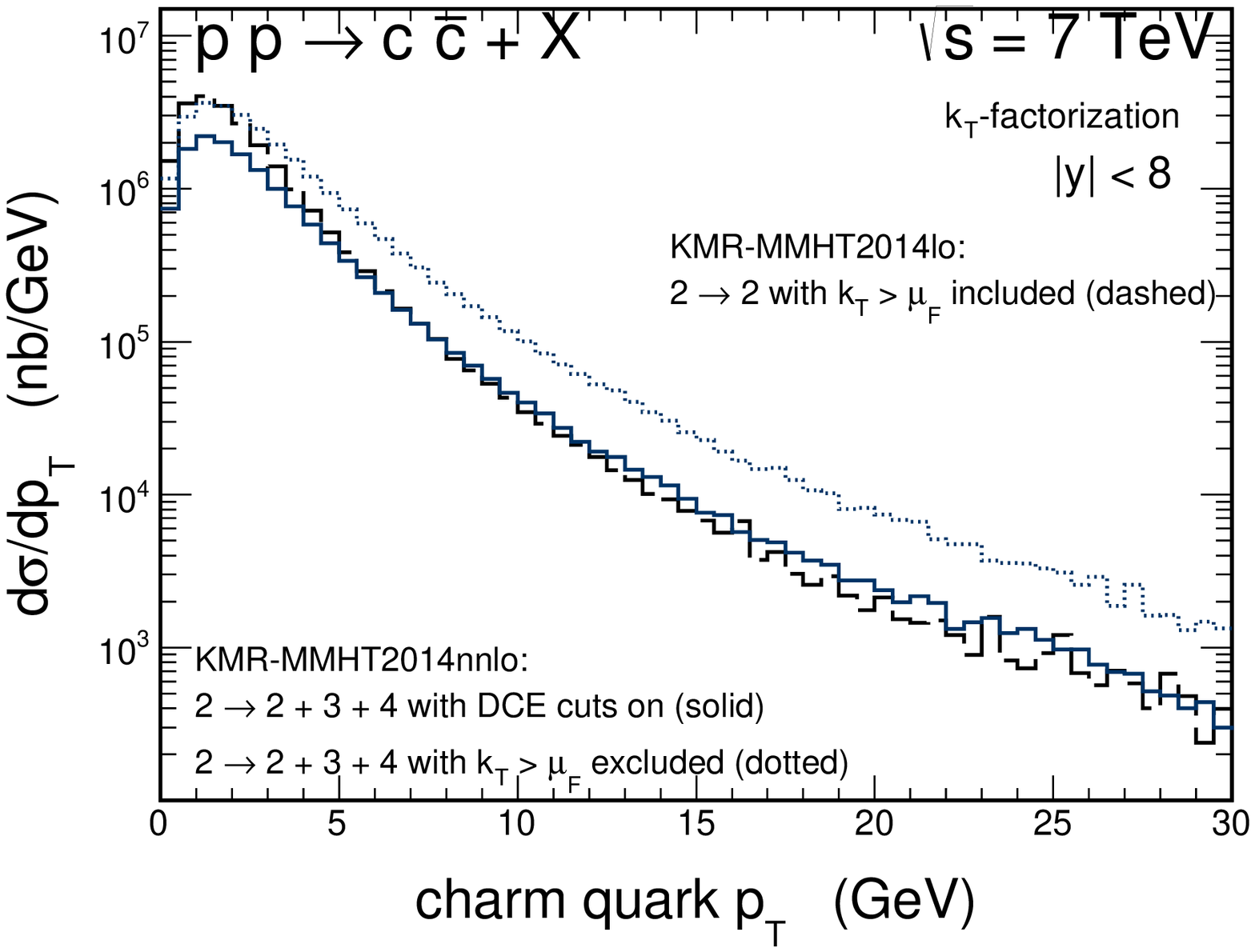}}
\end{minipage}
\begin{minipage}{0.47\textwidth}
 \centerline{\includegraphics[width=1.0\textwidth]{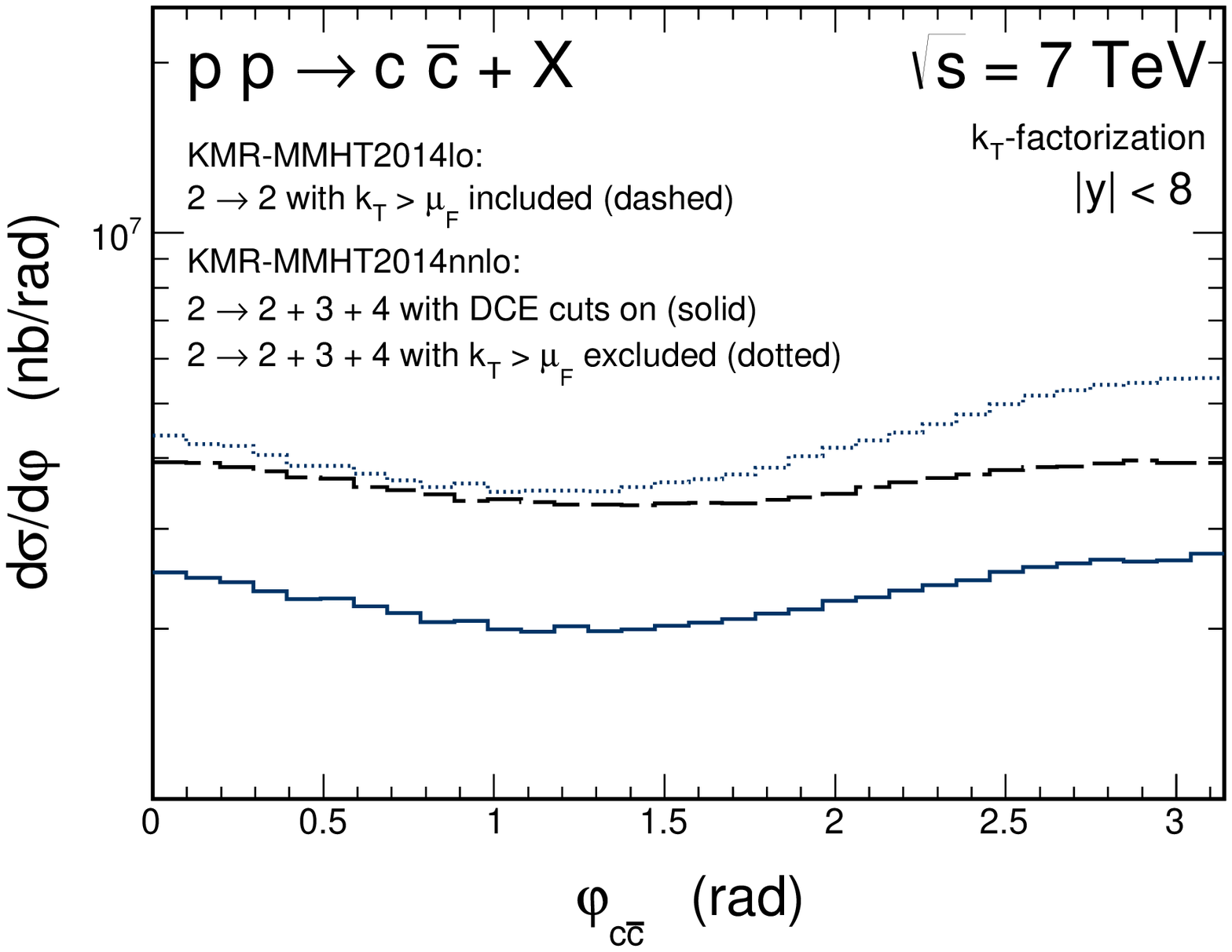}}
\end{minipage}
  \caption{
\small Transverse momentum distribution of charm quark (left) and azimuthal angle distributions $\varphi_{c\bar c}$ between $c$ quark and $\bar c$ antiquark (right) for $\sqrt{s}= 7$ TeV. Here, we compare the KMR-MMHT2014lo $g^*g^* \to c\bar c$ results with the $k_{T} > \mu_{F}$ contribution and the KMR-MMHT2014nnlo results for summed contributions from the $g^*g^* \to c\bar c$, $g^*g^* \to gc\bar c$ and $g^*g^* \to ggc\bar c$ mechanisms with extra conditions.
}
\label{fig:8}
\end{figure}

Now we wish to present the results of the proposed new scheme that includes higher-order corrections explicitly in comparison to the standard (leading-order) KMR calculations.
In Fig.~\ref{fig:8} we show the standard $2\to 2$ KMR calculations with the $k_{T} > \mu_{F}$ included (dashed histograms) and the results obtained within the proposed scheme for $2\to 2 + 3 + 4$ calculations. For the latter case, here we show both results, with only the $k_{T} < \mu_{F}$ limitations (dotted histograms) and with the DCE cuts (solid histograms). We clearly see that the DCE cuts are necessary for $2\to 2 + 3 + 4$ calculations to reproduce the successful standard $2\to 2$ KMR calculations. The calculations with the $k_{T} < \mu_{F}$ limitations would also lead to a significant overestimation of the LHC charm data.
The calculations within the $2\to 2 + 3 + 4$ scheme with the DCE cuts almost coincides with the standard $2\to 2$ calculations in the broad range of the considered transverse momenta of charm quark. Some discrepancy appears only at small $p_T$'s. The reason could be a different collinear PDFs used in both calculations. It is not clear for the $2\to 2$ case whether the LO, NLO or NNLO PDFs should be used, so there we keep the LO PDF as a default while in the case of the $2\to 2 + 3 + 4$ we assume that the NNLO PDFs are the most appropriate.
Our new scheme also leads to a very similar azimuthal angle distribution as in the standard $2\to 2$ calculations.

\subsubsection{The Parton-Branching uPDF}

\begin{figure}[!h]
\centering
\begin{minipage}{0.47\textwidth}
  \centerline{\includegraphics[width=1.0\textwidth]{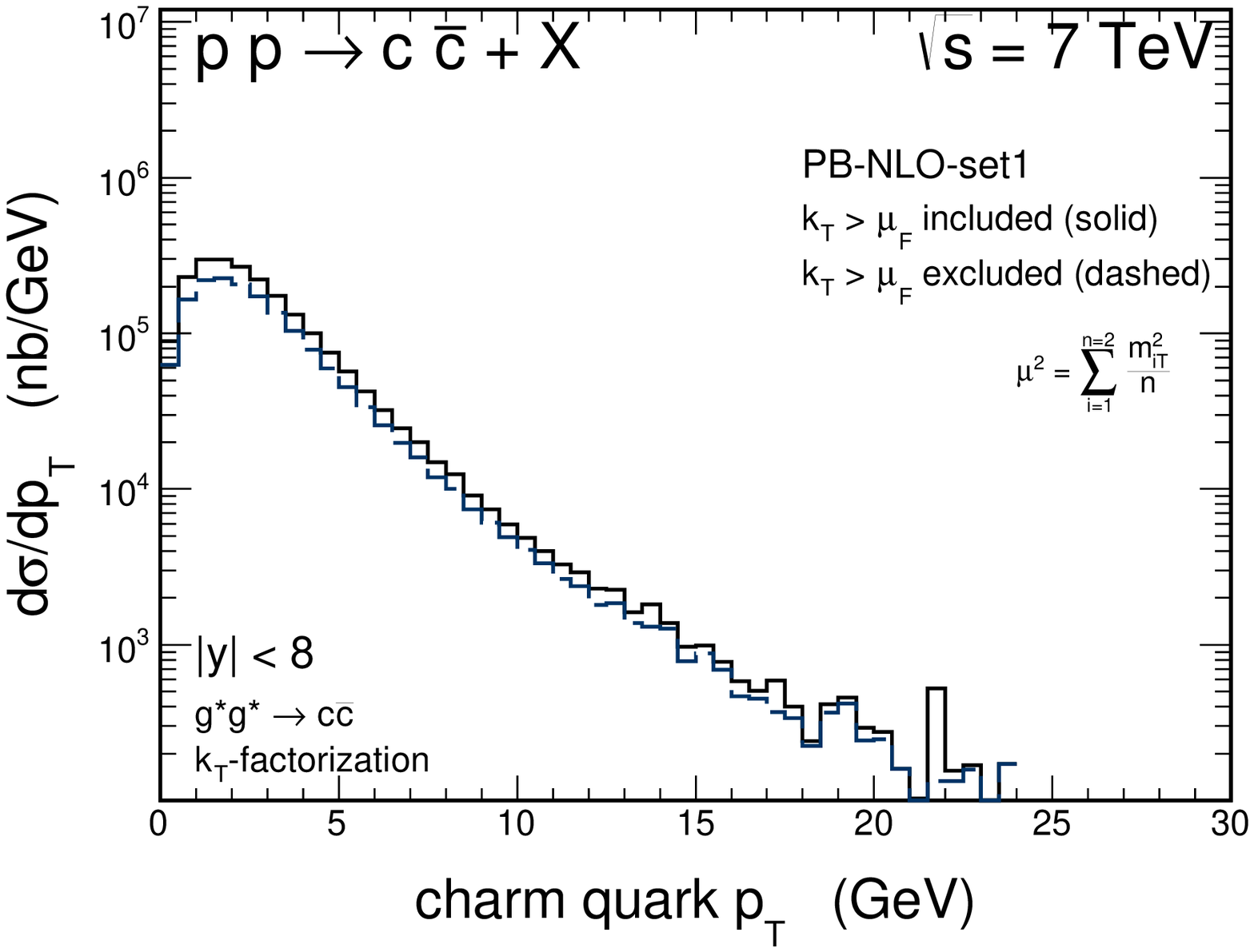}}
\end{minipage}
\begin{minipage}{0.47\textwidth}
 \centerline{\includegraphics[width=1.0\textwidth]{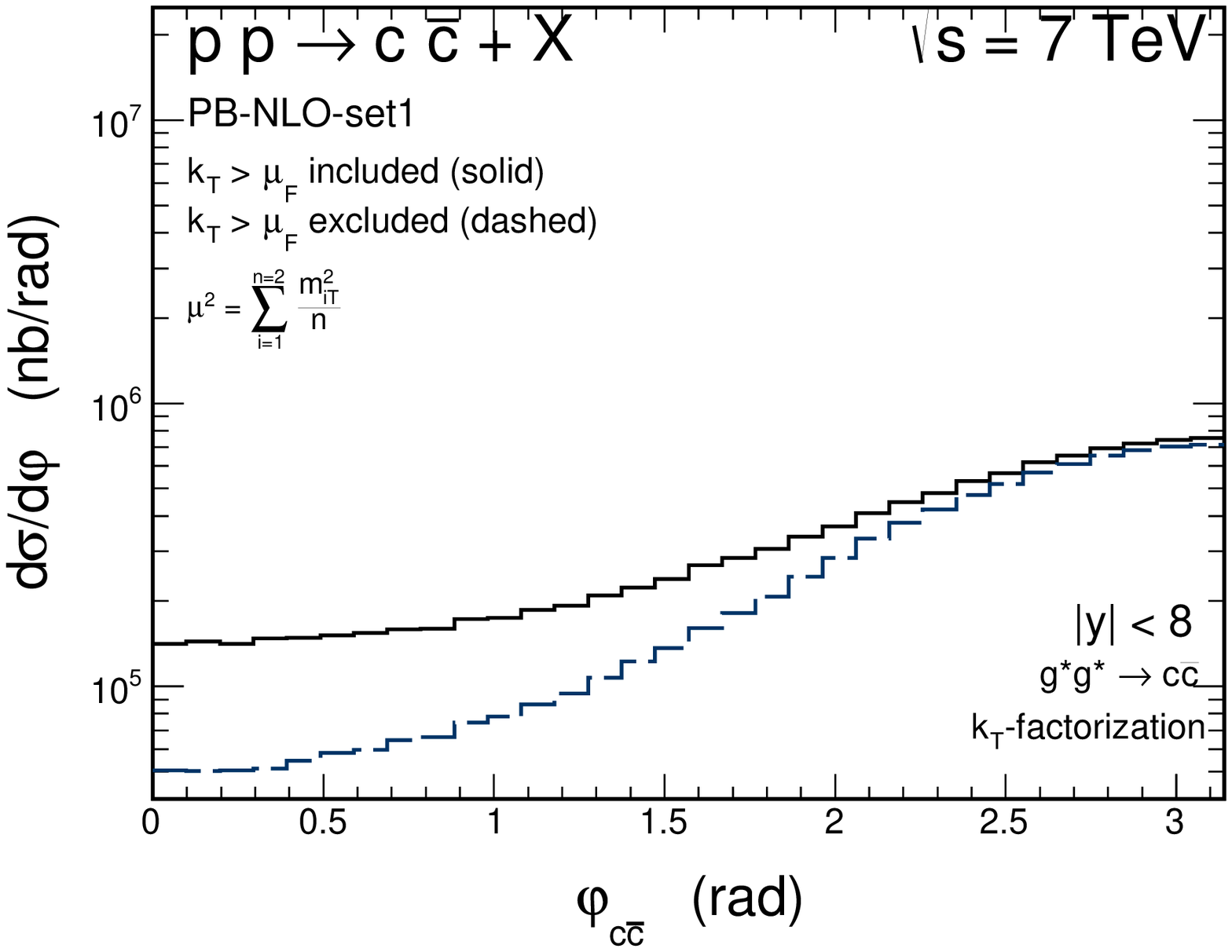}}
\end{minipage}
  \caption{
\small Transverse momentum distribution of charm quark (left) and azimuthal angle distributions $\varphi_{c\bar c}$ between $c$ quark and $\bar c$ antiquark (right) for $\sqrt{s}= 7$ TeV. Here, we compare the PB-NLO-set1 results with and without the contributions from the $k_{T} > \mu_{F}$ region for the $g^*g^* \to c\bar c$ mechanism.
}
\label{fig:9}
\end{figure}

\begin{figure}[!h]
\centering
\begin{minipage}{0.47\textwidth}
  \centerline{\includegraphics[width=1.0\textwidth]{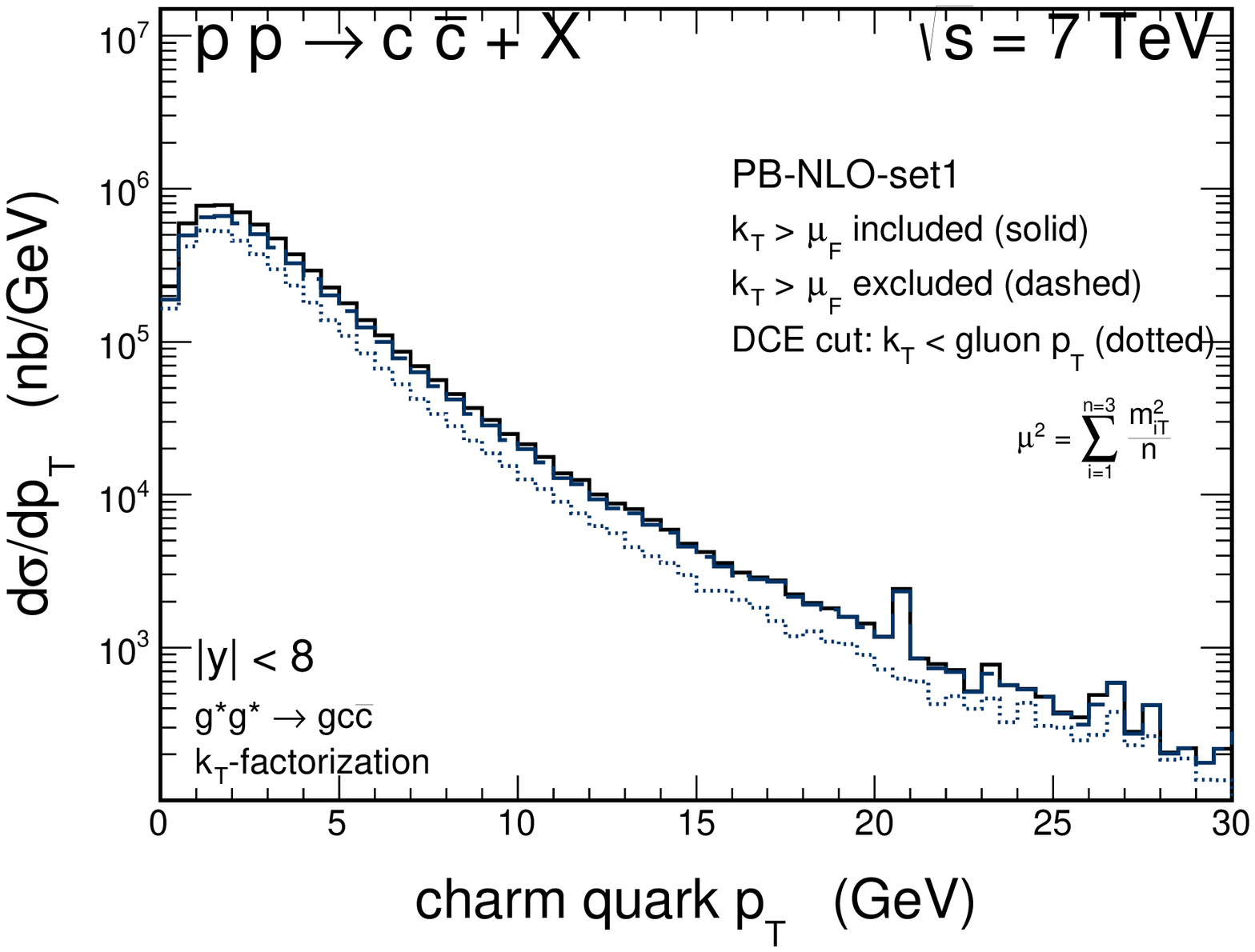}}
\end{minipage}
\begin{minipage}{0.47\textwidth}
 \centerline{\includegraphics[width=1.0\textwidth]{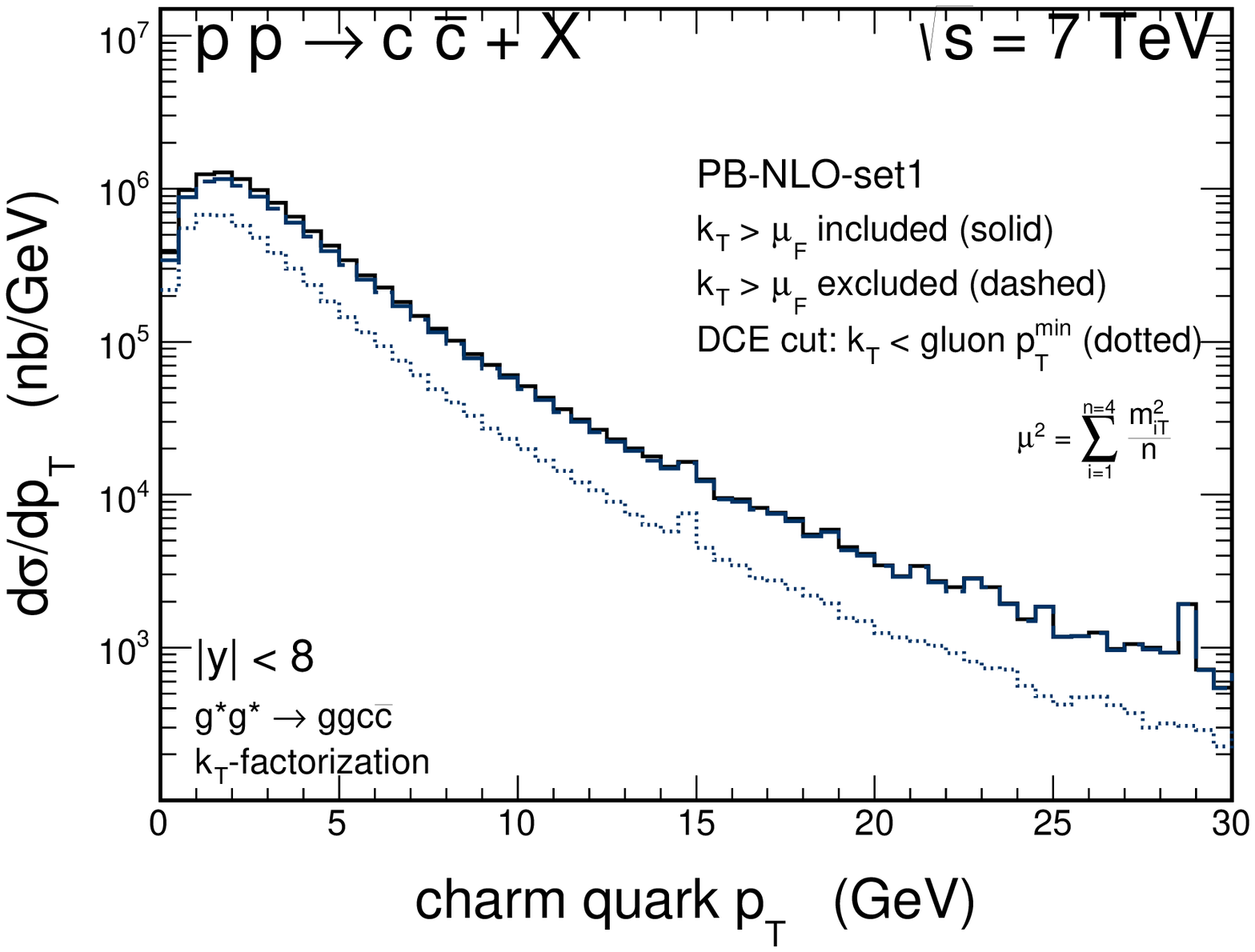}}
\end{minipage}
  \caption{
\small Transverse momentum distribution of charm quark for $\sqrt{s}= 7$ TeV. Here, we compare the PB-NLO-set1 results with and without the contributions from the $k_{T} > \mu_{F}$ region as well as with the extra double-counting-exclusion cuts for the $g^*g^* \to gc\bar c$ (left) and $g^*g^* \to ggc\bar c$ (right) mechanisms.
}
\label{fig:10}
\end{figure}

\begin{figure}[!h]
\centering
\begin{minipage}{0.47\textwidth}
  \centerline{\includegraphics[width=1.0\textwidth]{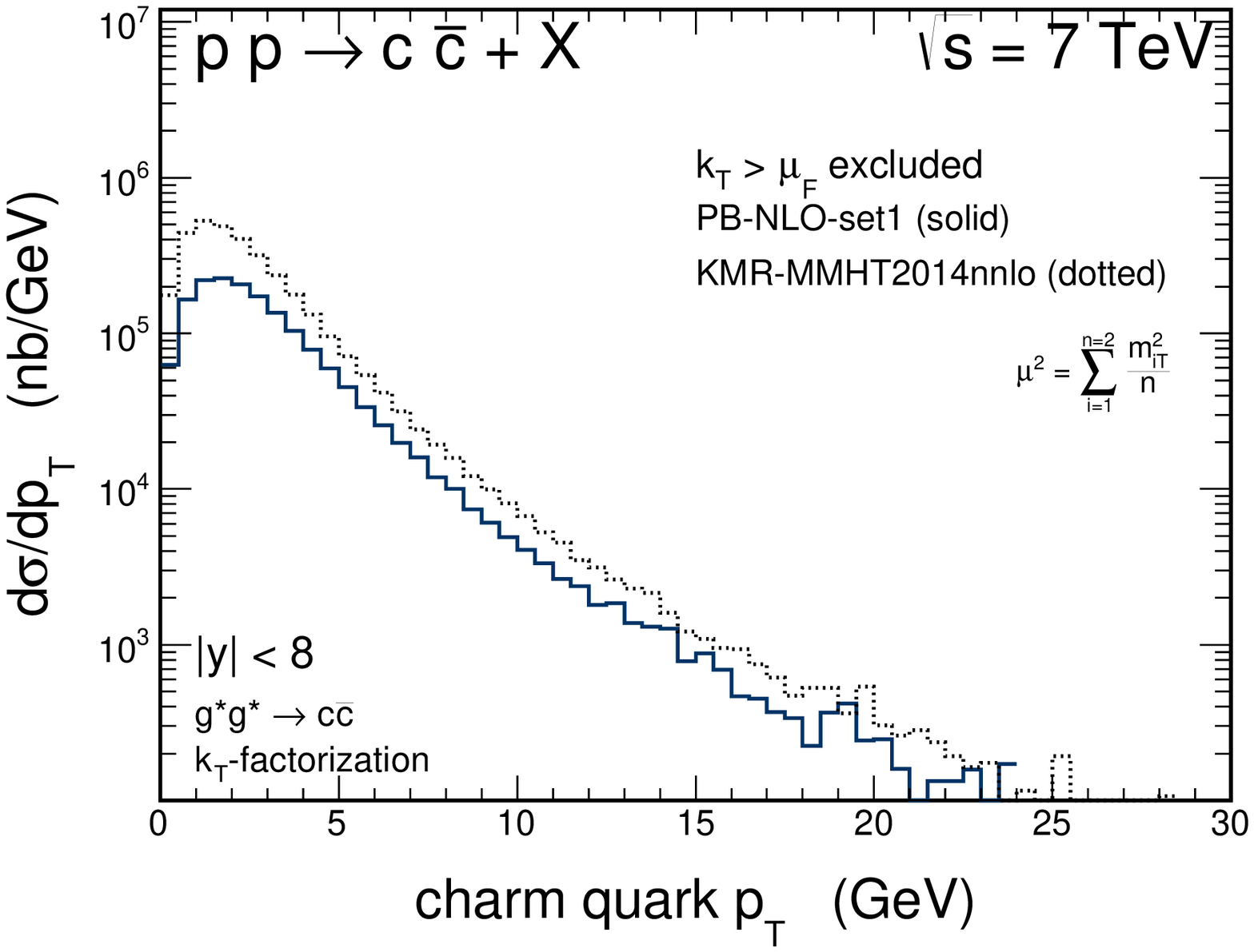}}
\end{minipage}
\begin{minipage}{0.47\textwidth}
 \centerline{\includegraphics[width=1.0\textwidth]{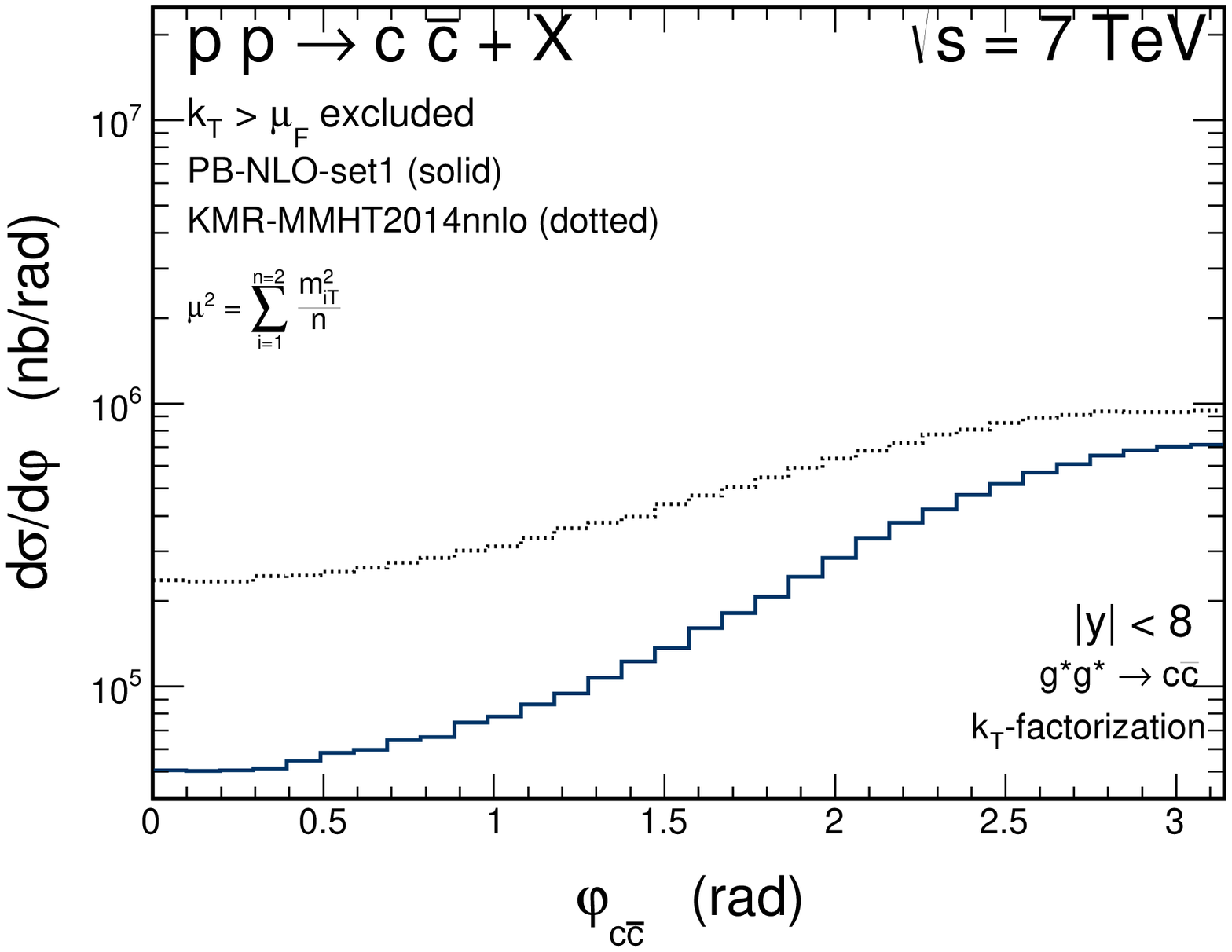}}
\end{minipage}\\
\begin{minipage}{0.47\textwidth}
  \centerline{\includegraphics[width=1.0\textwidth]{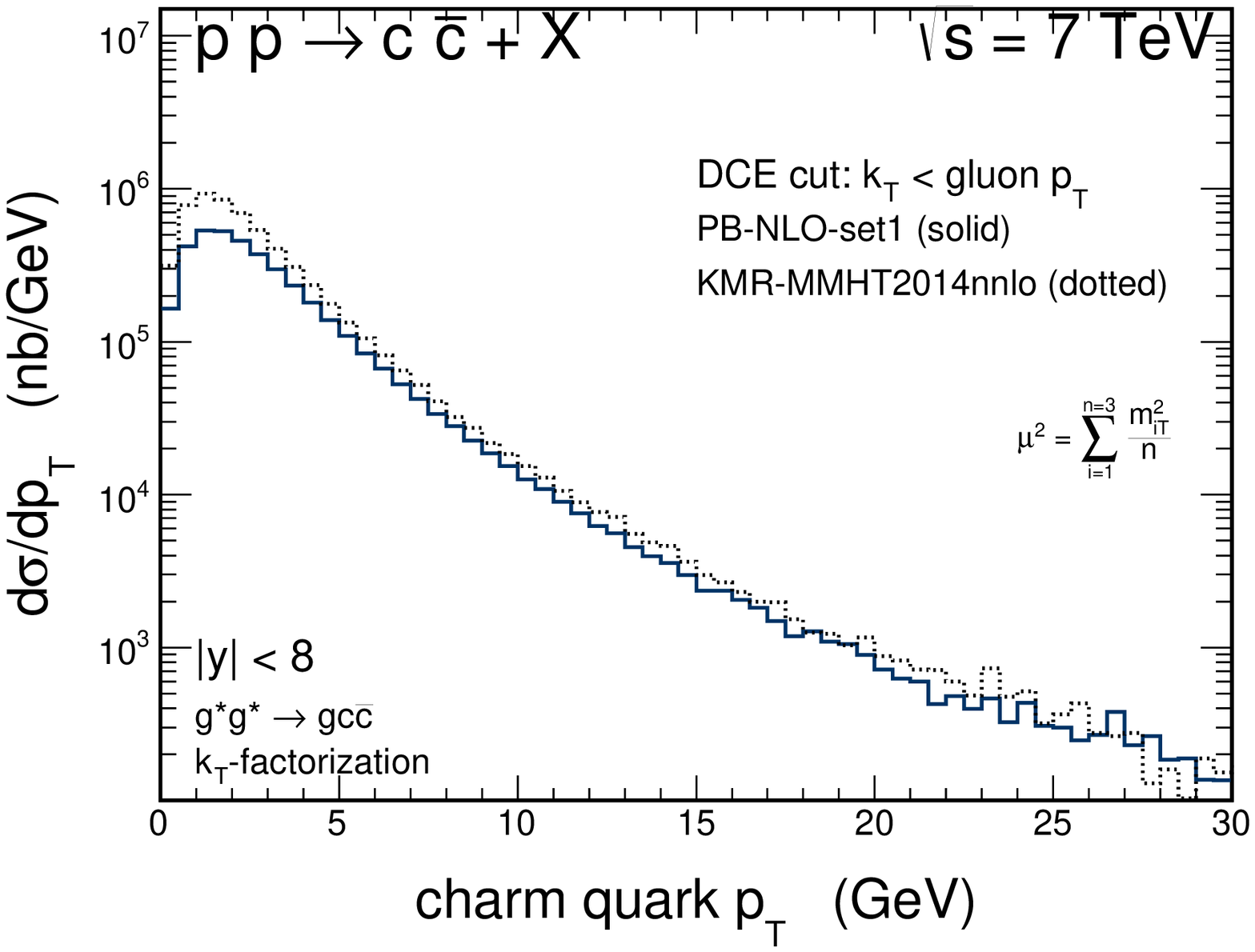}}
\end{minipage}
\begin{minipage}{0.47\textwidth}
 \centerline{\includegraphics[width=1.0\textwidth]{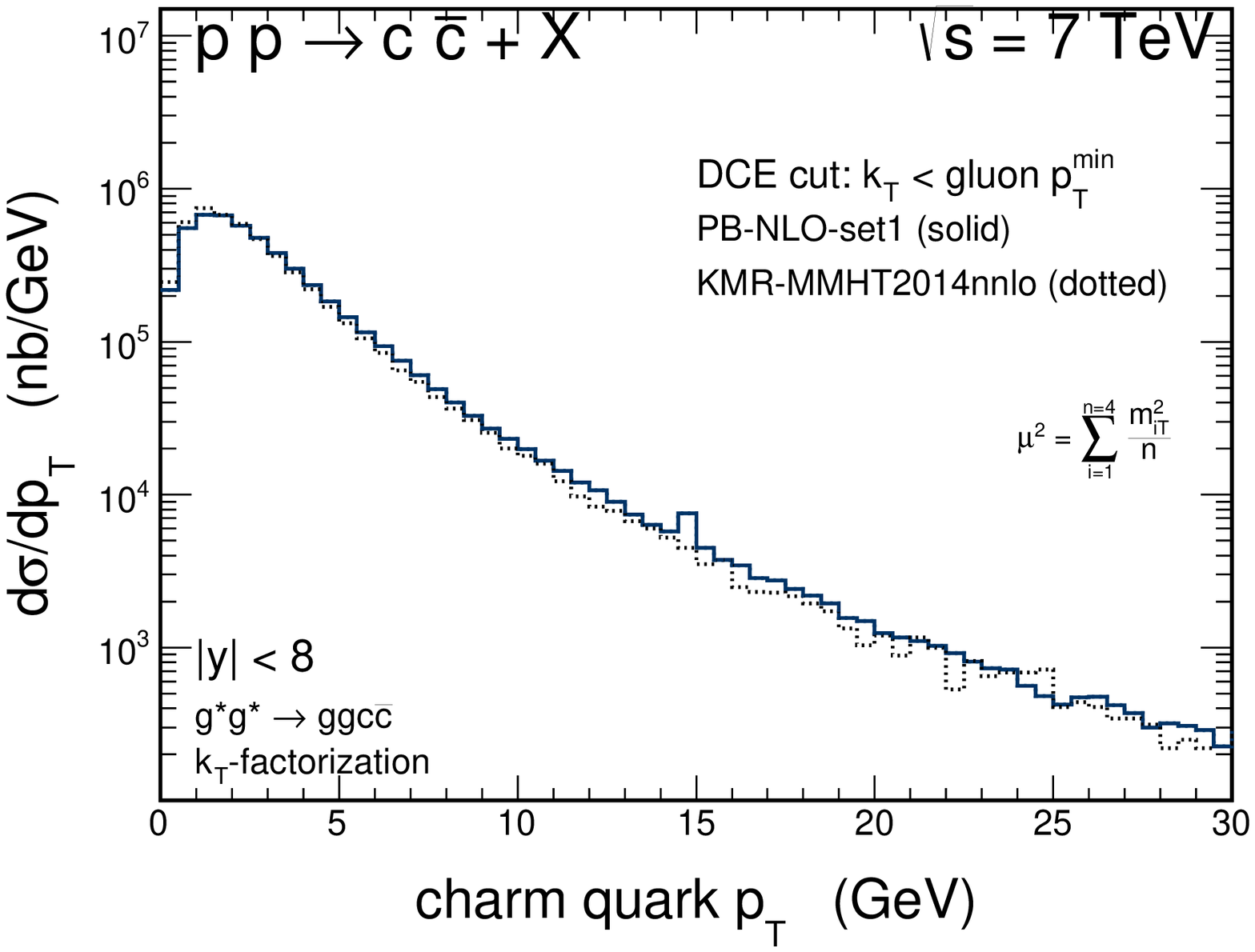}}
\end{minipage}
  \caption{
\small A comparison of the PB-NLO-set1 and the KMR-MMHT2014nnlo results for the $g^*g^* \to c\bar c$ (upper), $g^*g^* \to gc\bar c$ (bottom left) and $g^*g^* \to ggc\bar c$ (bottom right) mechanisms. The extra conditions used in the calculations are specified in the figures.
}
\label{fig:11}
\end{figure}

In this subsection we basically repeat the above studies for the KMR uPDF but here we apply the Parton-Branching uPDFs \cite{Hautmann:2017fcj,Martinez:2018jxt}. As we observe from Figs.~\ref{fig:9} and \ref{fig:10} in the case of the PB uPDFs the $k_{T} > \mu_{F}$ contributions are very small for the $2\to 2$ mechanism
and almost negligible for the $2\to 3$ and $2\to 4$ higher-orders (see almost coinciding solid and dashed histograms in Fig.~\ref{fig:10}). Therefore, it is imposible to decribe the LHCb open charm data within the PB uPDFs when considering only the $g^*g^* \to c\bar c$ mechanism. Here, the effects of the DCE cuts are still sizeable, however, much smaller than in the case of the KMR uPDF. In Fig.~\ref{fig:11}, for a more general comparison, we show
the $2\to 2 + 3 + 4$ results with the DCE cuts for the two uPDFs together on the same plots.
In the case of the $2\to2$ mechanism some differences bewteen the two results are obtained. For the $2\to3$ and $2\to4$ mechanisms, when the DCE cuts are imposed the results almost coincide.  

\begin{figure}[!h]
\centering
\begin{minipage}{0.47\textwidth}
  \centerline{\includegraphics[width=1.0\textwidth]{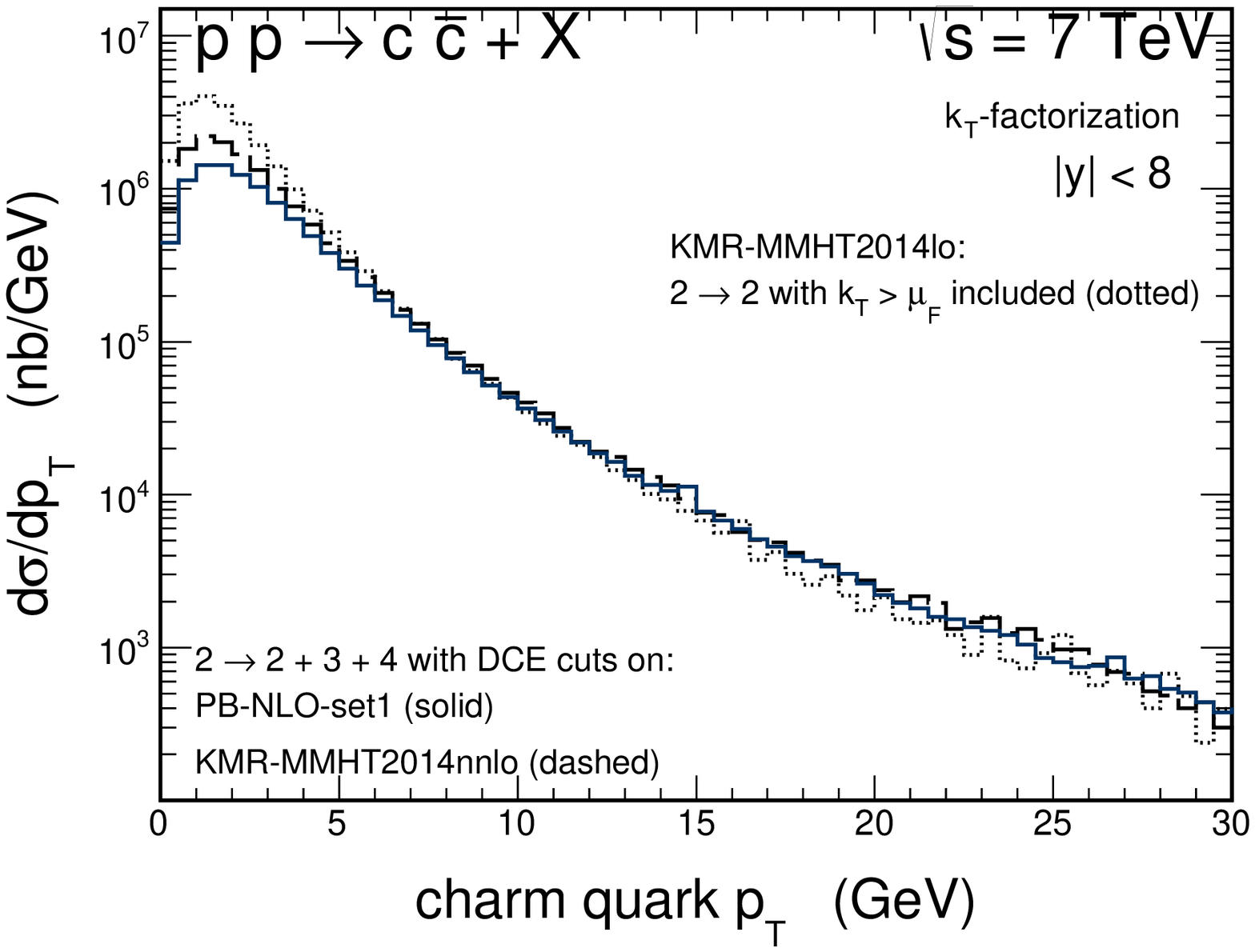}}
\end{minipage}
\begin{minipage}{0.47\textwidth}
 \centerline{\includegraphics[width=1.0\textwidth]{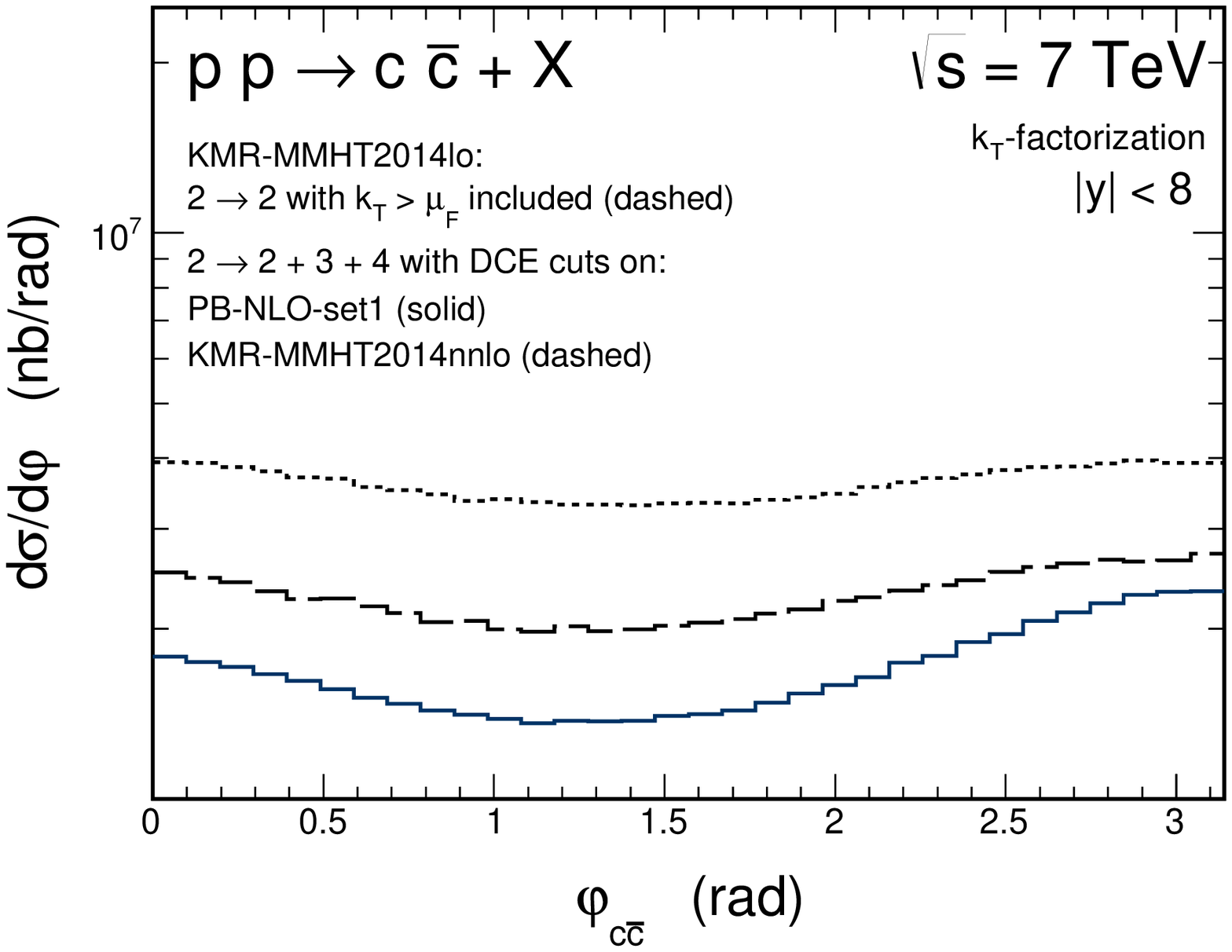}}
\end{minipage}
  \caption{
\small Transverse momentum distribution of charm quark (left) and azimuthal angle distributions $\varphi_{c\bar c}$ between $c$ quark and $\bar c$ antiquark (right) for $\sqrt{s}= 7$ TeV. Here, we compare the KMR-MMHT2014lo $g^*g^* \to c\bar c$ results with $k_{T} > \mu_{F}$ with the KMR-MMHT2014nnlo and PB-NLO-set1 results for summed contributions from $g^*g^* \to c\bar c$, $g^*g^* \to gc\bar c$ and $g^*g^* \to ggc\bar c$ mechanisms with the extra conditions.
}
\label{fig:12}
\end{figure}

In Fig.~\ref{fig:12} we again show the standard $2\to 2$ KMR calculations with the $k_{T} > \mu_{F}$ included (dashed histograms) and the results obtained within the proposed scheme for the $2\to 2 + 3 + 4$ calculations but with the KMR (dashed histograms) and the PB uPDFs (solid histograms) in addition. One can observe that the $2\to 2 + 3 + 4$ calculations for the two different uPDFs lead to very similar results.  
The proposed procedure is the only scheme which allows for a reasonable prediction for heavy flavour production within the PB uPDFs. Further improvement can be done only by the full NLO/NNLO $k_{T}$-factorization calculations.  

\subsubsection{Numerical representation of the double-counting exclusion cuts}

\begin{figure}[!h]
\centering
\begin{minipage}{0.32\textwidth}
  \centerline{\includegraphics[width=1.0\textwidth]{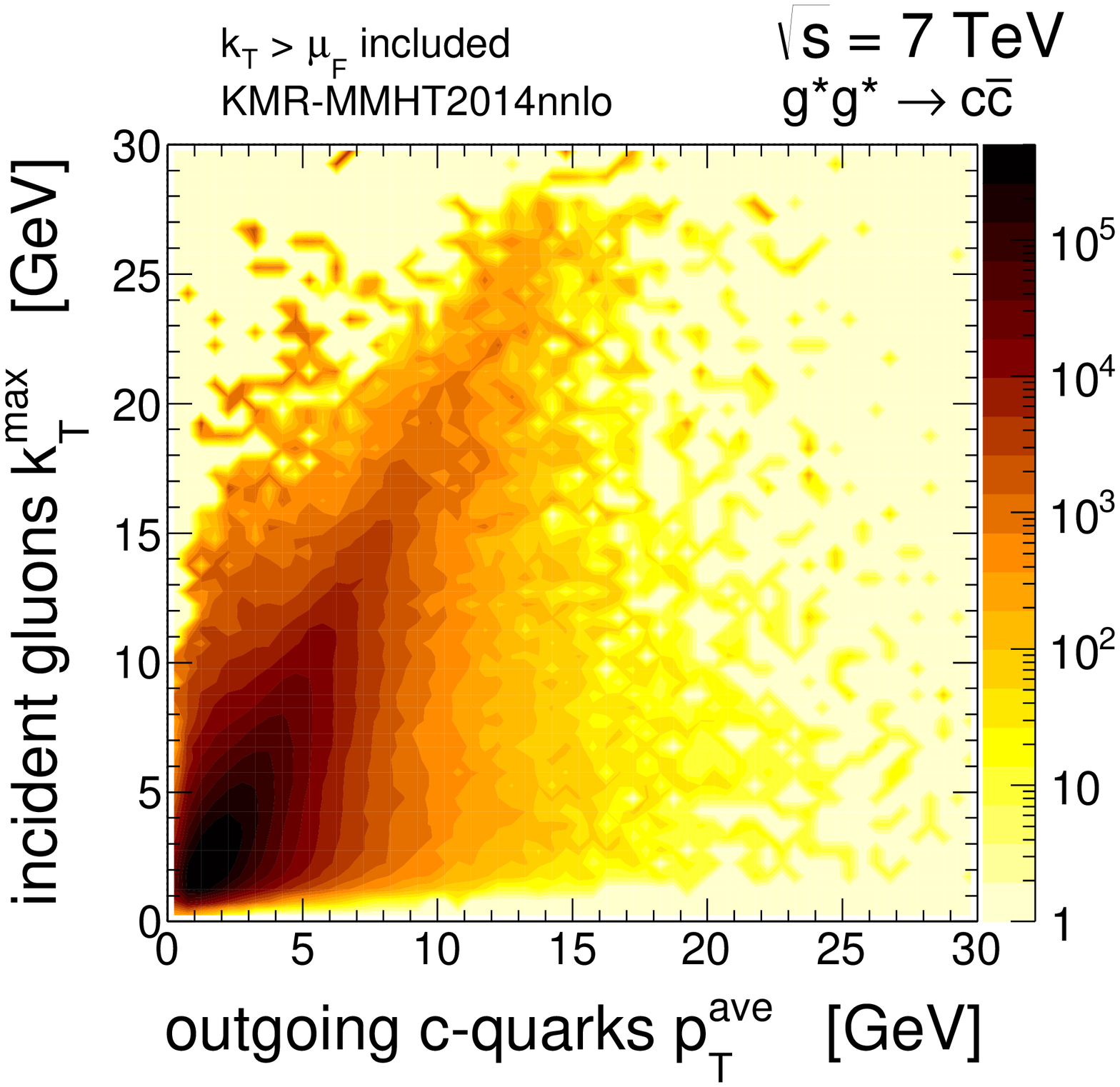}}
\end{minipage}
\begin{minipage}{0.32\textwidth}
 \centerline{\includegraphics[width=1.0\textwidth]{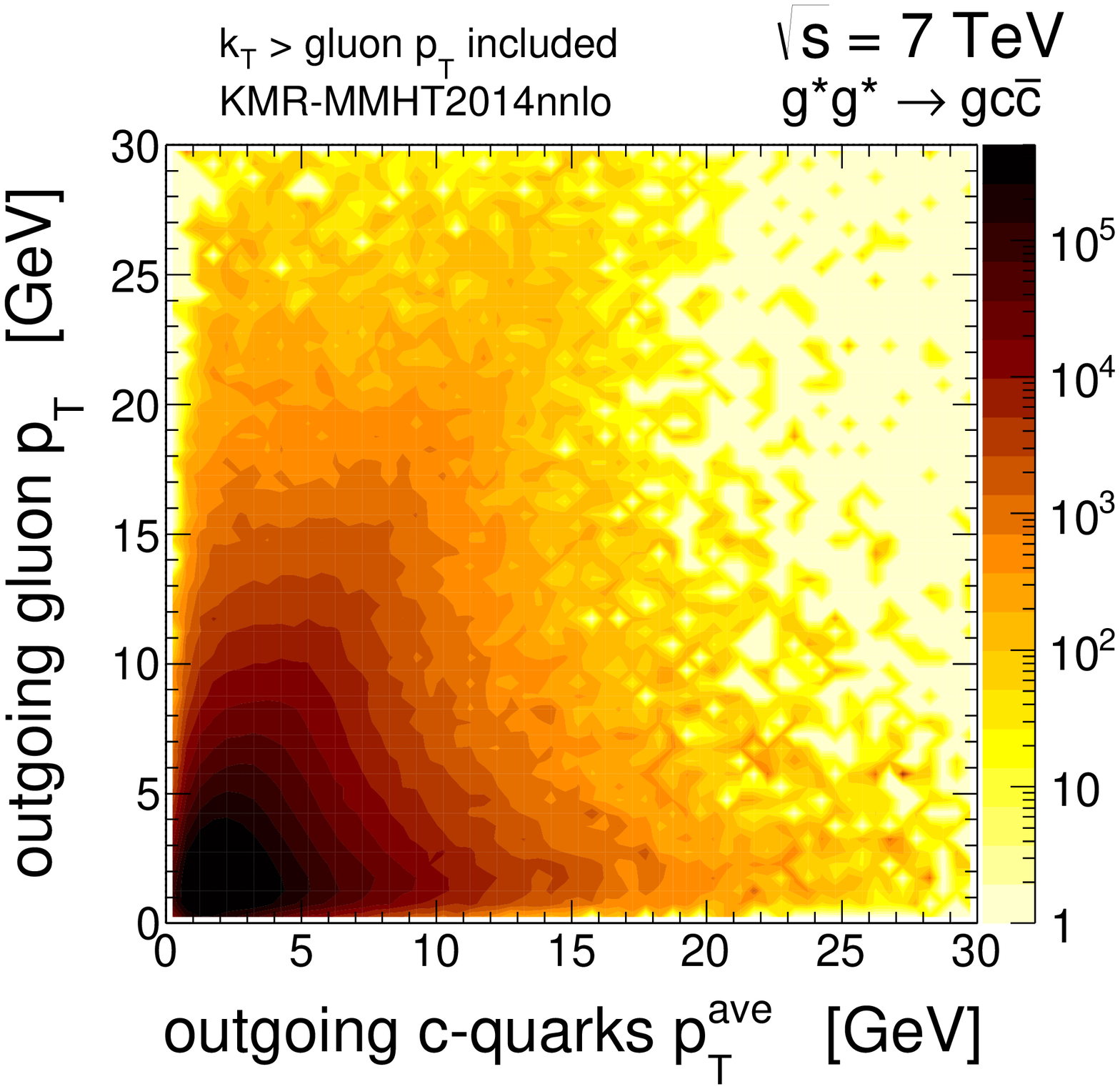}}
\end{minipage}
\begin{minipage}{0.32\textwidth}
 \centerline{\includegraphics[width=1.0\textwidth]{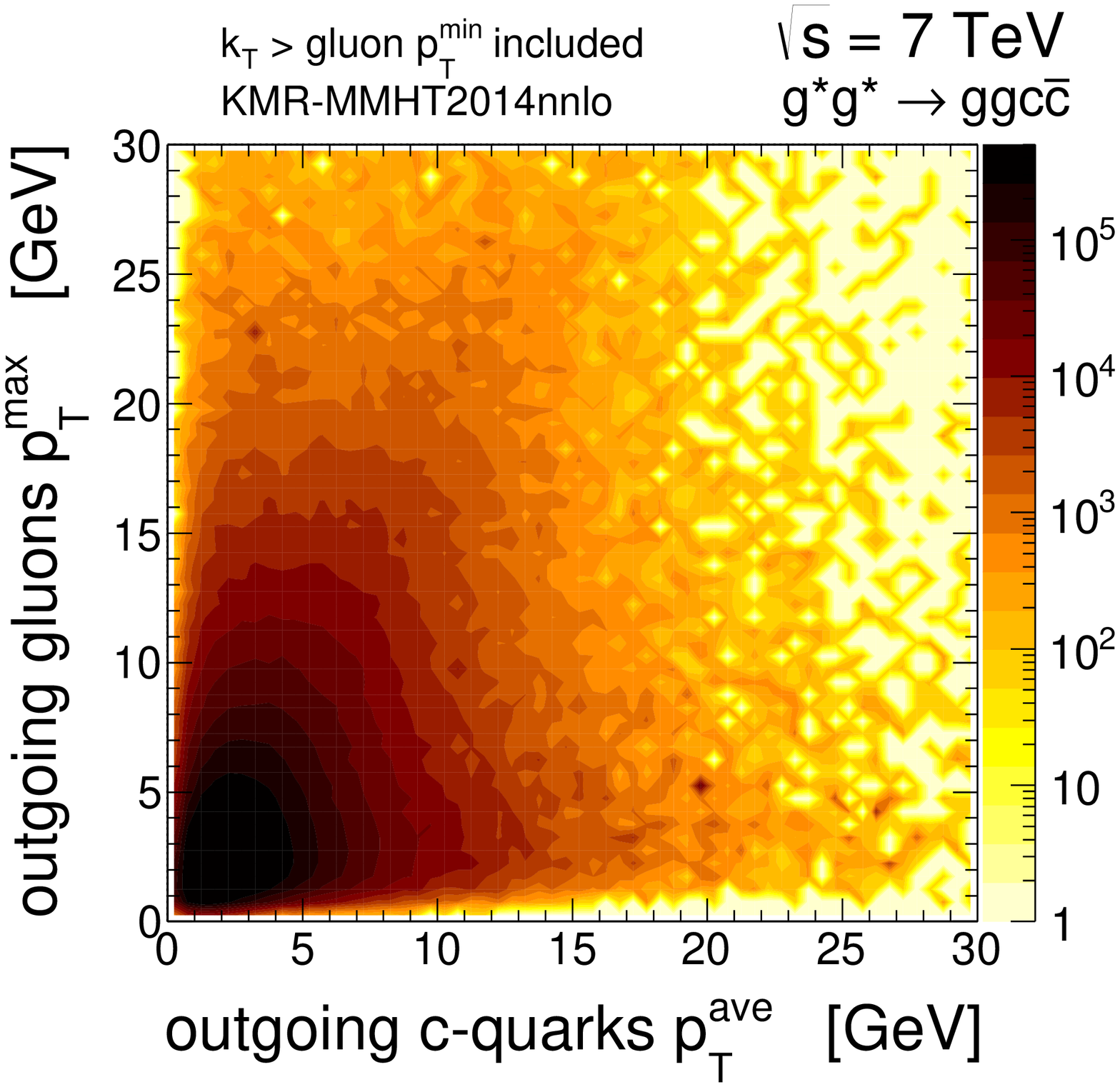}}
\end{minipage}\\
\begin{minipage}{0.32\textwidth}
  \centerline{\includegraphics[width=1.0\textwidth]{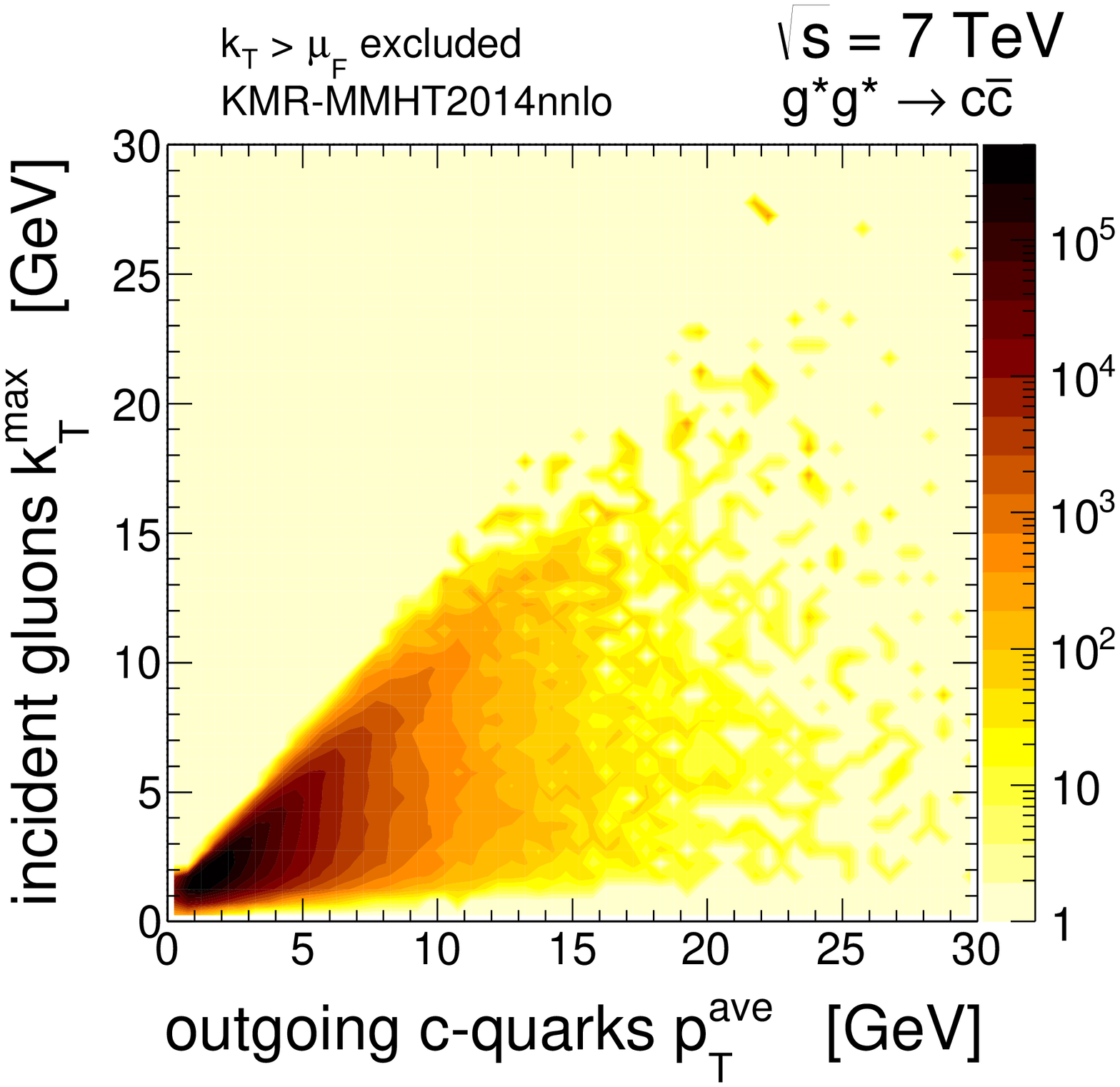}}
\end{minipage}
\begin{minipage}{0.32\textwidth}
 \centerline{\includegraphics[width=1.0\textwidth]{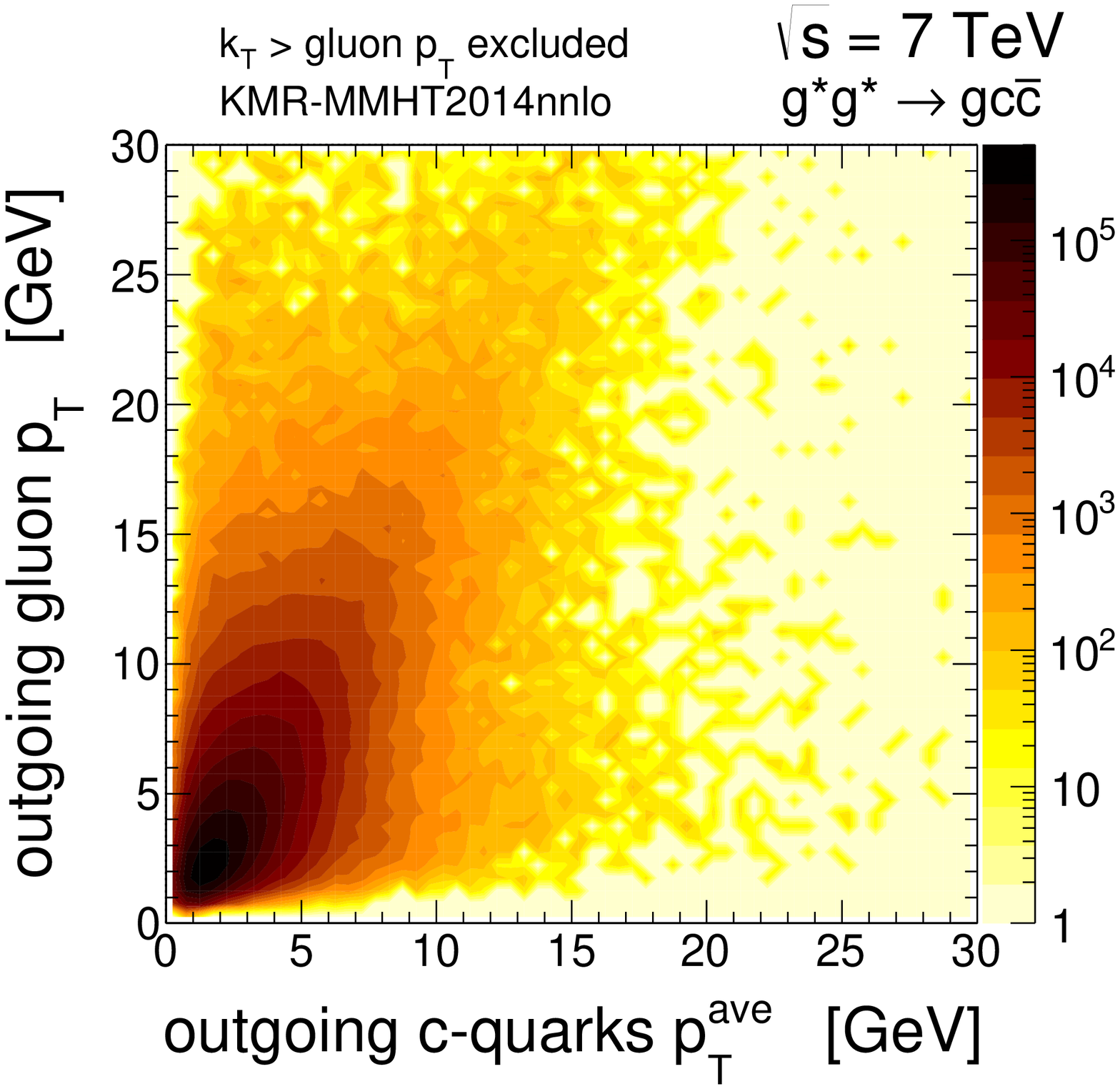}}
\end{minipage}
\begin{minipage}{0.32\textwidth}
 \centerline{\includegraphics[width=1.0\textwidth]{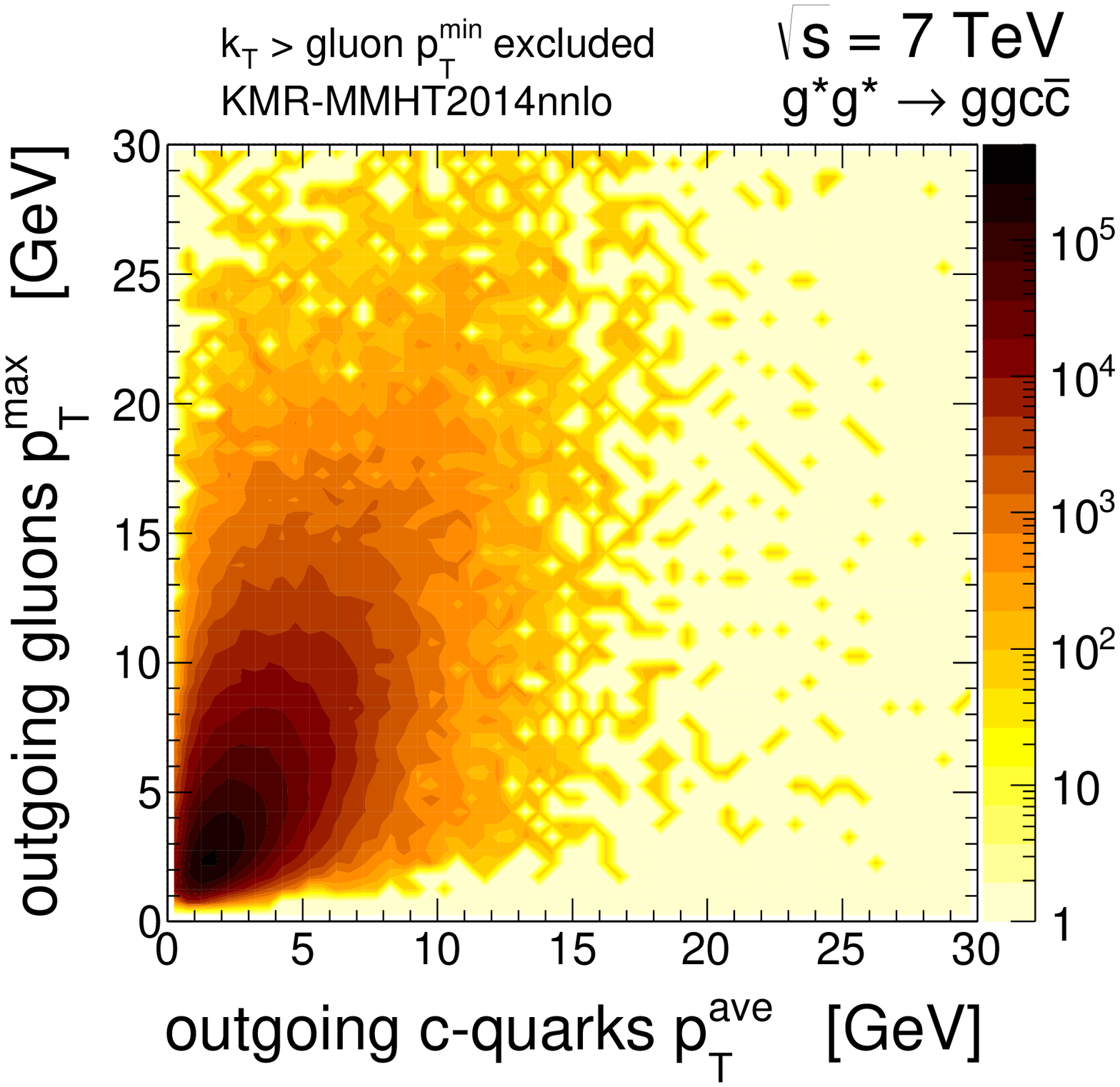}}
\end{minipage}
  \caption{
\small Correlations of the transverse momentum of the leading gluon jet and averaged transverse momentum of the charm quark and antiquark for the $2\to 2$ (left panels), $2\to 3$ (middle panels) and $2\to 4$ (right panels) events. The top panels correspond to the direct calculations without the DCE cuts. The bottom panels correspond to the calculations with the DCE cuts included. In the latter case a kind of a separation of the leading- and higher-order contributions is obtained. 
}
\label{fig:DCEcuts-2dim}
\end{figure}

Here, we wish to illustrate whether the proposed double-counting exclusion cuts
can really result in separation of the leading- and higher-order contributions. In Fig.~\ref{fig:DCEcuts-2dim} we plot the two-dimensional distributions 
as a function of the leading gluon jet $p_T$ and averaged transverse momentum of the charm quark and antiquark in the $2\to 2$ (left panels), $2\to 3$ (middle panels) and $2\to 4$ (right panels) events that could be helpful in schematic illustration of the complementarity of phase spaces for the leading- and higher-order contributions. The top panels correspond to the direct calculations without the DCE cuts while the bottom panels correspond to the calculations with the DCE cuts included. As we observe, the DCE cuts remove from the $2\to 2$ calculations the contributions of the $2\to 3$ and $2\to 4$ type, as well as
suppress the $2\to 3$ and $2\to 4$ components in the region populated by the $2\to 2$ mechanism. Although, the separation is not sharp which may be related to the chosen scales, the main tendency of the applied procedure is clear and seems to support the applied procedure. Similar conclusions were drawn in Ref.~\cite{Karpishkov:2016hnx}
in the case of $b\bar b$-pair production.

\subsubsection{Double-counting and two-dimensional distributions}

\begin{figure}[!h]
\centering
\begin{minipage}{0.32\textwidth}
  \centerline{\includegraphics[width=1.0\textwidth]{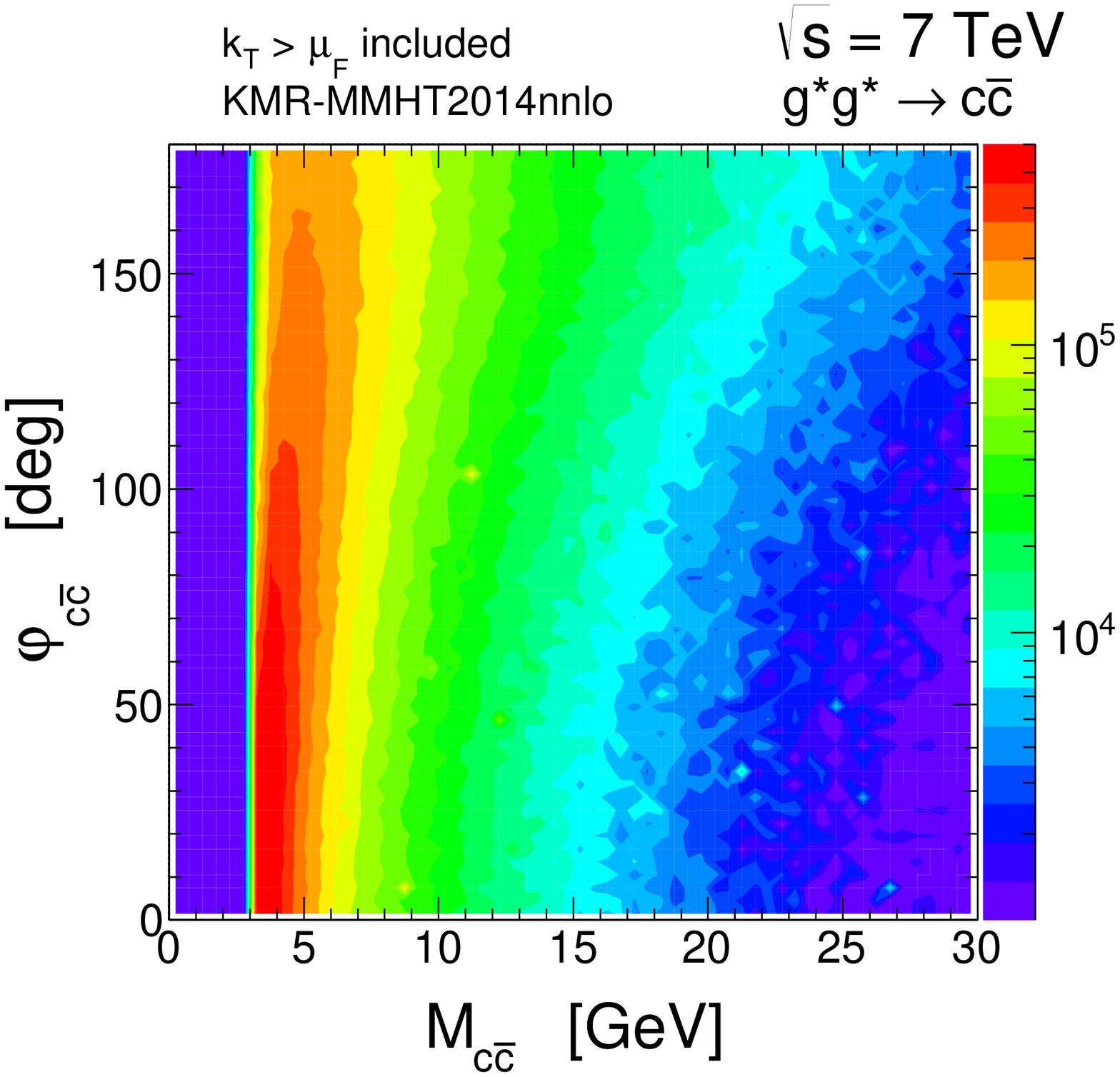}}
\end{minipage}
\begin{minipage}{0.32\textwidth}
 \centerline{\includegraphics[width=1.0\textwidth]{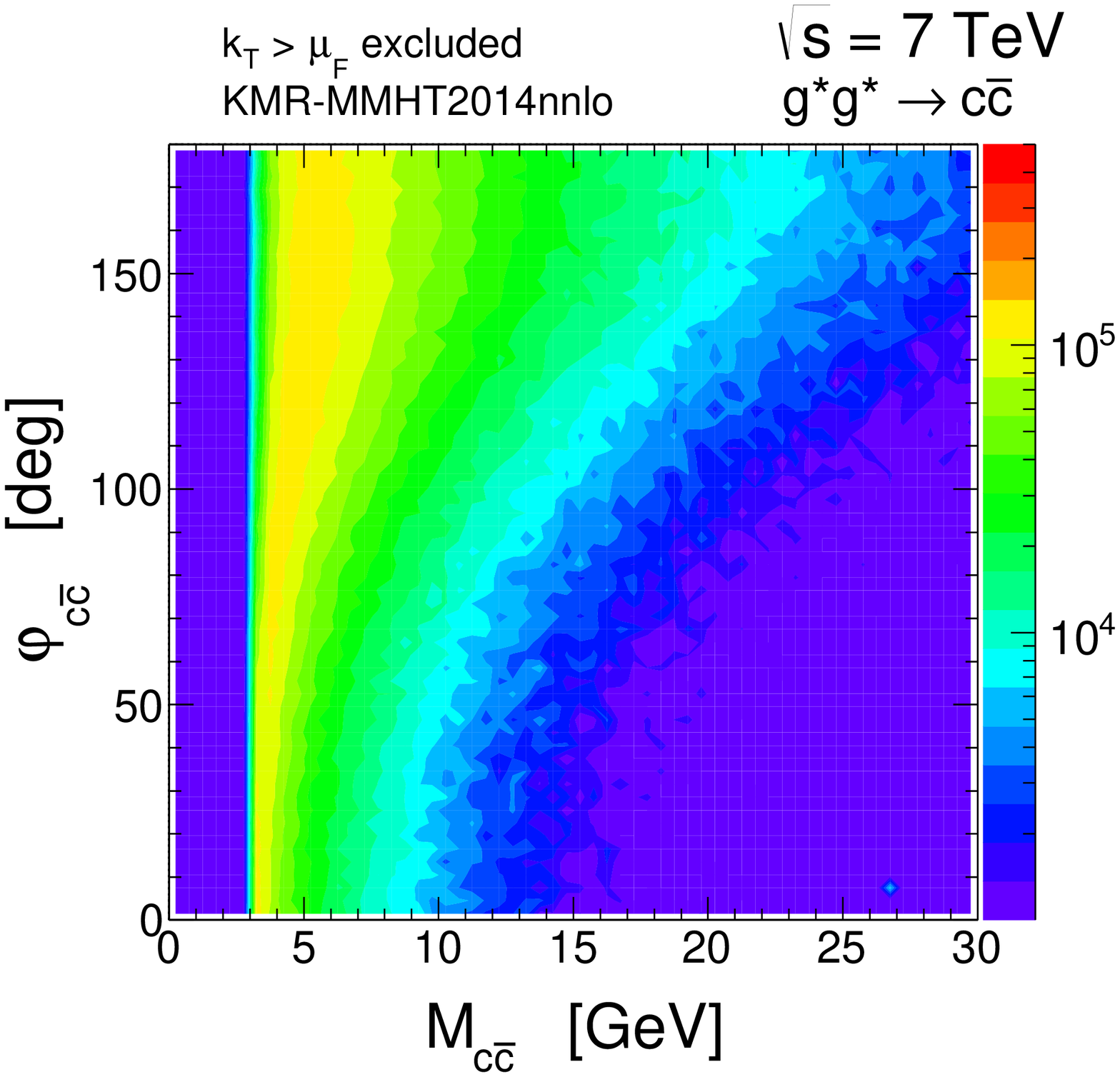}}
\end{minipage}
\begin{minipage}{0.32\textwidth}
 \centerline{\includegraphics[width=1.0\textwidth]{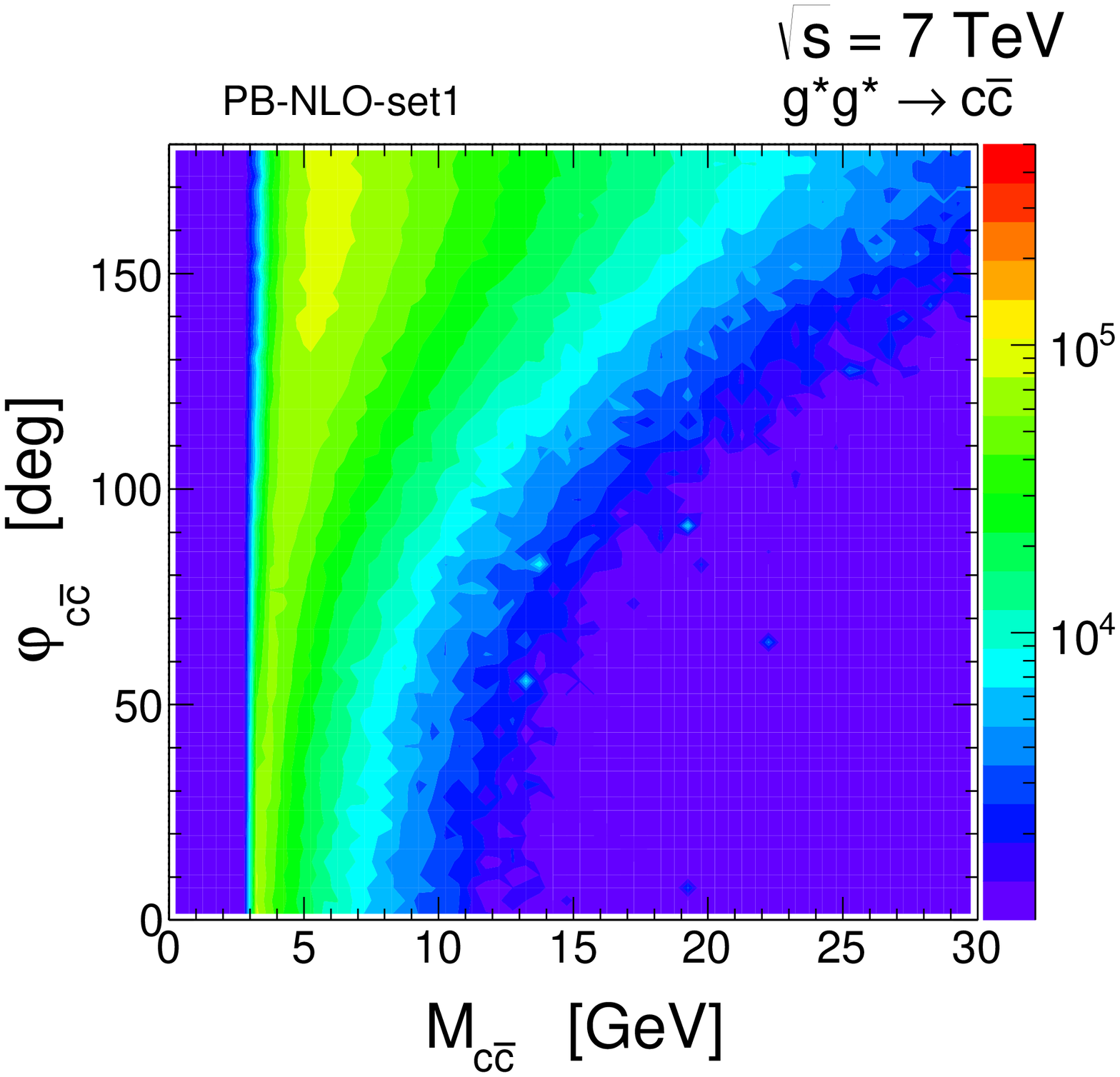}}
\end{minipage}
  \caption{
\small Correlations of the $M_{c \bar c}$ and $\phi_{c \bar c}$ in the case of the $2\to 2$ mechanism.
The left and middle panels correspond to the calculations for the KMR-MMHT2014nnlo uPDF with and without the DCE cut, respectively.
The right panel corresponds to the calculations for the PB-NLO-set1 uPDF.
}
\label{fig:1-2dim}
\end{figure}

We wish to discuss also how our prescription devoted to avoid double counting 
works for example for a two-dimensional distribution in
$(M_{c \bar c}, \phi_{c \bar c})$.
In Fig.~\ref{fig:1-2dim} we show results for the standard
KMR prescription (left panel) and when applying $k_T < \mu_F$ cut
(middle panel). For comparison we show also results obtained with the
PB-NLO-set1 (right panel).
As one can see for the KMR approach the cut removes the strength at small 
invariant masses and small $\varphi_{c \bar c}$. This region is
much less populated when using PB-NLO uPDF.

\begin{figure}[!h]
\centering
\begin{minipage}{0.32\textwidth}
  \centerline{\includegraphics[width=1.0\textwidth]{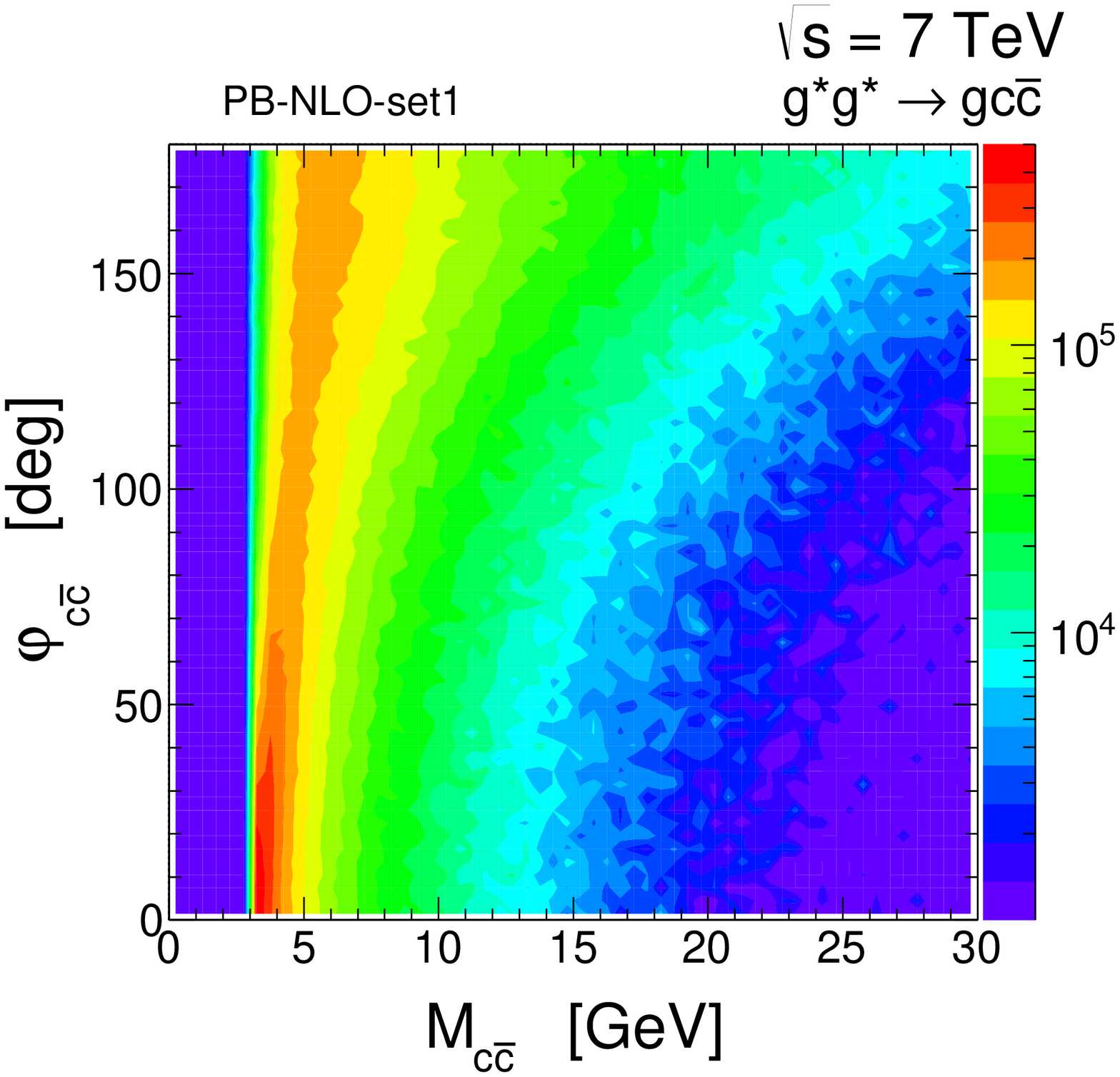}}
\end{minipage}
\begin{minipage}{0.32\textwidth}
 \centerline{\includegraphics[width=1.0\textwidth]{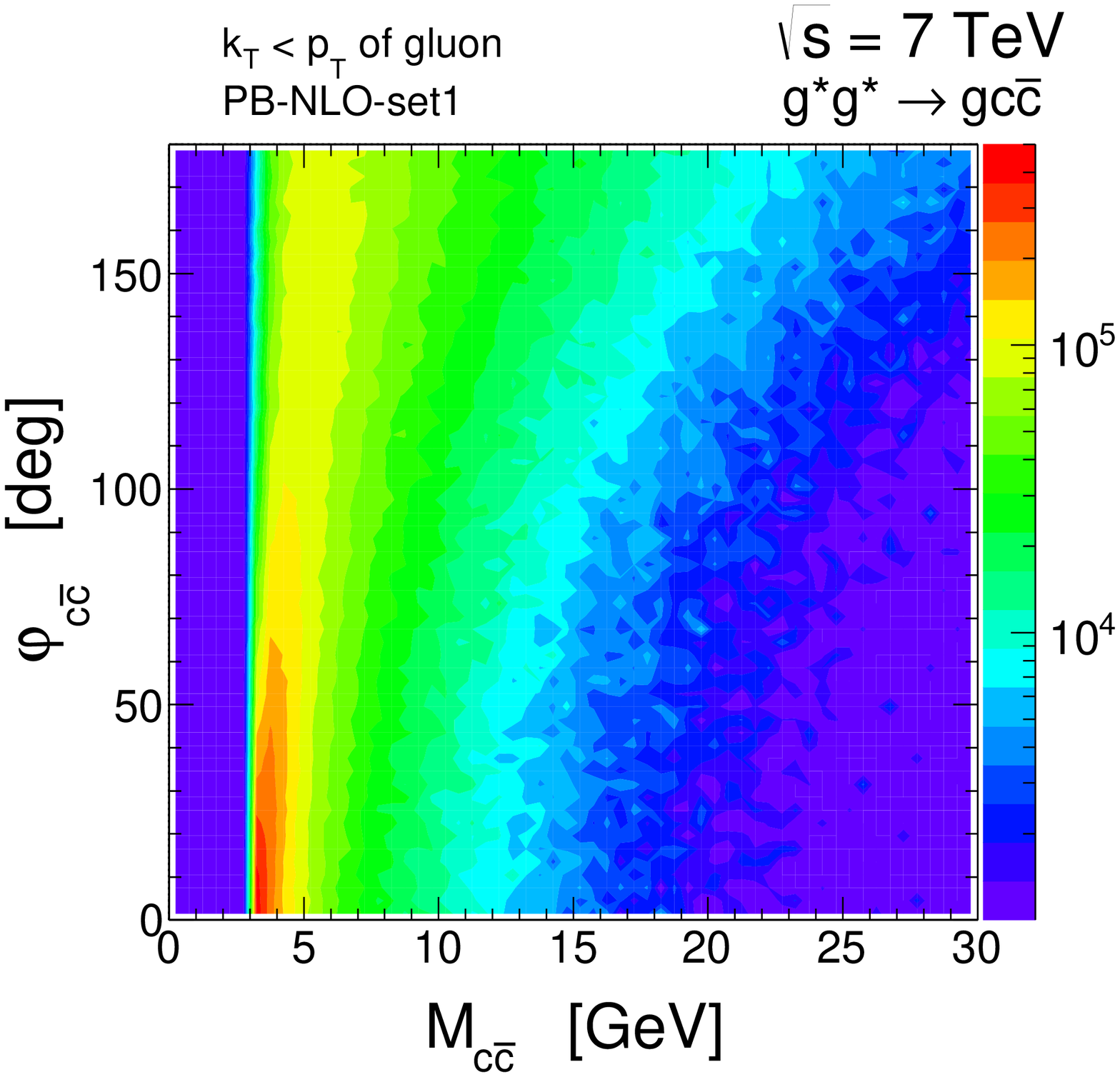}}
\end{minipage}
\begin{minipage}{0.32\textwidth}
 \centerline{\includegraphics[width=1.0\textwidth]{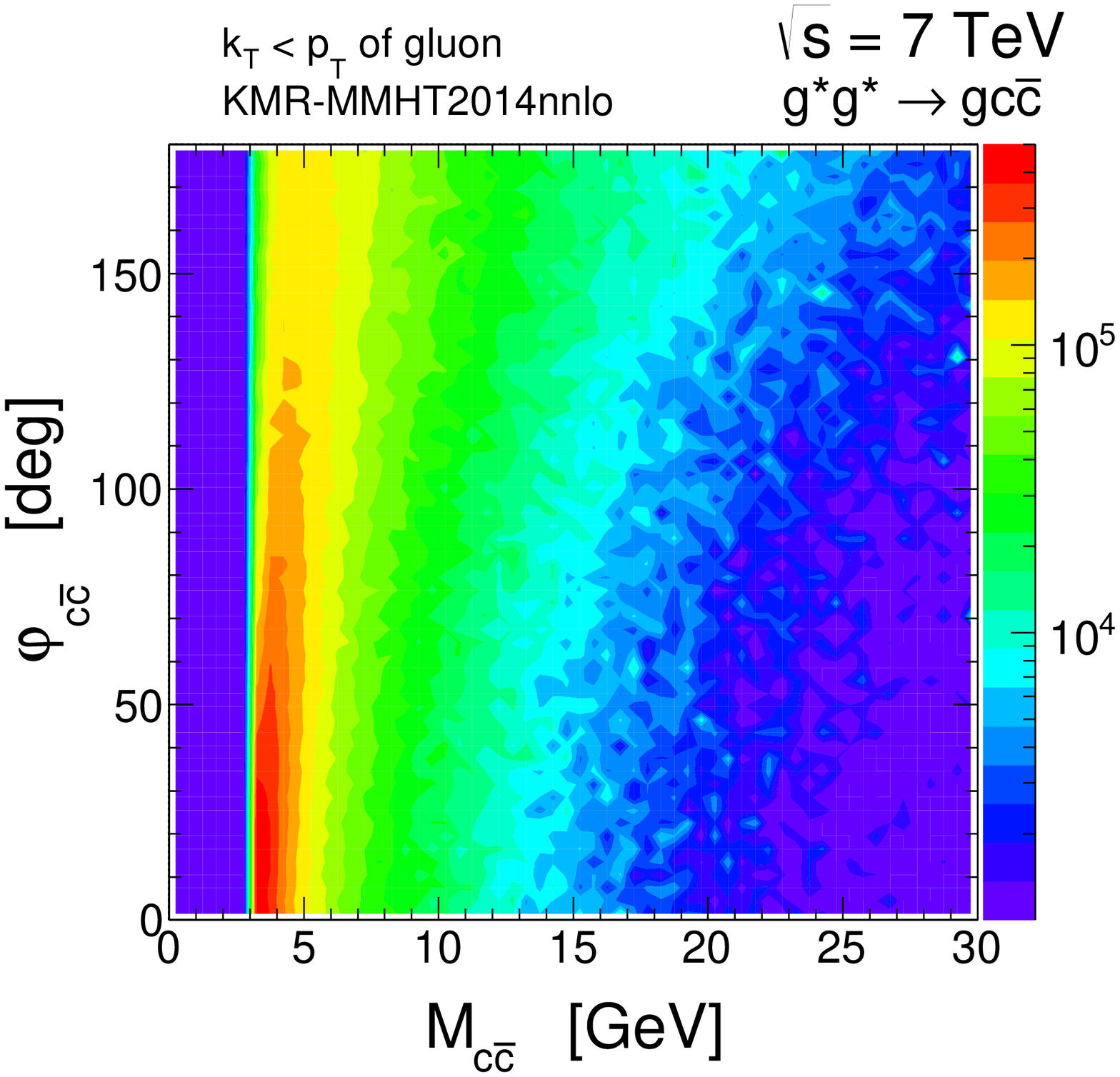}}
\end{minipage}
\begin{minipage}{0.32\textwidth}
  \centerline{\includegraphics[width=1.0\textwidth]{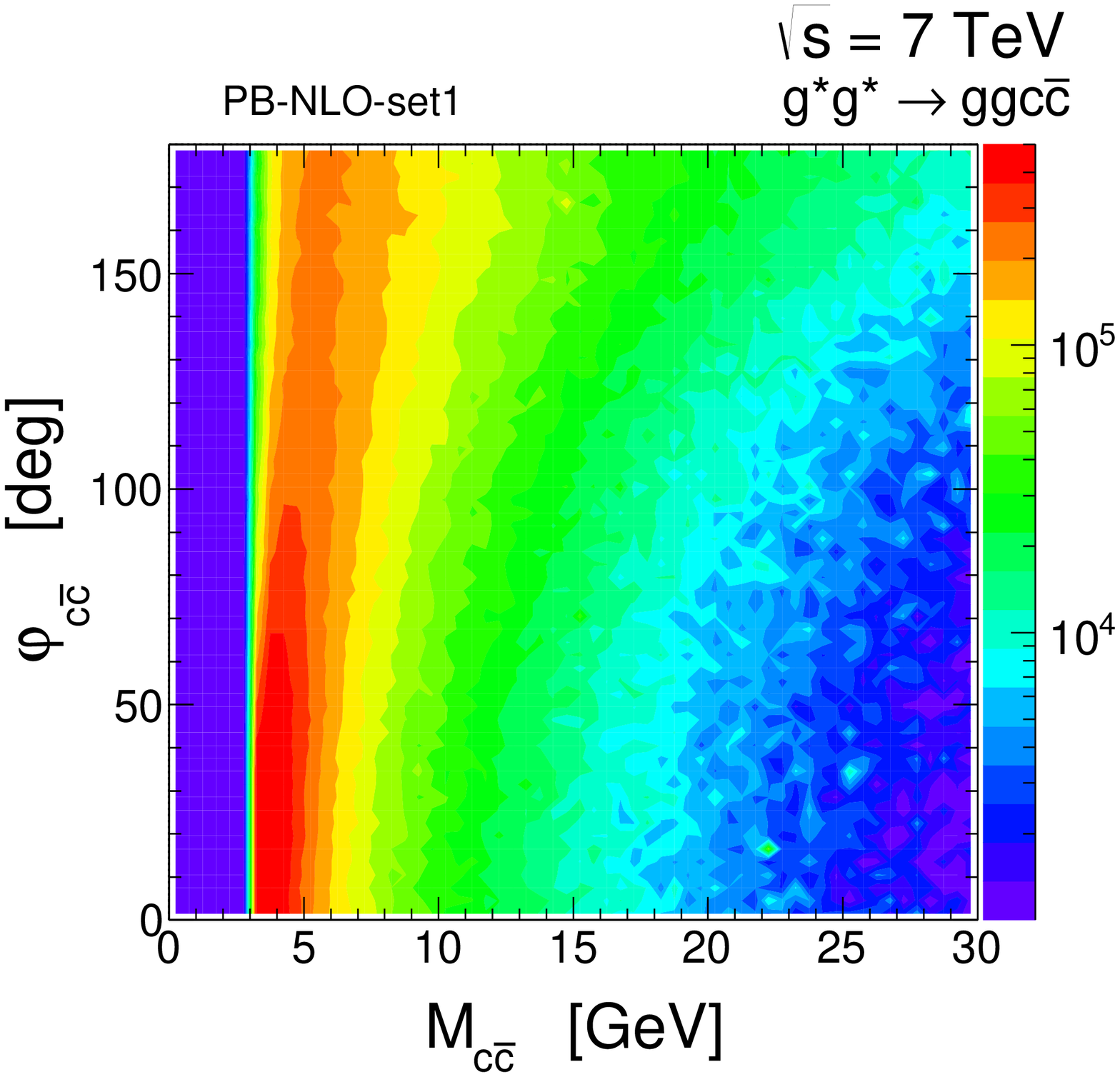}}
\end{minipage}
\begin{minipage}{0.32\textwidth}
 \centerline{\includegraphics[width=1.0\textwidth]{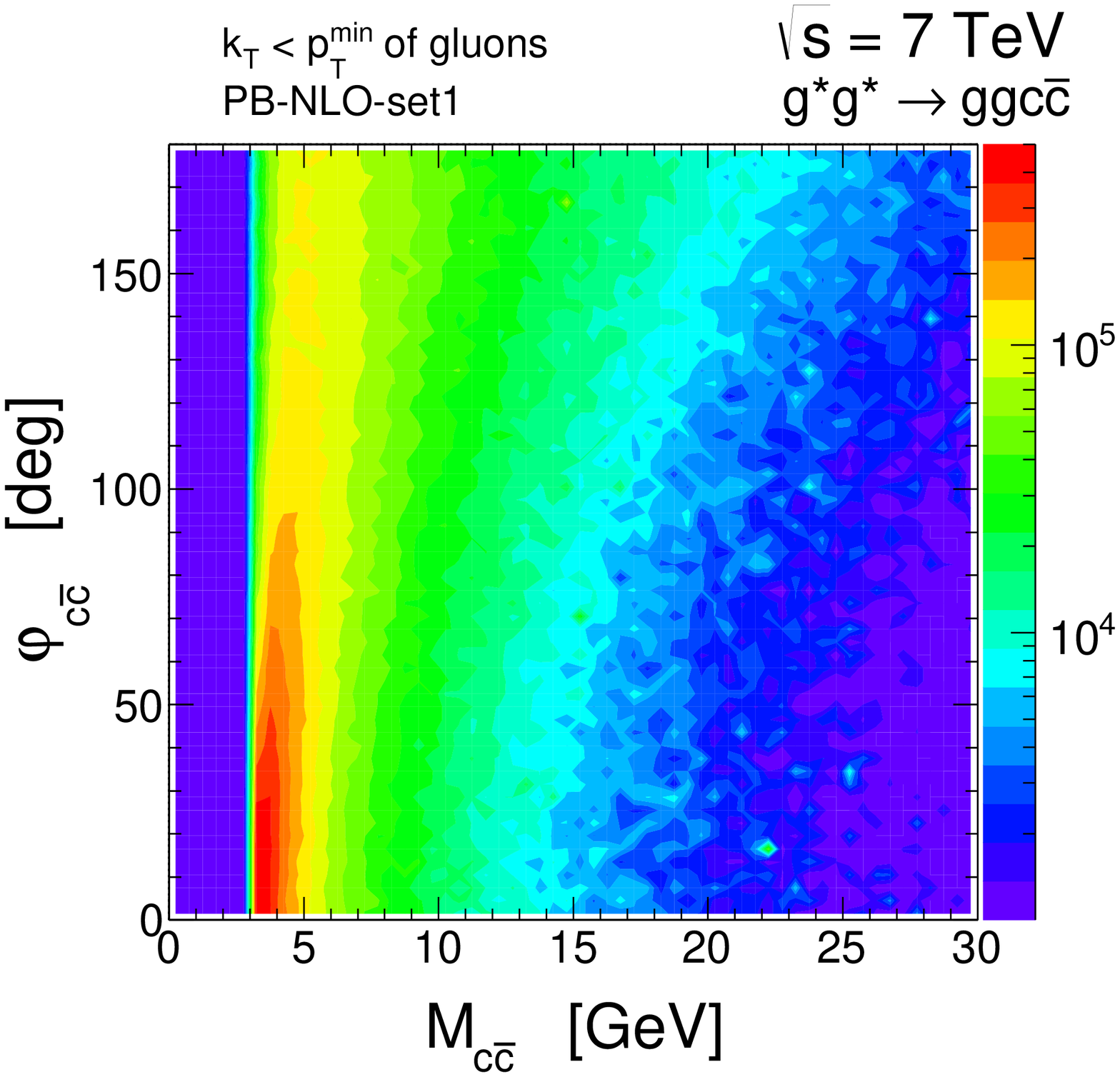}}
\end{minipage}
\begin{minipage}{0.32\textwidth}
 \centerline{\includegraphics[width=1.0\textwidth]{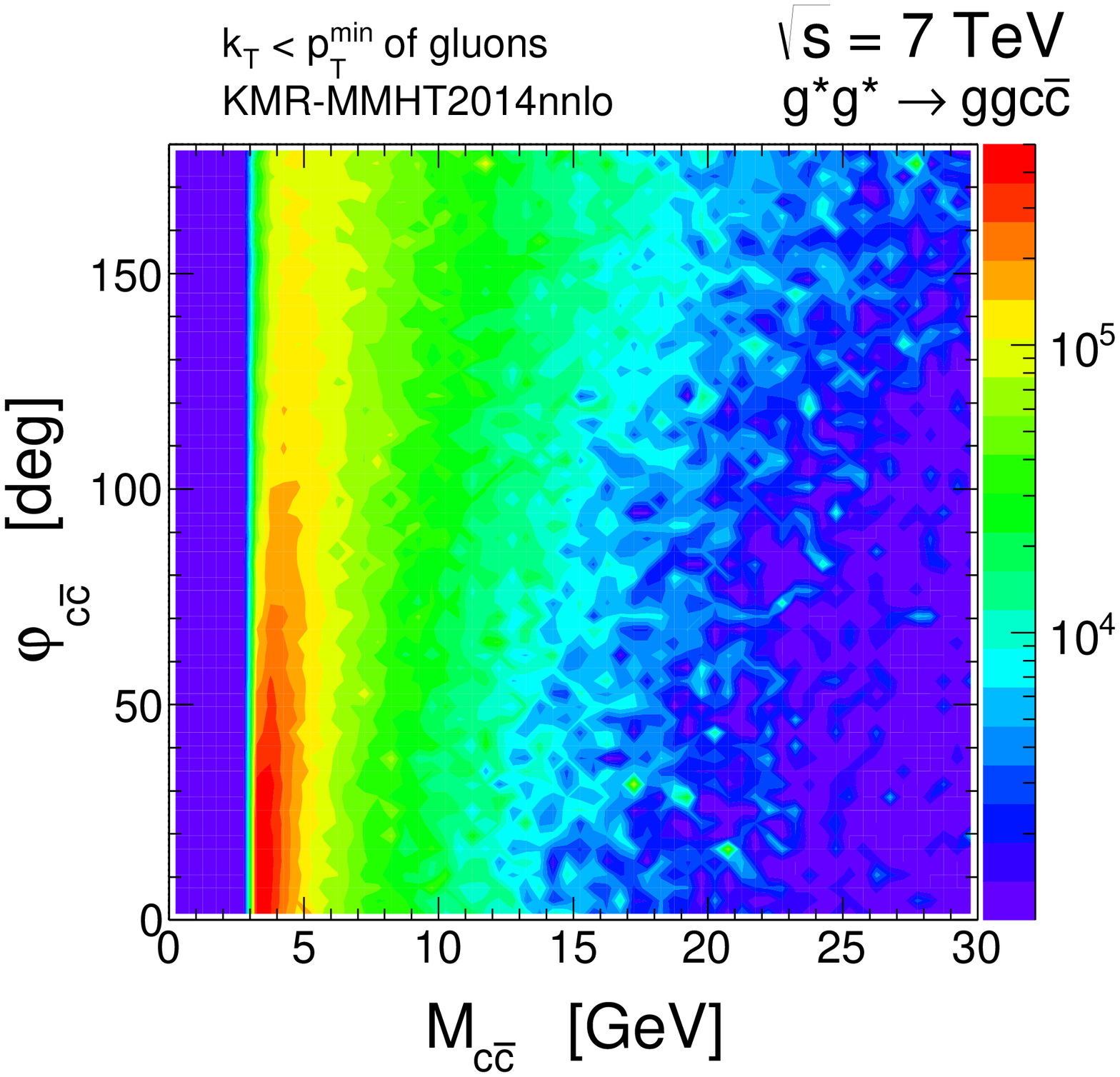}}
\end{minipage}
  \caption{
\small Correlations of the $M_{c \bar c}$ and $\phi_{c \bar c}$ in the case of the $2\to 3$ (top) and $2\to4$ (bottom) mechanisms.
The left and middle panels correspond to the calculations for the PB-NLO-set1 uPDF with and without the DCE cuts, respectively.
The right panel corresponds to the calculations for the KMR-MMHT2014nnlo uPDF with the DCE cuts.
}
\label{fig:2-2dim}
\end{figure}

What about higher orders? In Fig.~\ref{fig:2-2dim} we show
similar distributions for $2 \to 3$ (upper panels)
and $2 \to 4$ (bottom panels). We start with the PB-NLO-set1 distributions (left and middle panels). 
We see that the removed (for the KMR) regions 
reappear in the higher-order corrections. The left panels are
results of direct calculation, whereas the middle panels
include the DCE cuts. The direct calculations (left panels) lead 
to a significant contributions for back-to-back configurations already 
included in the $2 \to 2$ processes. We observe that our DCE cuts allow to avoid double
counting. In the right panels we show for comparison results with the DCE
cuts but for the KMR uPDF.

We see that our DCE cut fulfills the necessary requirements
supporting its practical applicability.

\subsubsection{Comparison to the calculations based on the collinear approach}

\begin{figure}[!h]
\centering
\begin{minipage}{0.47\textwidth}
  \centerline{\includegraphics[width=1.0\textwidth]{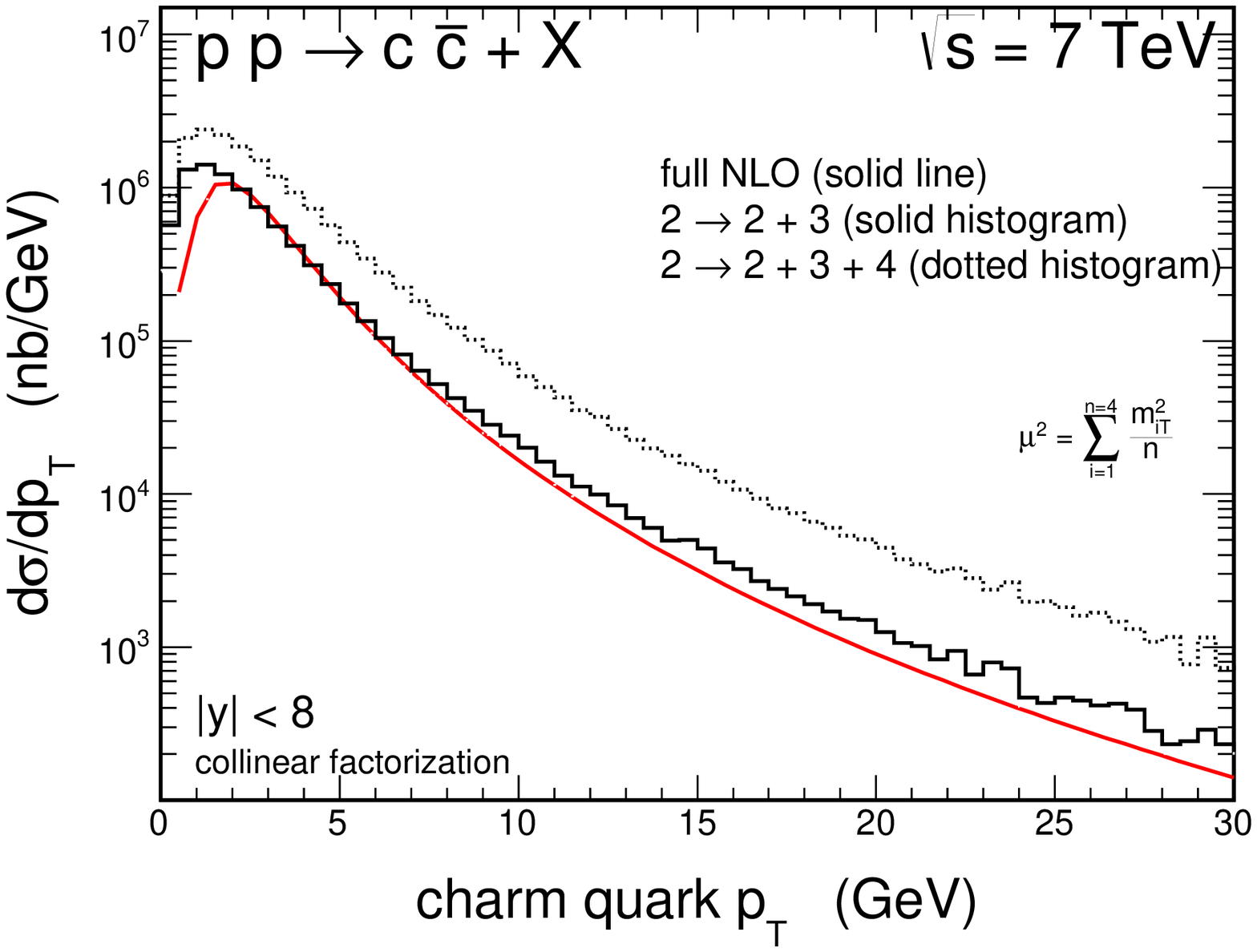}}
\end{minipage}
\begin{minipage}{0.47\textwidth}
 \centerline{\includegraphics[width=1.0\textwidth]{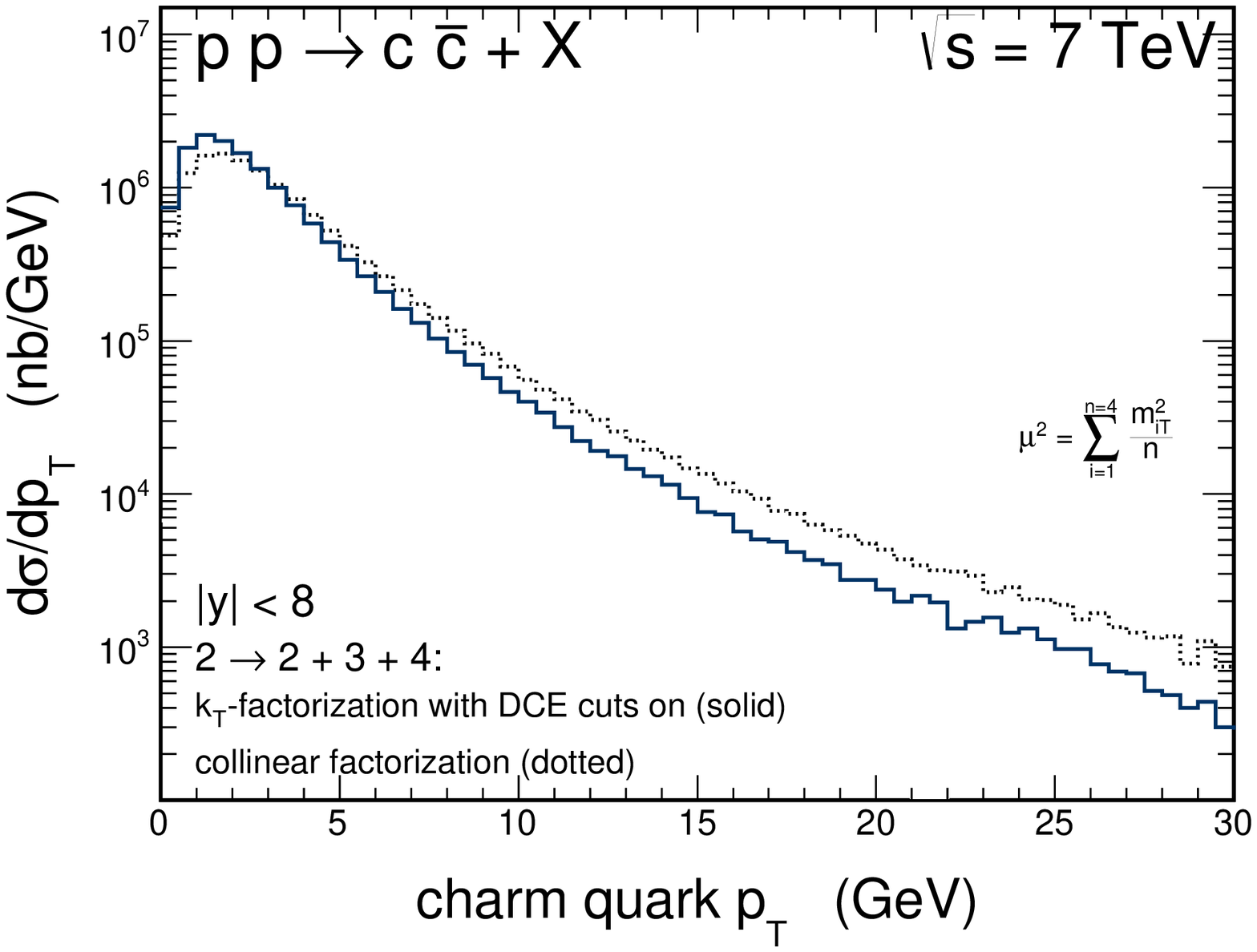}}
\end{minipage}
  \caption{
\small The transverse momentum distributions of charm quarks for $\sqrt{s} = 7$ TeV. Left: comaprison of the full NLO (solid line),  $2\to 2+3$ (solid histogram), and $2\to 2+3+4$ (dotted histogram) calculations in the collinear approach. Right: comparison of the $2\to 2+3+4$ calculations in the collinear (dotted histogram)
and $k_{T}$-factorization (solid histogram) tree-level approach.
}
\label{fig:NLO}
\end{figure}

Here, we wish to demonstrate how the $2\to 2 +3$ calculations at the tree-level with the DCE cuts differs from the full NLO approach.
This can be done only in the collinear approximation because of the lack of the NLO $k_{T}$-factorization framework.
In the left panel of Fig.~\ref{fig:NLO} we compare results for the transverse momentum distribution of charm quark
obtained in the full NLO framework (solid line) and calculated by summing the $2\to 2$ and $2\to 3$ contributions at the tree-level (solid histogram).
The two approaches almost coincide in the broad range of considered $p_{T}$'s. Significant differences appear only at very small transverse momenta
where the effects related to the loop-corrections are expected to be of special importance. Here, we plot in addition the $2\to 2+3+4$ contribution (dotted histogram). We observe a huge contribution of the NNLO-type to the transverse momentum distribution of charm quarks. It is not taken into account in the state-of-art calculations of the FONLL and GM-VFNS frameworks. Here, within our more pragmatic model we only wish to pay attention to the importance of the NNLO corrections to differential distributions of heavy quarks. Definite conclusions about their size are strongly limited since the NNLO collinear framework is not available for differential distributions of charm and bottom quarks.  

For completeness, in the right panel of Fig.~\ref{fig:NLO} we compare the $2\to 2+3+4$ contributions calculated in both, the collinear and $k_{T}$-factorization tree-level approach with the DCE cuts.

\subsection{The \bm{$D$}-meson cross section with the Parton-Branching uPDF, beyond the leading-order}

\begin{figure}[!h]
\centering
\begin{minipage}{0.47\textwidth}
  \centerline{\includegraphics[width=1.0\textwidth]{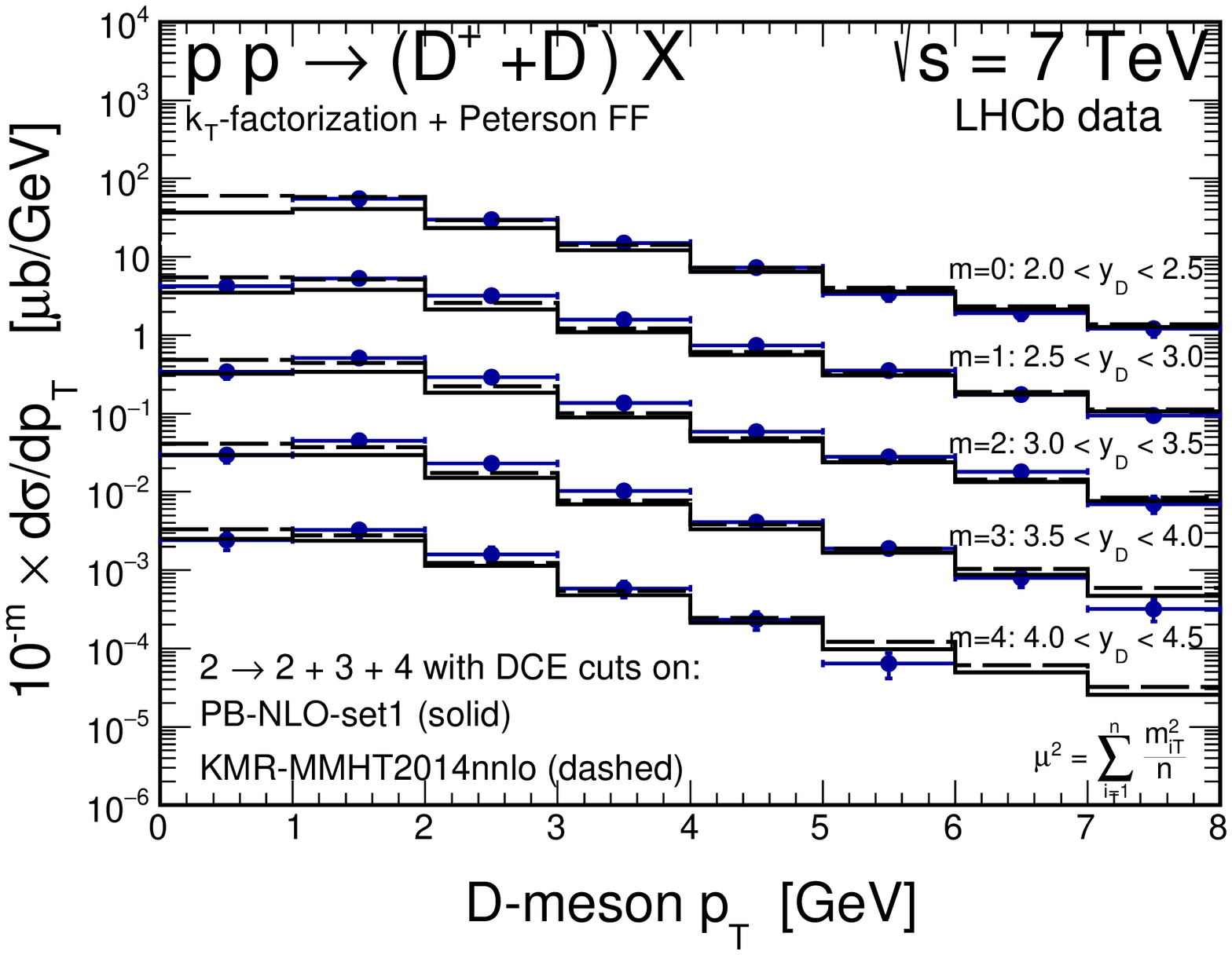}}
\end{minipage}
\begin{minipage}{0.47\textwidth}
 \centerline{\includegraphics[width=1.0\textwidth]{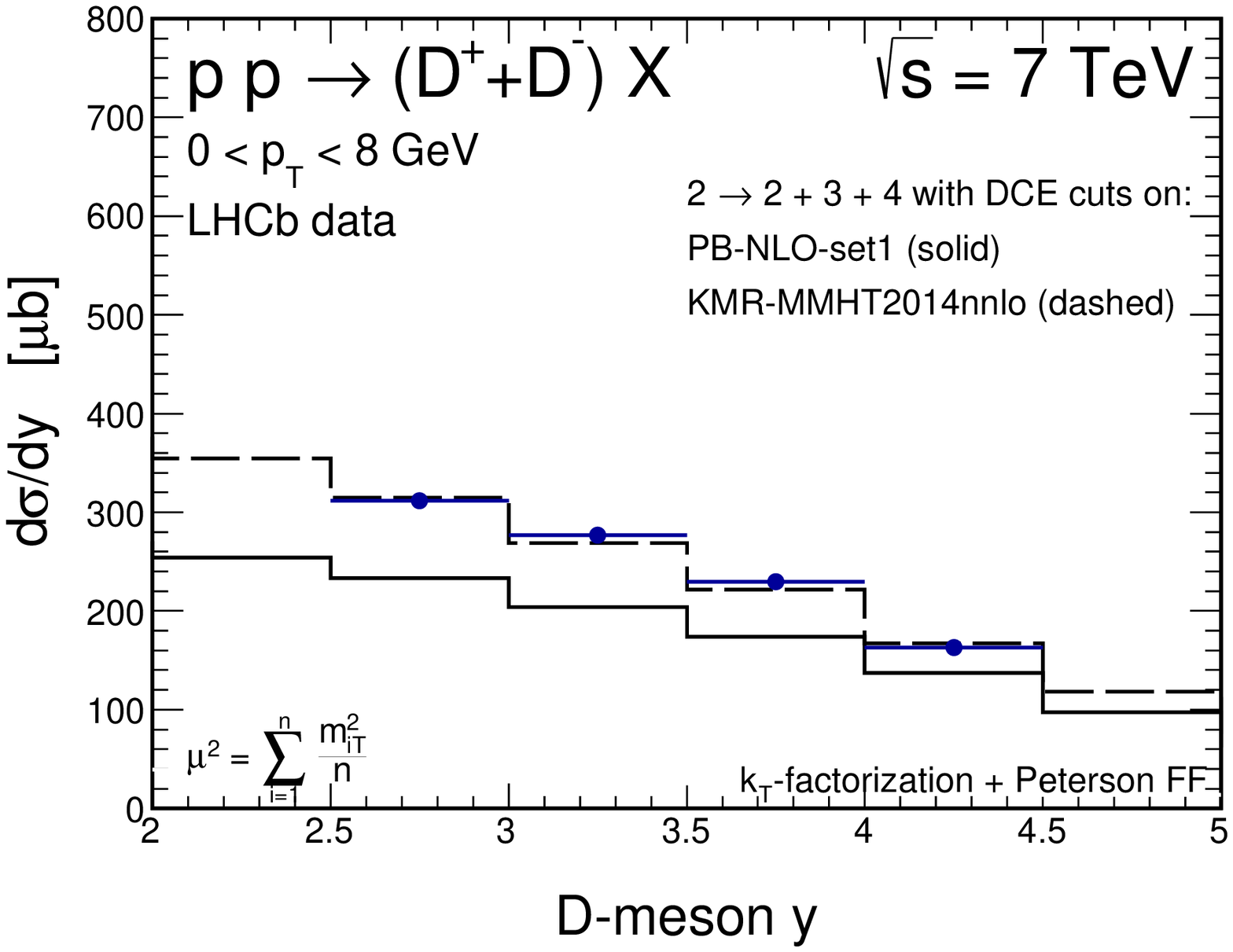}}
\end{minipage}
  \caption{
\small Transverse momentum (left) and rapidity (right) distributions of charged $D$-meson for $\sqrt{s}= 7$ TeV together with the LHCb data \cite{Aaij:2013mga}. Here, we show the PB-NLO-set1 and the KMR-MMHT2014nnlo results for summed contributions of $g^*g^* \to c\bar c$, $g^*g^* \to gc\bar c$ and $g^*g^* \to ggc\bar c$ mechanisms with the extra conditions.
}
\label{fig:13}
\end{figure}

\begin{figure}[!h]
\centering
\begin{minipage}{0.47\textwidth}
  \centerline{\includegraphics[width=1.0\textwidth]{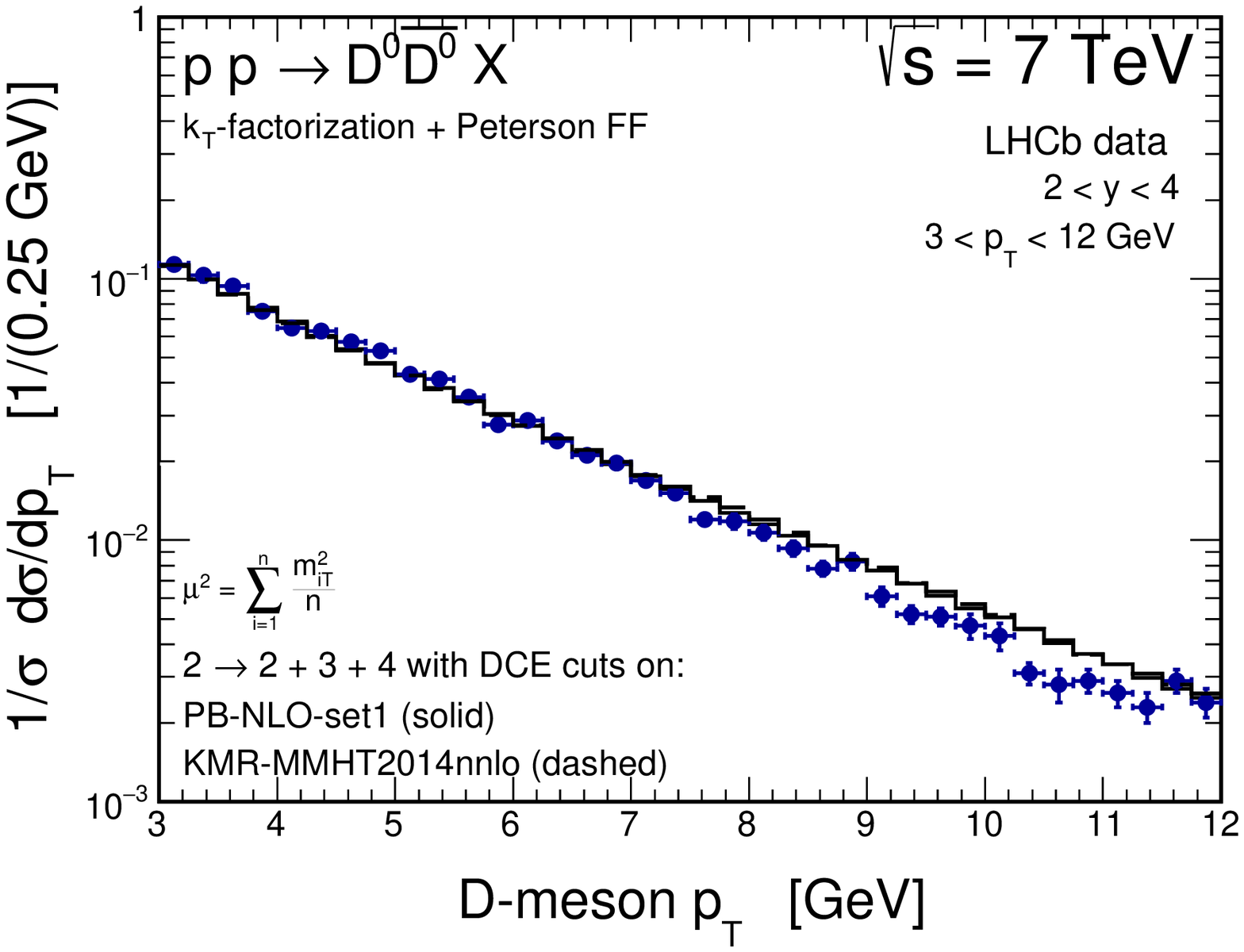}}
\end{minipage}
\begin{minipage}{0.47\textwidth}
 \centerline{\includegraphics[width=1.0\textwidth]{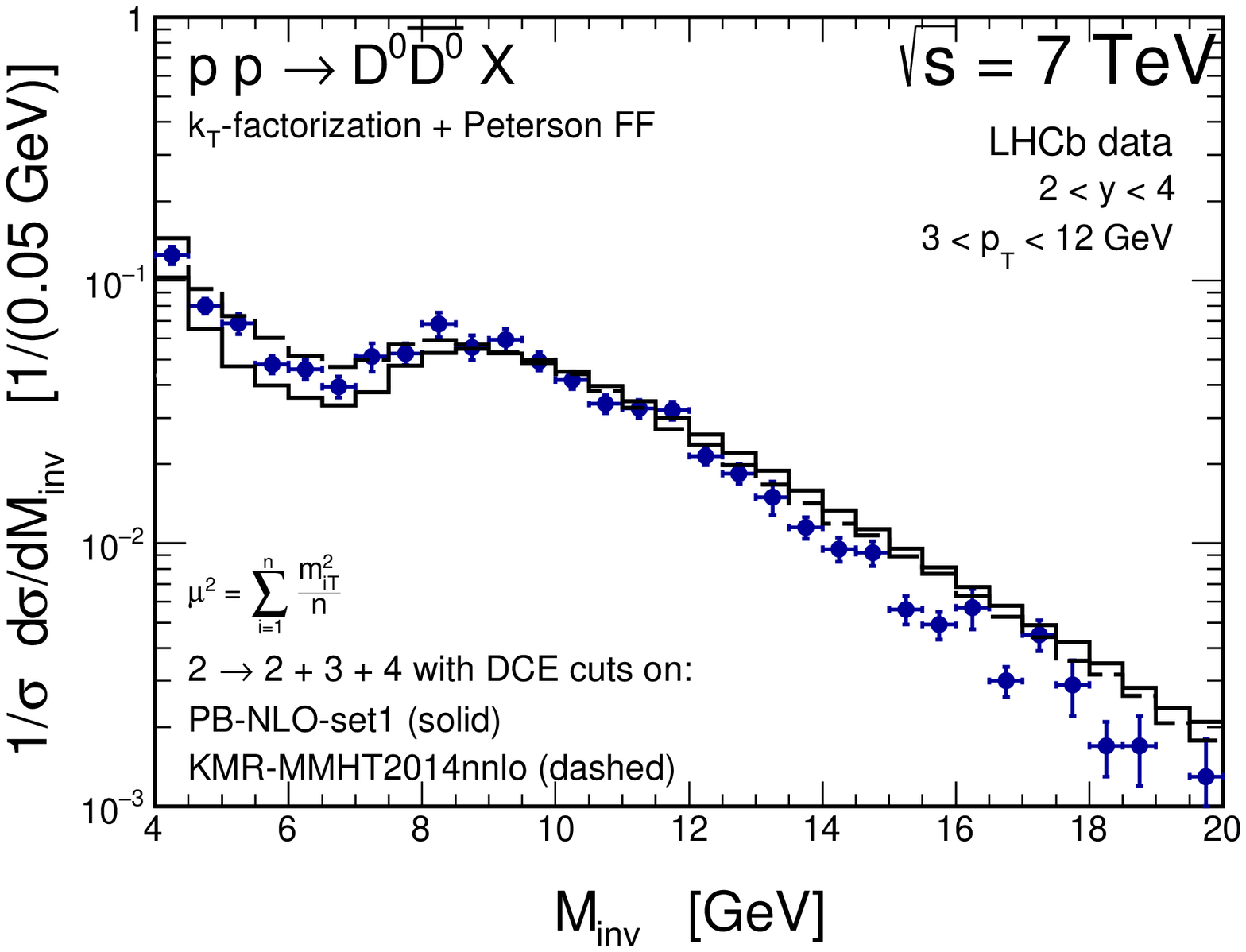}}
\end{minipage}\\
\begin{minipage}{0.47\textwidth}
  \centerline{\includegraphics[width=1.0\textwidth]{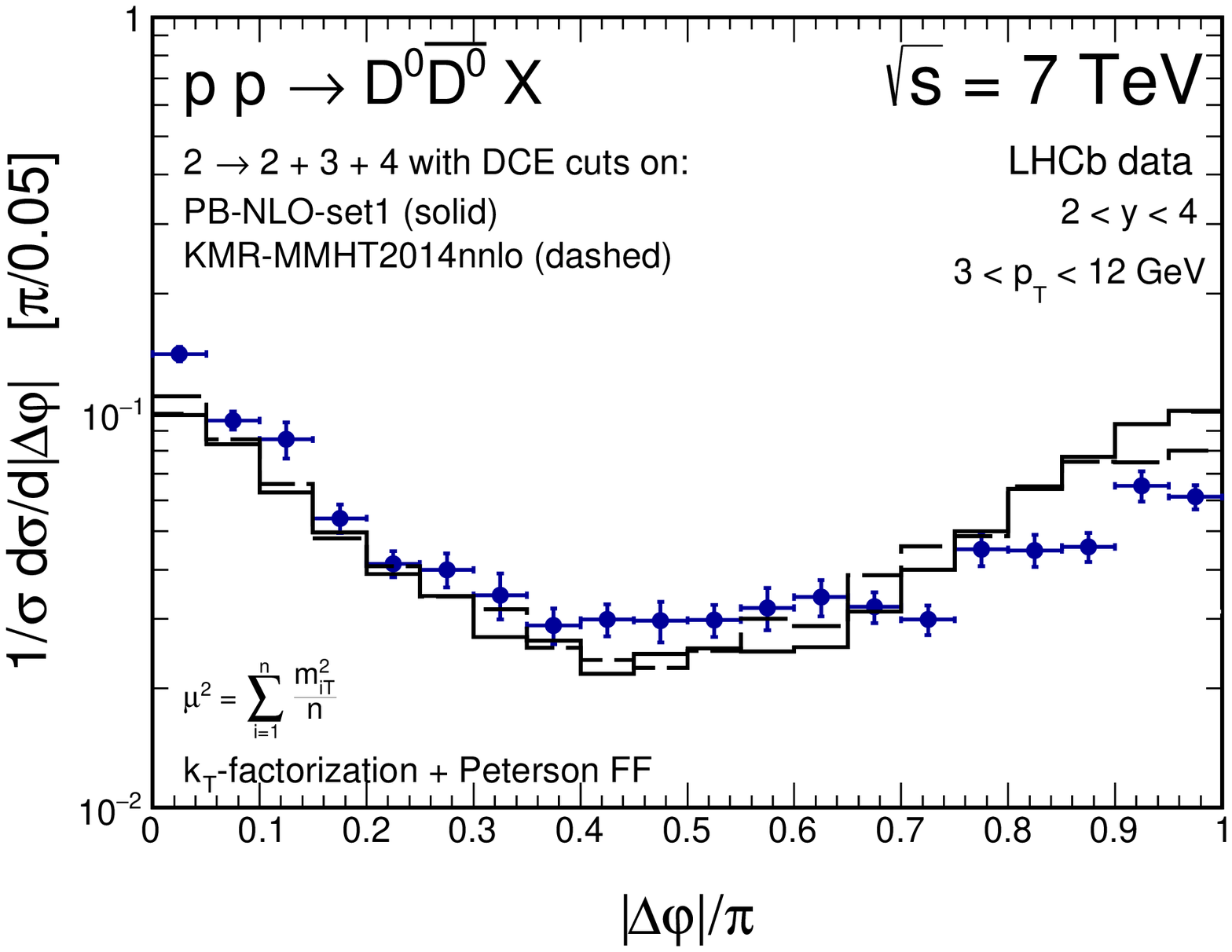}}
\end{minipage}
\begin{minipage}{0.47\textwidth}
 \centerline{\includegraphics[width=1.0\textwidth]{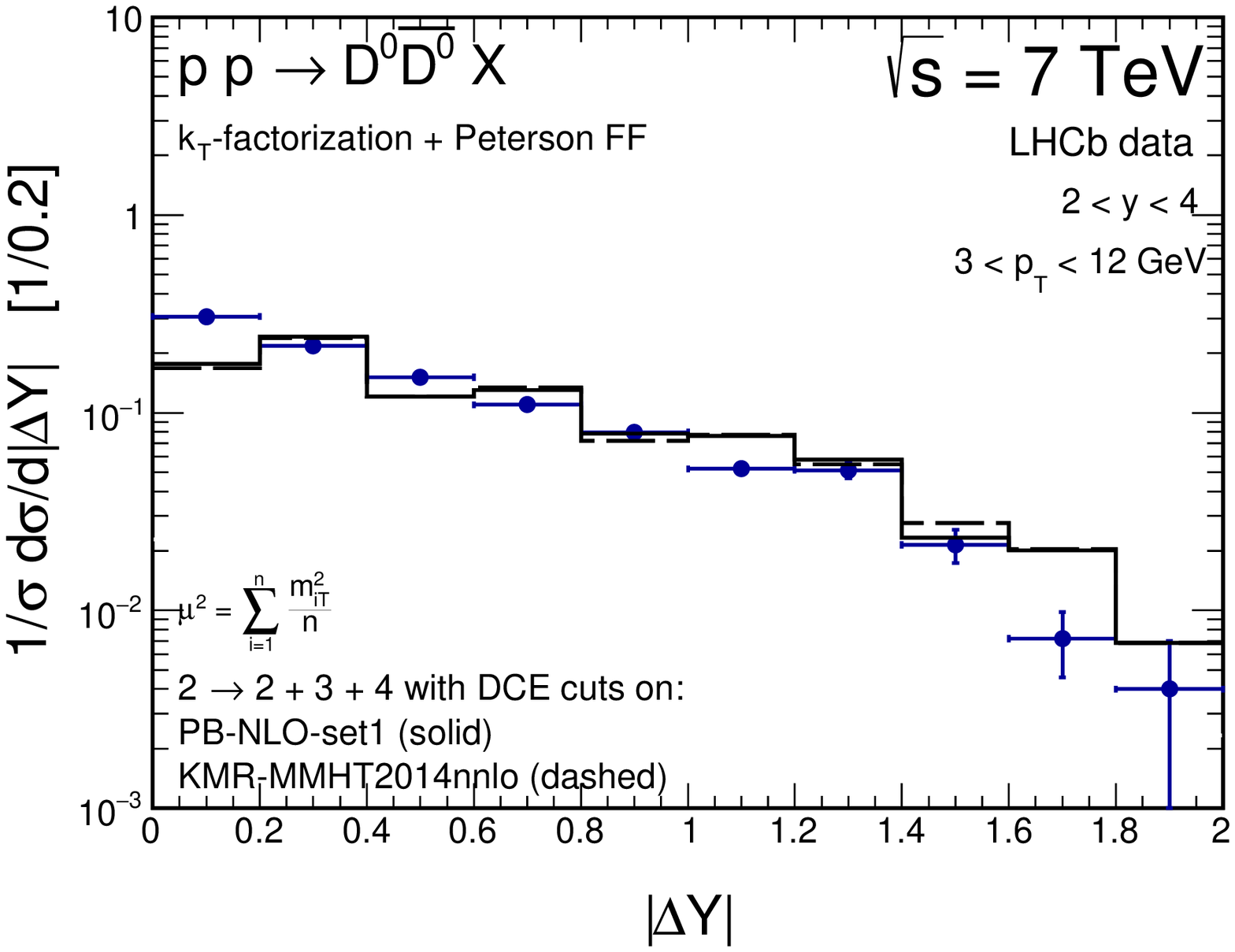}}
\end{minipage}
  \caption{
\small Transverse momentum (top-left), invariant mass (top-right), azimuthal angle (bottom-left) and rapidity distance (bottom-right) distributions for charged $D \bar D$ meson-antimeson pair production at $\sqrt{s}= 7$ TeV together with the LHCb data \cite{Aaij:2012dz}. Here, we show results for the PB-NLO-set1 and KMR-MMHT2014nnlo results for summed contributions from $g^*g^* \to c\bar c$, $g^*g^* \to gc\bar c$ and $g^*g^* \to ggc\bar c$ mechanisms with the extra conditions.
}
\label{fig:14}
\end{figure}

Finally, we wish to verify the results obtained in our $2\to 2+3+4$ scheme with both uPDFs 
against the LHCb open charm data. In Figs.~\ref{fig:13} and \ref{fig:14} we compare our results with inclusive $D$-meson and $D\bar D$ correlation LHCb data, respectively. The solid histograms correspond to the PB uPDF while the dashed ones to the KMR-MMHT2014nnlo uPDFs. In both cases we get the description of the experimental data of the same quality as in the case of the standard $2\to 2$ $k_{T}$-factorization calculations with the KMR uPDF. 

\section{Conclusions}

In the present paper we have considered charm production at the LHC within the $k_{T}$-factorization approach beyond the standard leading-order $g^*g^* \to c \bar c$ partonic mechanism. For the first time we have included in this context next-to- and next-to-next-to-leading order mechanisms for the differential distributions. 
We have proposed a new scheme for calculating the charm quark cross section including in addition the $2\to 3$ and $2\to 4$ higher-order contributions at the tree-level.
The calculations of the $g^*g^* \to g c \bar c$ and $g^*g^* \to g g c \bar c$ mechanisms have been done also in the framework of the $k_{T}$-factorization, with off-shell initial state partons, for two different unintegrated gluon densities from the literature -- Kimber-Martin-Ryskin and recent Parton Branching uPDFs. We have proposed special conditions in order to avoid the problem of double counting when calculating the higher-order corrections at the tree-level.

We have made a detailed comparison of the results for charm production obtained in the standard $2 \to 2$ calculations with the KMR uPDF and those from the proposed $2 \to 2+3+4$ scheme with the higher-order contributions taken into account at the tree-level. Both approaches were found to lead to very similar results. This conclusion applies exclusively for the KMR uPDF model. The analogous analysis have been done also for the PB-NLO-set1 uPDFs. In the latter case, the $2 \to 2$ calculations leads to a significant underestimation of the charm cross section at the LHC. Within this model of the gluon uPDF the experimental data can be described 
only in the $2 \to 2+3+4$ scheme, with higher-order contributions taken into account at the level of hard-matrix elements. This observations may be also valid for other models of the uPDFs from the literature, including different CCFM-fits, that do not allow for extra hard emissions encoded in their evolution.

Several differential distributions, including correlations observables, for open charm mesons for the LHCb experiment have been analyzed.   
Within the proposed $2 \to 2+3+4$ calculational scheme a good quality description of the data have been obtained for both the KMR and the Parton-Branching unintegrated gluon densities.

\section*{Acknowledgements}
We are indebted to Hannes Jung for
interesting discussions.
This study was partially
supported by the Polish National Science Center grant
DEC-2014/15/B/ST2/02528, by the Center for Innovation and
Transfer of Natural Sciences and Engineering Knowledge in
Rzesz{\'o}w. 


\end{document}